\PassOptionsToPackage{unicode}{hyperref}
\PassOptionsToPackage{hyphens}{url}
\documentclass[11pt]{article}
\usepackage[a4paper,margin=22mm]{geometry}
\usepackage{xcolor,amsmath,amssymb,fontspec,unicode-math}
\defaultfontfeatures{Ligatures=TeX}
\newfontfamily\symbolfont{DejaVuSans.ttf}

\usepackage{microtype,longtable,booktabs,array,caption,calc,etoolbox,enumitem,needspace,fancyhdr,ragged2e}
\usepackage{hyperref}
\IfFileExists{xurl.sty}{\usepackage{xurl}}{}
\hypersetup{hidelinks,breaklinks=true,pdfcreator={XeLaTeX},pdftitle={D-PIN SCU: From Nominal Capacity to Deliverable Compute}}
\providecommand{\tightlist}{\setlength{\itemsep}{1.5pt}\setlength{\parskip}{0pt}}
\setlist[enumerate]{leftmargin=1.7em,label=\arabic*.,itemsep=2pt,topsep=4pt,parsep=1pt}
\usepackage[english]{babel}
\usepackage{titlesec}
\titleformat{\section}{\fontsize{13}{16}\selectfont\bfseries}{}{0pt}{}
\titleformat{\subsection}{\fontsize{11.5}{14}\selectfont\bfseries}{}{0pt}{}
\titleformat{\subsubsection}{\normalsize\bfseries}{}{0pt}{}
\titleformat{\paragraph}[block]{\normalsize\bfseries}{}{0pt}{}
\titlespacing*{\section}{0pt}{13pt}{6pt}
\titlespacing*{\subsection}{0pt}{10pt}{4pt}
\titlespacing*{\subsubsection}{0pt}{8pt}{3pt}
\titlespacing*{\paragraph}{0pt}{6pt}{3pt}
\pretocmd{\section}{\Needspace{5\baselineskip}}{}{}
\pretocmd{\subsection}{\Needspace{4\baselineskip}}{}{}
\pretocmd{\subsubsection}{\Needspace{3\baselineskip}}{}{}
\pretocmd{\paragraph}{\Needspace{3\baselineskip}}{}{}
\allowdisplaybreaks[1]
\BeforeBeginEnvironment{longtable}{\par\begingroup\fontsize{9.5}{11.5}\selectfont\setlength{\tabcolsep}{2.8pt}\renewcommand{\arraystretch}{1.12}}
\AfterEndEnvironment{longtable}{\endgroup\par}
\makeatletter
\patchcmd\longtable{\par}{\if@noskipsec\mbox{}\fi\par}{}{}
\makeatother

\author{}\date{}
\begin{document}

\thispagestyle{plain} 

\begin{center}\setlength{\parskip}{0pt}
{\fontsize{18}{21}\selectfont\bfseries From Nominal Capacity to Deliverable Compute: Design, Calibration and
Validation of the D-PIN SCU Dynamic Metering System\par}
\vspace{5pt}
{\fontsize{11}{13}\selectfont D-PIN Standardized Compute Units for Distributed Het\-ero\-ge\-neous AI
Compute Networks\par}
\vspace{6pt}
{\small
\textbf{Authors}: Cheng Li, Jiexiong Liu, Shikun Wang, Feiyu Ren\par\vspace{1pt}
\textbf{Affiliation}: KunlunMeta Research, Shanghai, China\par}
\end{center}

\section{Abstract}\label{abstract}

\textmd{Context.} Cross-architecture comparability of heterogeneous
compute has become the central bottleneck for the market-oriented
operation of computing power networks. Two largely disconnected
technical routes exist today: (i) \emph{statistical and conversion
conventions}, represented by GB/T 47514---2026 and YD/T 6693---2026,
which serve national-scale aggregation and grid interconnection; and
(ii) \emph{accounting and pricing units}, represented by AWS ECU, Azure
ACU, SLURM TRESBillingWeights, and the standardized compute unit (SCU)
and ``One-Degree Compute'' (DCU), which serve quota management and
contractual settlement. We note that the term SCU already carries
several semantically distinct public uses in this field, whose work line
dates back to 2024; our scope is therefore defined in layers (see the
originality statement and Appendix C.3), and our own unit is named the
\textmd{D-PIN SCU} to distinguish it from these homonyms (Sections 2.2
and C.3). Both rest on the same premise: what is measured is the
\emph{nominal capacity} promised by a chip's datasheet.

\textmd{Problem.} When compute networks shift from centralized
datacenters to distributed physical infrastructure --- building-level
edge nodes, heterogeneous co-hosting, compute--electricity co-scheduling
--- this premise breaks down. Three systematic gaps separate nominal
capacity from deliverable compute: \emph{topology gaps} (the
communication ridge imposed by multi-chip interconnect), \emph{runtime
gaps} (thermal throttling, fault degradation, occupancy), and
\emph{behavioural gaps} (over-claiming, location spoofing). Static units
cannot observe any of them, and therefore cannot support real-time
scheduling or a verifiable, auditable distributed market.

\textmd{Method.} We formalize the \textmd{D-PIN SCU dynamic metering
system} (dynamic, topology-aware, runtime-corrected). (1) Compute
resources are defined by a five-dimensional vector \((F,\allowbreak  M,\allowbreak  B,\allowbreak  I,\allowbreak  E)\);
the normalization baseline is defined as a \textmd{method rather than a
specific hardware model} --- fixed by a spec\-i\-fi\-ca\-tion of dimensional
conventions, value-sourcing rules and re-anchoring rules --- with H100
SXM serving only as its current \textmd{realization}. We prove
\textmd{origin invariance} (uniform re-anchoring leaves all relative
metering results unchanged), so the baseline instance is replaceable
without loss of comparability. (2) A \textmd{topology-aware factor}
\(\eta_{\text{topo}}\) is derived from a cluster-scale Roofline model,
with a closed-form communication ridge \(\text{COI}^{*}\) and an
empirical calibration procedure. (3) \textmd{Runtime corrections}
\(\delta_T\) (thermal), \(\delta_F\) (fault) and \(\delta_L\) (load) are
modelled temporally; we prove that metering bias under nominal
spec\-i\-fi\-ca\-tions is unobservable (Proposition 2). (4) A
simplex-constrained \textmd{weight calibration algorithm} with
exponentiated-gradient (entropy mirror-descent) updates is proposed,
together with an identifiability analysis under dimensional
collinearity. (5) D-PIN SCU declarations are coupled to a three-layer
Proof-of-Physical-Work (PoPW), yielding a \textmd{no-arbitrage
condition} for node over-claiming. (6) Unit price is coupled to
electricity cost, green-energy ratio and reputation, closing the
compute--electricity loop.

\textmd{Validation.} An end-to-end recomputation is performed on the
Zhangjiang Science Gate T1/T2 production nodes (ten heterogeneous 8-GPU
servers --- four imported: 2×NVIDIA H20, 1×AMD MI300X, 1×AMD MI308X; six
domestic: 2×Muxi C500, 4×Ascend-910B --- 59.5 D-PIN SCU in total, 180 kW
supply-and-distribution design capacity, \ensuremath{\approx}54.5 kW measured at the IT
boundary). The billed throughput of 33,000 D-PIN SCU·hours is
self-consistent with the \textmd{Physical Upper Bound Principle}
(\(59.5 \times 720 \times 78\%\)), and two new recomputable metrics are
derived: \textmd{3.02 kW/D-PIN SCU} and \textmd{0.26 D-PIN SCU·h per kWh
delivered}. Appendix B discloses the chip baseline under \textmd{two
parallel conventions} --- the published value in an \emph{empirically
calibrated} convention (declared non-reproducible from vendor
datasheets) versus a \emph{datasheet-projected} convention --- which
diverge materially (H200 +10.5\%, MI300X +74.2\%, Ascend-910B −31.1\%,
Muxi C500 −30.9\%). Item-by-item verification in Appendix B.3 further
confirms the key inputs for MI300X and H200 and closes the key inputs of
Muxi C500 through vendor public channels, ruling out \emph{parameter
error} as the alternative explanation. We present this divergence as
direct evidence that static conventions are insufficient to meter
deliverable compute, and that an empirical calibration loop is a
necessary rather than optional component; minimum reproducibility (A6)
requirements are specified.

\textmd{Conclusion.} Compute does not lack units; it lacks a measure of
\emph{deliverable} compute. The contribution of D-PIN SCU is not ``yet
another standardized compute unit'', but the advancement of the metering
object from \emph{nominal capacity} to \emph{deliverable compute}, such
that a single quantity is simultaneously schedulable, verifiable, and
priceable --- a metering-theoretic construction integrating metering,
scheduling, verification and pricing. At the industry level, D-PIN SCU
can serve as a \emph{scheduling-level metering middle layer} between
statistical and accounting conventions, supporting grid interconnection,
distributed pricing, compute financialization, and compute--electricity
co-scheduling (Section 10.1). We make no claim to priority over the term
SCU, nor over the individual elements of normalized weighted metering;
our originality claim is confined to the integrated system --- runtime
identifiability, closed-form topology ridge, and verification--pricing
coupling --- taken as a whole.

\textmd{Keywords}: D-PIN standardized compute unit (D-PIN SCU);
heterogeneous compute metering; topology-aware; runtime correction;
proof of physical work; compute--electricity co-scheduling; computing
power network

\section{Statement on Data Conventions and
Originality}\label{statement-on-data-conventions-and-originality}

\textmd{Writing discipline of this paper}:

\textmd{1.} \textmd{No fabrication of quantitative data.} All capacity,
power, latency and utilization data come from D-PIN Zhangjiang node
trial-operation records (August 2026) and vendor spec\-i\-fi\-ca\-tions, with
sources identified individually.

\textmd{2.} \textmd{Derived values and illustrative calculations are
labelled separately.} Arithmetic results derived from existing data,
such as kW/D-PIN SCU and D-PIN SCU·h/kWh, are labelled ``derived
values''. Calculations involving prices, costs or investment are
labelled ``illustrative calculations'' and do not constitute
commitments.

\textmd{3.} \textmd{Unverified information is excluded.} Parameters,
standard identifiers and provisions without authoritative primary
sources are not presented as facts through approximation.

\textmd{4.} \textmd{Originality conventions (a layered statement revised
on 2026-10-05).} Originality is strictly distinguished at three levels,
and claims must not extend beyond their level:

\textmd{(a) Terminology: no claim to priority.} ``Standard /
Standardized Compute Unit'' (SCU) has \textmd{at least five publicly
documented uses with different meanings} in AI computing: the Standard
Compute Unit of Symmetric Research for financial collateral valuation
(normalizing heterogeneous GPU clusters and explicitly including
workload mix, multi-GPU scaling overhead and energy draw; its
spec\-i\-fi\-ca\-tion is machine-verified in Lean and its calibration is
anchored to public MLCommons benchmarks; the \textmd{underlying
mathematical line of work} appeared in arXiv:2406.19261 in 2024-06,
where the unit is Compute Hours (CH); the SCU term and four-part
spec\-i\-fi\-ca\-tion appear on the institutional research page, so 2024 denotes
the \textmd{start of the line of work}) {[}33{]}; the CME / Silicon Data
SCU for compute futures (1 SCU = 1×H100 FP16·h, with other chips
converted using preset coefficients, 2026-05) {[}10{]}{[}34{]};
Kohlenberg's static SCU for contract clearing (a four-dimensional
weighted dot product, H100 SXM5 = 1.00, 2026-05) {[}5{]}; and the
\textmd{``Standard Compute Unit (SCU)'' of Tianyi Cloud (China Telecom)}
(1 SCU defined by the product of single-GPU compute capacity and
bandwidth for a reference GPU, with task-dependent dynamic conversion)
{[}49{]}. There are also \textmd{two uses of the related term NCU
(Normalized Compute Unit)} (Revenue Model {[}50{]} and COIL {[}51{]}).
\textmd{This paper claims no priority for the concept or terminology},
and recommends consistently using \textmd{D-PIN SCU} in external
communication (Appendix C.3).

\textmd{(b) Individual components: no claim to priority.}
\textmd{Components} including weighted normalization, energy-efficiency
dimensions, scenario-dependent weights, runtime metering, topology-aware
scheduling and metering-pricing coupling have precedents. The normalized
``information-energy-time'' score of ZTE Communications
\(S=w_1(I/I_{\max})+w_2(\eta/\eta_{\max})+w_3(T_{\min}/T)\) already
includes energy efficiency and scenario-dependent weights {[}35{]}.
DataCanvas's ``one degree of compute'' (DCU) uses \textmd{1 DCU = 312
TFLOPS(FP16) × 1 hour} to convert heterogeneous chip inputs and charges
for actual consumption {[}9{]}. Alibaba's OSDI 2026 production study
documents ``high GPU demand does not imply high effective utilization''
and GPU undeliverability (stranded GPUs, CPU bottlenecks and topology
constraints) {[}36{]}. \textmd{We claim no priority for any individual
component.}

\textmd{(c) Integrated system: a claim to the complete system and
theoretical contribution.} As of the search cutoff of 2026-10-05, no
identified work jointly treats the closed-form derivation of topology
efficiency (Eqs. (9)-(13)), stochastic modelling and identifiability of
three runtime factors (Proposition 2), economic binding of metering and
verification (the no-arbitrage condition, Eq. (25)), and coupling of
metering to electricity-market pricing signals (Eq. (28)) as
\textmd{inseparable parts of one metering unit}. \textmd{An explicit
qualification is necessary}: the complete SLR in Appendix D newly
identified two \textmd{formally adjacent} precedents, Tianyi Cloud's SCU
{[}49{]} (multidimensional weighting and task-dependent dynamic
conversion) and Agora {[}44{]} (runtime metering of three-dimensional
resources using hardware performance counters, with usage-based
pricing). These precedents \textmd{further narrow} our claim. Novelty is
based \textmd{only on the integration of all four components},
particularly the closed-form topology ridge and identifiability
argument, rather than on the \textmd{formal feature} of multidimensional
weighting with runtime or task-dependent conversion. The contribution is
this \textmd{complete dynamic metering system for distributed real-time
scheduling}, together with the theoretical \textmd{identifiability
proposition (Proposition 2)}: runtime deviations are unidentifiable from
nominal spec\-i\-fi\-ca\-tions, making runtime metering \textmd{necessary for
identifiability rather than merely an engineering optimization}.

\textmd{Discipline of expression}: the manuscript and accompanying
external materials \textmd{must not use ``first'', ``original'' or ``the
first''} to qualify SCU or a metering system named SCU. Use ``no
corresponding work was identified as of the search cutoff'', with the
comparison table and search-scope statement in Appendix C. The
originality claim is governed by the formal comparison in Appendix C.

\textmd{5.} \textmd{Terminological consistency.} Throughout the paper,
use D-PIN SCU (D-PIN Standardized Compute Unit), PoPW (Proof of Physical
Work), \(\eta_{\text{topo}}\) (topology efficiency factor), KunlunKVS
(heterogeneous distributed scheduling kernel), and GPT-Factory (AI Infra
operating system). Do not mix abbreviated forms.

\section{1 Introduction}\label{introduction}

\subsection{1.1 Background: compute as a measurable and tradable factor
of
production}\label{background-compute-as-a-measurable-and-tradable-factor-of-production}

Structural growth in AI compute demand has transformed computing from an
internal IT cost into a factor of production with financial attributes.
In May 2026, CME Group and Silicon Data announced compute futures
settled against GPU rental-price indices, using the \textmd{Standardized
Compute Unit (SCU)} for cross-chip conversion (1 SCU = one H100 running
FP16 for one hour; other chips use preset coefficients). The final
contract specifies \textmd{730 GPU-hours per contract}, the average
monthly full-load hours of one GPU (\(365/12\times24\approx730\)),
listed under NYMEX GPU1 / GPU2 (H100 and B200 rental indices,
respectively), cash-settled and pending regulatory approval {[}10{]}. On
2026-09-03, Nodal Exchange and Compute Desk announced a competing
compute-futures plan based on Compute Desk's IOSCO/BMR-compliant GPU
rental index, covering Hopper and Blackwell, with cash settlement, Nodal
Clear clearing and CFTC oversight {[}56{]}. In China,
\textmd{DataCanvas} developed ``one degree of compute'' (DCU; 1 DCU =
312 TFLOPS(FP16) × 1 hour), selected among the Ministry of Industry and
Information Technology's 2025 exemplary cases of coordinated
new-information-infrastructure development and already used for
consumption-based billing in its own and managed AI compute centres
{[}9{]}. Meanwhile, the compute network has been included among national
strategic infrastructure for the 15th Five-Year Plan;
compute-electricity coordination appears in the government work report,
establishing integration of compute and electricity dispatch as a policy
direction.

The technical prerequisite is that \textmd{compute must be measurable}.
Without a common, recomputable and trustworthy unit, there is no basis
for price discovery, risk hedging or settlement in distributed compute
markets.

\subsection{1.2 Problem: four discontinuities caused by a mismatch in
the metering
object}\label{problem-four-discontinuities-caused-by-a-mismatch-in-the-metering-object}

Existing metering is sufficient in \textmd{centralized clouds}: a single
operator manages resources, hardware models are relatively uniform, and
scheduling and billing share trusted telemetry. With \textmd{distributed
physical infrastructure}, where compute nodes occupy urban buildings,
heterogeneous vendor hardware is co-hosted, and electricity is a
first-class matching resource, four discontinuities change these
assumptions:

\textmd{Discontinuity 1: statistical conventions ≠ scheduling
conventions.} Statistical conventions (FP32 theoretical peaks converted
to EFLOPS, with random checks of electricity and traffic meters) serve
\textmd{aggregate-scale reporting}. Quar\-ter\-ly reporting differs from
millisecond-to-second scheduling by 7-10 orders of magnitude in update
frequency and temporal granularity. Sched\-ul\-ing from such statistics is
equivalent to managing live inventory with annual accounts.

\textmd{Discontinuity 2: accounting conventions ≠ delivery conventions.}
Accounting units (ECU, ACU, SLURM billing weights and standardized
compute units) take \textmd{nominal spec\-i\-fi\-ca\-tions} as input and measure
promised resources. Tasks consume \textmd{actual deliverable compute}:
interconnect topology determines the multi-GPU communication ridge,
thermal throttling and hardware degradation reduce available capacity,
and multitenant competition changes instantaneous availability. These
differences are systematic bias rather than noise.

\textmd{Discontinuity 3: capability declarations ≠ verifiable facts.} In
permissionless or weakly permissioned distributed environments, nodes
have incentives to overstate resources. Without continuing spot checks,
metering becomes \textmd{self-reported data}, whose credibility declines
as network openness increases.

\textmd{Discontinuity 4: compute metering ≠ economic settlement.}
Electricity accounts for one-third to one-half of compute costs. A unit
detached from real-time electricity prices, renewable-energy shares and
load conditions cannot respond to electricity-market signals.
Compute-electricity coordination then remains at the aggregate
scheduling level and cannot close an economic loop for individual tasks.

\subsection{1.3 Contributions}\label{contributions}

We study the D-PIN SCU metering system of D-PIN (Distributed Physical
Infrastructure Network for AI Compute) {[}28{]} and develop seven
contributions (C1-C7) addressing these discontinuities. They form
\textmd{a progressive chain rather than a parallel list}: determine what
to meter, how to construct the measurement, how to establish trust, and
how to interoperate. The four stages are \textmd{foundation \ensuremath{\rightarrow} accurate
measurement \ensuremath{\rightarrow} trust \ensuremath{\rightarrow} integration}. The table summarizes the claims,
differences from precedents and verification anchors.

{\def\LTcaptype{none} 
\begin{longtable}[]{@{}
  >{\RaggedRight\arraybackslash}p{(\linewidth - 8\tabcolsep) * \real{0.1200}}
  >{\RaggedRight\arraybackslash}p{(\linewidth - 8\tabcolsep) * \real{0.1327}}
  >{\RaggedRight\arraybackslash}p{(\linewidth - 8\tabcolsep) * \real{0.2889}}
  >{\RaggedRight\arraybackslash}p{(\linewidth - 8\tabcolsep) * \real{0.3258}}
  >{\RaggedRight\arraybackslash}p{(\linewidth - 8\tabcolsep) * \real{0.1327}}@{}}
\toprule\noalign{}
\begin{minipage}[b]{\linewidth}\RaggedRight
Layer
\end{minipage} & \begin{minipage}[b]{\linewidth}\RaggedRight
Contribution
\end{minipage} & \begin{minipage}[b]{\linewidth}\RaggedRight
Central claim
\end{minipage} & \begin{minipage}[b]{\linewidth}\RaggedRight
Difference from precedents
\end{minipage} & \begin{minipage}[b]{\linewidth}\RaggedRight
Anchor
\end{minipage} \\
\midrule\noalign{}
\endhead
\bottomrule\noalign{}
\endlastfoot
Foundation & \textmd{C0} Baseline as a spec\-i\-fi\-ca\-tion & Define
normalization as a reproducible \textmd{method} rather than hardware,
and prove \textmd{origin invariance}: replacement of the baseline
instance preserves all relative results & CME/Silicon Data SCU
{[}10{]}{[}34{]}, Kohlenberg SCU {[}5{]} and Tianyi Cloud SCU {[}49{]}
anchor to specific models, \textmd{without separating spec\-i\-fi\-ca\-tion from
realization} or establishing re-anchoring invariance & Proposition 3;
Eqs. (6)(7); Section 4.2 \\
Foundation & \textmd{C1} Redefining the object & Runtime deviations are
unidentifiable from static spec\-i\-fi\-ca\-tions, making runtime metering
\textmd{necessary for identifiability} & Static SCU, DCU and ECU/ACU
expressions have no time-varying terms and \textmd{cannot in principle}
estimate these deviations & Proposition 2; Eqs. (3)-(4); Section 3.3 \\
Mea\-sure\-ment & \textmd{C2} Closed-form topology awareness & Derive
topology efficiency from the Roofline ridge in a form that supports
derivation, calibration and extrapolation & Symmetric Research SCU
declares multi-GPU overhead but gives no public analytic form;
scheduling research keeps corrections internal rather than defining them
in the unit & Eqs. (9)-(13); Section 4.3 \\
Mea\-sure\-ment & \textmd{C3} Runtime correction and separate conventions &
The \textmd{Physical Upper Bound Principle} constrains billed
quantities; multiplexing cannot exceed it & No identified precedent
states the physical billing ceiling as a recomputable hard constraint &
Proposition 1; Eqs. (5)(16)-(20) \\
Mea\-sure\-ment & \textmd{C4} Calibratable weights & Calibrate weights using
simplex-constrained exponentiated gradient (entropy mirror descent),
with mandatory uncertainty reporting & Prior weights are generally
empirical or have undisclosed, non-reproducible calibration & Eqs.
(21)-(22); Algorithm 1; Section 5 \\
Trust & \textmd{C5} Metering-verification coupling & Bind spot-check
frequency to the present value of over-claiming gains through no
arbitrage, decoupling verification overhead from scale & Existing
verification addresses correctness (ZK/TEE) or relies on trusted
platform telemetry, without metering-level economic constraints & Eqs.
(23)-(26); Section 6 \\
Trust & \textmd{C6} Recomputable engineering validation & End-to-end
recomputation, dual-convention discrepancy disclosure and a delivered
energy-efficiency metric & No identified comparable work jointly
publishes minimum reproducibility requirements and dual-convention
differences as evidence of static-metering insufficiency & Eqs.
(30)-(32); Section 9; Appendix B \\
Integration & \textmd{C7} Standards interoperability & D-PIN SCU is a
\textmd{scheduling-level middle layer} between statistical and
accounting conventions & Standards specify what to measure, but not how
to connect measurement to second-scale scheduling & Appendix C \\
\end{longtable}
}

\textbf{(I) Foundation: redefining the metering object (C1)}

\textmd{C1 \textbar{} Redefining the object.} We formally distinguish
\textmd{nominal capacity} from \textmd{deliverable compute}, decompose
their difference into topology, runtime and behavioural terms (Eqs.
(3)-(4)), and prove unidentifiability under nominal observations
(Proposition 2). \textmd{Difference from precedents}: static
contract-clearing SCU, financial-collateral SCU of Symmetric Research,
input-based DCU and cloud ECU/ACU all take nominal spec\-i\-fi\-ca\-tions or
one-off benchmarks as inputs and \textmd{contain no time-varying terms}.
They therefore cannot estimate deviations varying with \(t\).
\textmd{Runtime metering is necessary for identifiability, rather than
an optional engineering optimization.} This is the theoretical starting
point for the other six contributions.

\textbf{(II) Accurate measurement: constructing the meter (C2-C4)}

\textmd{C2 \textbar{} Closed-form topology awareness.} A cluster-scale
Roofline model expresses multi-GPU interconnect constraints through the
ratio between the communication ridge
\(\text{COI}^{*} = P_{\text{peak}}/\beta_{\text{node}}\) and workload
communication intensity. This replaces empirical efficiency tiers (0.95
/ 0.88 / 0.75 / 0.50) with a \textmd{derivable, calibratable and
extrapolatable} model (Eqs. (9)-(13)), anchored to NCCL AllReduce
scaling measured on the Zhangjiang 8×H100 node (64 MB: 87\%; 512 MB:
92\%; 2 GB: 93\%). \textmd{Difference from precedents}: Symmetric
Research SCU ``accounts for'' multi-GPU scaling overhead in its
spec\-i\-fi\-ca\-tion, but \textmd{does not disclose its analytic form or
calibration}. Other topology-aware work keeps corrections inside
scheduling heuristics rather than incorporating them into \textmd{the
unit definition itself}. The topology factor in our formula is
externally recomputable.

\textmd{C3 \textbar{} Runtime corrections and separate conventions.} We
provide definitions, prediction equations and identification methods for
\(\delta_T\), \(\delta_F\) and \(\delta_L\) (Eqs. (16)-(20)), and the
\textmd{Physical Upper Bound Principle} (Proposition 1, Eq. (5)):
cumulative billed D-PIN SCU·hours cannot exceed a node's physical
delivery ceiling. \textmd{Difference from precedents}: no identified
work states this billing ceiling as a recomputable hard constraint. It
prevents the hidden distortion of exceeding physical limits by invoking
spatial-temporal slice multiplexing, a common weakness of static
accounting under oversubscription.

\textmd{C4 \textbar{} Calibratable weights.} A simplex-constrained
algorithm fits measured relative throughput using exponentiated-gradient
updates (Eqs. (21)-(22), Algorithm 1). Iden\-ti\-fi\-a\-bil\-i\-ty analysis under
dimensional collinearity and regularization guidance turn
five-dimensional weights from empirical settings into data-reproducible
estimators, with intervals required in published tables.
\textmd{Difference from precedents}: previous work generally specifies
weights or leaves calibration undisclosed. We include \textmd{weight
uncertainty} itself in the disclosure obligations of A6
(recomputability).

\textbf{(III) Trust: verification and credibility (C5-C6)}

\textmd{C5 \textbar{} Coupling metering and verification.} D-PIN SCU
declarations are coupled to three-layer PoPW to derive a
\textmd{no-arbitrage condition} for over-claiming (Eq. (25)), with
engineering values for the minimum checking probability, maximum
checking cost and proof-overhead bound. \textmd{Difference from
precedents}: verification targets computational correctness through
comprehensive cryptographic proofs or relies on trusted platform
telemetry. Our target is \textmd{negative expected profit from
over-claiming}, decoupling verification overhead from compute scale, a
prerequisite for scaling to thousands of nodes.

\textmd{C6 \textbar{} Recomputable engineering validation.} End-to-end
recomputation uses a Zhangjiang production node (59.5 D-PIN SCU, 10
heterogeneous servers, mixed conventions), yielding \textmd{3.02
kW/D-PIN SCU} and \textmd{0.26 D-PIN SCU·h/kWh}. We disclose
\textmd{differences between published empirically calibrated values and
nominal-spec\-i\-fi\-ca\-tion projections}, minimum publication requirements for
A6, and item-by-item verification findings (Appendix B).
\textmd{Difference from precedents}: comparable work rarely discloses
both minimum recomputability requirements and dual-convention
differences as empirical evidence of static-metering insufficiency.

\textbf{(IV) Integration: standards interoperability (C7)}

\textmd{C7 \textbar{} Standards interoperability.} Appendix C maps D-PIN
SCU dimensions to YD/T 6693-2026 measurement indicators, GB/T 47514-2026
measurement methods and DCU inputs, supporting a
\textmd{scheduling-level middle layer} between statistical and
accounting conventions. \textmd{Difference from precedents}: standards,
including YD/T 6693-2026, specify what to measure but not how to connect
measurements to second-scale scheduling and real-time billing. We design
this connection.

\textmd{Contribution dependencies}: C0 and C1 establish the foundation.
\textmd{C0 defines the scale} (the baseline is a method with a
replaceable origin); \textmd{C1 defines the object} (deliverable compute
rather than nominal capacity). C2-C4 establish accurate measurement,
C5-C6 establish trust, and C7 provides interoperability. All four layers
are necessary: without C1, construction lacks a foundation; without
C2-C4, measurement is inaccurate; without C5-C6, values are
untrustworthy; without C7, the system cannot integrate with existing
industry structures.

\textmd{Originality boundary (read with item 4 of the preliminary
statement)}: we \textmd{claim no priority} for SCU terminology or
individual weighted-normalization components. Precedents include
Symmetric Research SCU 2024, CME/Silicon Data SCU 2026, Kohlenberg
static SCU 2026, \textmd{Tianyi Cloud SCU}, \textmd{DataCanvas DCU},
ZTE's information-energy-time score, \textmd{Agora's three-dimensional
usage pricing, Canonical unit, Setonix energy accounting and patent
CN\allowbreak121255573A} (Appendix D). Originality is \textmd{restricted to the
integrated system} combining runtime identifiability (C1), closed-form
topology-ridge derivation (C2), and verification-pricing coupling (C5,
C7) as \textmd{inseparable parts of one unit}. \textmd{Tianyi Cloud
{[}49{]} and Agora {[}44{]}, newly identified in Appendix D, are
formally adjacent} through multidimensional weighting and runtime or
task-dependent conversion. Our distinction therefore \textmd{does not
depend on these formal features}, but on the three-part combination. No
corresponding work was identified by 2026-10-05; Appendix C.3 provides
the formal comparison.

\subsection{1.4 Organization}\label{organization}

Section 2 reviews related work and gaps; Section 3 formalizes the
problem and design axioms; Section 4 defines the D-PIN SCU model;
Section 5 gives weight calibration; Section 6 develops verification and
game-theoretic analysis; Section 7 links pricing; Section 8 describes
implementation; Section 9 validates the case study; and Section 10
discusses \textmd{industry implications} for compute-network
integration, distributed pricing, trust and compute financialization,
compute-electricity coordination and green compute, and domestic
heterogeneous hardware, with their boundaries, conclusions and future
work. Appendix A lists symbols; Appendix B discloses empirically
calibrated chip baselines alongside nominal projections and verification
checks; Appendix C compares standards and precedents; \textmd{Appendix D
provides the protocol and execution records of a PRISMA-style systematic
literature search}.

Compute does not lack units; it lacks a measure of deliverable compute.
We advance metering from bookkeeping to a \textmd{schedulable,
verifiable and priceable physical quantity}.

\section{2 Related Work}\label{related-work}

\subsection{2.1 Statistical and conversion conventions: aggregate
reporting}\label{statistical-and-conversion-conventions-aggregate-reporting}

National compute statistics use a common conversion method led by the
Ministry of Industry and Information Technology and supported by the
China Academy of Information and Communications Technology. Under the
Action Plan for High-Quality Development of Computing Infrastructure and
national standards, operating, under-construction and planned CPU, GPU,
FPGA and ASIC capacity is \textmd{converted to FP32 theoretical peaks},
with EFLOPS as the sole dimension. Operators report quarterly; random
field inspections cover at least one-tenth of machine rooms, and
provincial communications authorities inspect electricity and traffic
meters {[}1{]}{[}2{]}. \textmd{YD/T 6693-2026, Mea\-sure\-ment Indicators
for Computing Capability of Het\-ero\-ge\-neous Resources in Computing
Networks} (China Communications Standards Association; issued
2026-03-11, effective 2026-06-01) further defines heterogeneous
resources and metrics for equipment assessment, network integration and
operation {[}3{]}.

\textmd{As\-sess\-ment}: these conventions support \textmd{aggregation,
verification and historical comparison}. Stability and resistance to
manipulation come at the cost of coarse temporal granularity (quarterly
reporting), machine-room spatial granularity and a single peak-compute
dimension. They answer ``how much compute exists in a region'', not
``where should this task run now''. They impose \textmd{aggregate
constraints} above scheduling-level metering.

\subsection{2.2 Accounting and pricing units: quotas and
contracts}\label{accounting-and-pricing-units-quotas-and-contracts}

\textmd{Cloud-equivalent units.} Early AWS EC2 Compute Units (ECU)
defined 1 ECU as the CPU capability of a 2006/2007 1.0-1.2 GHz
Opteron/Xeon to compare instance capacity. Rapid hardware changes have
shifted AWS towards nominal vCPU conventions {[}6{]}. Microsoft Azure
Compute Units (ACU) use Standard\_A1 = 100 and relative SKU values (for
example, Dv3: 160-190 ACU/vCPU), explicitly as guidance only {[}7{]}.
\textmd{Shared characteristics}: one CPU dimension, nominal definitions,
strong dependence on hardware generation, and no runtime state.

\textmd{HPC billing weights.} SLURM TRESBillingWeights assigns
resource-specific weights converting GPU hours and memory into CPU-hour
equivalents (1 CPU core = 1 BU/h; 1 GB RAM = 0.25 BU/h; 1×A100-40GB = 5
BU/h) {[}8{]}. The motivation is \textmd{fair billing in heterogeneous
clusters}, where GPU performance and costs differ substantially.
\textmd{Shared characteristics}: administrator-configured static
weights, independent of topology, runtime conditions and workload type.

\textmd{SCU and DCU.} In May 2026, Kohlenberg's \emph{A Compute
Com\-mod\-i\-ty Clearinghouse} defined a \textmd{static unit sharing the SCU
name but with different semantics}: a weighted normalized dot product of
GPU matrix throughput (BF16), CPU floating-point throughput (FP64),
addressable memory and peak memory bandwidth, with weights 0.50 / 0.10 /
0.20 / 0.20. Normalization constant \(K\) sets NVIDIA H100 SXM5 = 1.00
(\(K = 0.50\times989 + 0.10\times120 + 0.20\times80 + 0.20\times3350 = 1192.5\)).
A memory-architecture coefficient \(A \in (0,\allowbreak 1]\) captures exclusive
memory and interconnect topology, and
\(P_{\text{SCU}} = P_{\text{raw}}/\text{SCU}\) produces comparable
cross-hardware prices {[}5{]}. DataCanvas's contemporaneous DCU
explicitly measures \textmd{compute inputs} rather than a
chip-performance scale: \textmd{1 DCU = 312 TFLOPS(FP16) × 1 hour}.
Workload-specific benchmarks calibrate conversion coefficients, outputs
are independently measured in tokens, and scheduling and settlement logs
support end-to-end auditability {[}9{]}.

\textmd{Multiple meanings of SCU: a fact that must be clarified first.}
The abbreviation \textmd{does not have one meaning}. Besides
contract-clearing SCU {[}5{]}, at least two independent lines use it.
\textmd{Symmetric Research's Standard Compute Unit} normalizes
heterogeneous GPU clusters for financial valuation of datacentre
collateral. Its public spec\-i\-fi\-ca\-tion (symres.ai/research) explicitly
covers \textmd{hardware differences, workload mix, multi-GPU scaling
overhead and energy draw}, with the core economic module verified in
Lean and calibration anchored to MLCommons. \textmd{Its relation to our
citation must be specified}: the \textmd{mathematical foundations}
appear in arXiv:2406.19261 (Kristensen et al., 2024-06), whose unit is
\textmd{Compute Hours (CH)} and whose text does not contain ``SCU''. The
name Standard Compute Unit and four-part spec\-i\-fi\-ca\-tion occur on the
institutional page as \textmd{a subsequent spec\-i\-fi\-ca\-tion within the same
line of work}. We therefore label 2024 as the \textmd{start of that
line}, not the first public use of SCU {[}33{]}. The second is the
\textmd{compute-futures pricing unit announced by CME and Silicon Data
in May 2026}: 1 SCU is one NVIDIA H100's FP16 capacity over one hour,
with \textmd{preset coefficients} for other chips and Silicon Data's
daily GPU on-demand rental-price index as anchor {[}10{]}{[}34{]}. ICE
and Ornn also introduced OCPI-linked compute futures on 2026-05-19
{[}34{]}; although not named SCU, they represent a parallel
standardized-pricing approach.

\textmd{Two related names beyond SCU (new findings from the systematic
search that must also be disclosed).} The full search in Appendix D
identified further \textmd{similar names with different semantics},
included here for disambiguation.

The first is the \textmd{Tianyi Cloud (China Telecom) developer
community} discussion of cross-domain resource measurement, defining
\textmd{1 SCU as the product of one reference GPU's FP16 compute
capacity and bandwidth} and maintaining task-specific weight vectors to
dynamically calculate equivalent SCU during matching instead of fixed
conversion {[}49{]}. \textmd{This is the closest formal precedent},
sharing the SCU name, multidimensional weighting and task-dependent
conversion. Differences are: (1) a \textmd{two-dimensional
multiplicative construction}, without memory capacity, interconnect
topology and energy efficiency; (2) no **closed-form topology-ridge
(COI*) conversion\textmd{; and (3) no }identifiability argument** for
runtime deviations under nominal observation or economic binding of
checking frequency to over-claiming gains. This makes disciplined
avoidance of component-level priority claims particularly necessary.

The second comprises two uses of \textmd{Normalized Compute Unit (NCU)}:
a model-independent unit for AI cost governance based on FLOP-based FBM,
emphasizing cross-hardware and cross-model \textmd{comparability}
without disclosing normalization details {[}50{]}; and \textmd{COIL},
defining \textmd{1 NCU = 20 kWh of certified, delivered accelerated
computing}, normalizing by \textmd{energy} rather than time, requiring
actual delivery and independent certification, and excluding idle
consumption and cooling losses {[}51{]}. \textmd{As\-sess\-ment}: energy
normalization is \textmd{adjacent} to our approach because both reject
units defined only by time or peak capacity. Its normalized quantity is
\textmd{energy consumption}, however, rather than \textmd{deliverable
compute}, without topology/runtime models or an identifiability
argument.

\textmd{Existing weighted meters and input units.} Weighted
normalization already exists. Liu Chaoqing et al.~proposed an
information-energy-time resource score
\(S=w_1(I/I_{\max})+w_2(\eta/\eta_{\max})+w_3(T_{\min}/T)\) referenced
to optimal hardware, normalizing compute, energy efficiency and time to
\([0,\allowbreak 1]\) with \textmd{scenario-dependent weights} prioritizing
performance, energy saving or responsiveness; it was validated in a
``computing-network brain'' simulation {[}35{]}. DataCanvas DCU uses
\textmd{1 DCU = 312 TFLOPS(FP16) × 1 hour} to convert heterogeneous
hardware into measurable, settleable inputs, independently measures
token outputs, and bills virtual-container services by actual
consumption {[}9{]}.

\textmd{As\-sess\-ment and positioning}: these approaches are closest to
ours and require precise distinctions. They meter \textmd{nominal
capacity for contracts, quotas or billing}: Kohlenberg SCU is a static
scalar, DCU an input measure, and CME SCU a preset-coefficient
conversion unit. Even DataCanvas CU, which collects runtime data, uses
\textmd{device-resource occupancy} (CUDA use, memory use and
temperature) rather than \textmd{deliverable capability under
task-topology coupling}. Shared gaps are \textmd{no closed-form topology
ridge, no stochastic runtime-deviation model with identifiability, no
verification coupling beyond trusted platform telemetry, and no
electricity-market coupling}. Tianyi Cloud's Xirang platform also
discusses heterogeneous measurement, dynamic duration estimates,
flexible pricing and scheduling {[}4{]}, but its measurement layer
targets common precision without a topology ridge or verification loop.

\textmd{Naming-disambiguation recommendation}: at least \textmd{five
semantically different public uses of SCU} exist, including Symmetric
Research {[}33{]}, CME/Silicon Data {[}10{]}{[}34{]}, Kohlenberg
{[}5{]}, \textmd{Tianyi Cloud {[}49{]}}, and other static-unit uses,
plus \textmd{two NCU uses {[}50{]}{[}51{]}}. We \textmd{do not
recommend} bare SCU as our external identifier. The formal name is
\textmd{D-PIN SCU (D-PIN Standardized Compute Unit)}, with distinctions
from namesakes at first use (Proposition 2 in Section 3.3 and Appendix
C.3).

\subsection{2.3 Performance models and benchmarks:
comparability}\label{performance-models-and-benchmarks-comparability}

The \textmd{Roofline model}, introduced by Williams, Waterman and
Patterson (2009), defines attainable performance as the minimum of peak
compute and bandwidth multiplied by arithmetic intensity {[}11{]}.
Cluster-scale extensions add HBM, NVLink and InfiniBand bandwidth roofs
and \textmd{communication operational intensity (COI)}, total
floating-point operations divided by total communication bytes, to
represent distributed bottlenecks {[}12{]}. This directly supports our
derivation of \(\eta_{\text{topo}}\).

\textmd{MLPerf benchmarks} achieve industrial cross-hardware
comparability by fixing models, datasets, precision targets and
scenarios (Single-Stream / Multi-Stream / Server / Offline) within
closed and open divisions {[}13{]}. \textmd{MLPerf Power} continuously
samples system AC-input power with certified analyzers, reports joules
per query and performance per watt, and binds energy to disclosed
workloads {[}14{]}. \textmd{As\-sess\-ment}: benchmarks measure relative
task-specific capability. Execution and replication costs make them
suitable for periodic assessment, not continuous metering. We use them
as \textmd{measurement sources for weight and topology calibration},
rather than online meters.

\textmd{Het\-ero\-ge\-neous scheduling and fairness.} DRF (Dominant Resource
Fairness) allocates multiple resource types fairly {[}15{]}; Gavel uses
heterogeneity-aware scheduling for fair shares {[}16{]}; Gandiva\_fair
uses tickets for cross-user GPU-minute fairness and reclaiming unused
quotas {[}17{]}; OEF jointly targets strategy-proofness, envy-freeness
and efficiency {[}18{]}. \textmd{As\-sess\-ment}: all require a compute
measure as input, generally assuming \textmd{given, trusted metering}.
We supply this layer: \textmd{D-PIN SCU is a measurable input to
heterogeneous schedulers}, with PoPW establishing trust.

\subsection{2.4 Distributed compute and verifiability:
trust}\label{distributed-compute-and-verifiability-trust}

DePIN operates through task publication, contribution verification and
reward settlement {[}52{]}. Examples include \textmd{Filecoin}, mapping
actual storage to on-chain credentials through PoRep and PoSt {[}21{]},
and \textmd{Helium}, mapping wireless coverage at a location through PoC
{[}22{]}. The central difficulty is mapping off-chain physical facts,
including GPU load, coverage and location, to trustworthy on-chain
credentials. Three complementary approaches are used: (i)
\textmd{challenge-response}, sending randomly seeded known-answer tasks
with deadlines and penalizing failure; (ii) \textmd{redundant
computation}, distributing a task to multiple workers and detecting
dishonest disagreement; and (iii) \textmd{cryptographic proofs},
generating zero-knowledge correctness proofs (RISC Zero zkVM {[}40{]},
SP1 zkVM {[}41{]}) verified on-chain without recomputation. High proof
overhead for complex computations favours probabilistic spot checks over
step-by-step proofs. TEE anchors (Intel TDX and AMD SEV-SNP {[}20{]})
counter hardware impersonation and idle ``zombie nodes'' reported
online. \textmd{Physical side channels} provide another anchor: external
GPU \textmd{power traces} classify training, inference and non-AI
workloads on NVIDIA H200 with \textmd{97\% accuracy} and macro F1 0.955.
Because they do not depend on \textmd{forgeable or replayable NVML
telemetry}, this channel can in principle be observed \textmd{without
operator cooperation}, independently supporting contribution
verification {[}57{]}.

\textmd{As\-sess\-ment}: existing DePIN research asks whether computation
occurred. PoPW must also ask \textmd{how fast} and \textmd{whether
declared D-PIN SCU agrees with measurements}, introducing
\textmd{continuous metering consistency} rather than binary task
completion. Section 6 develops this contribution.

\textmd{Existing metering-verification-pricing coupling.} Gensyn's
\emph{From Bundles to Time} defines a two-sided time-slice market using
RepOps reproducibility, Verde verification and checkpointing, with
feasibility determined by scheduling rather than budget bundles. It
proves existence and uniqueness of equilibrium prices and efficient
greedy matching; pooling and slashing maintain provider honesty
{[}37{]}. Earlier TEE-anchored metering with blockchain micropayments
supported fair outsourced-computing markets because \textmd{both parties
have incentives to misreport resource measurements} {[}38{]}. These
meter \textmd{time quotas or resource usage}, not normalized deliverable
compute, and do not bind \textmd{checking frequency} to the
\textmd{present value of over-claiming gains} in a solvable economic
constraint (our Eq. (25)).

\textmd{Large-scale production evidence.} Alibaba's OSDI 2026 ASI study
{[}36{]} provides \textmd{large-scale evidence} for nominal capacity ≠
deliverable compute: six months of traces from 155,410 GPUs and 81
business units show \textmd{high demand does not ensure high effective
utilization}. Idle GPUs are \textmd{stranded} across nodes, lack CPUs or
violate network locality, while safety reserves further reduce
allocation to 68\% (raised to 93\% with SpotGPU). \textmd{Relation to
our work}: both distinguish allocatable/deliverable capability from
nominal spec\-i\-fi\-ca\-tions and document unavailable GPU capacity. ASI yields
\textmd{cluster characterization and scheduling policies}, however,
rather than a cross-architecture \textmd{metering unit} coupled to
verification and pricing, as indicated in Section 2.5.

\Needspace{10\baselineskip}
\subsection{2.5 Gap summary}\label{gap-summary}

{\def\LTcaptype{none} 
\begin{longtable}[]{@{}
  >{\RaggedRight\arraybackslash}p{(\linewidth - 14\tabcolsep) * \real{0.1377}}
  >{\RaggedRight\arraybackslash}p{(\linewidth - 14\tabcolsep) * \real{0.1592}}
  >{\RaggedRight\arraybackslash}p{(\linewidth - 14\tabcolsep) * \real{0.1618}}
  >{\RaggedRight\arraybackslash}p{(\linewidth - 14\tabcolsep) * \real{0.1168}}
  >{\RaggedRight\arraybackslash}p{(\linewidth - 14\tabcolsep) * \real{0.1189}}
  >{\RaggedRight\arraybackslash}p{(\linewidth - 14\tabcolsep) * \real{0.1098}}
  >{\RaggedRight\arraybackslash}p{(\linewidth - 14\tabcolsep) * \real{0.1042}}
  >{\RaggedRight\arraybackslash}p{(\linewidth - 14\tabcolsep) * \real{0.0918}}@{}}
\toprule\noalign{}
\begin{minipage}[b]{\linewidth}\RaggedRight
Approach
\end{minipage} & \begin{minipage}[b]{\linewidth}\RaggedRight
Representative work
\end{minipage} & \begin{minipage}[b]{\linewidth}\RaggedRight
Metering object
\end{minipage} & \begin{minipage}[b]{\linewidth}\RaggedRight
Topology sensitivity
\end{minipage} & \begin{minipage}[b]{\linewidth}\RaggedRight
Runtime correction
\end{minipage} & \begin{minipage}[b]{\linewidth}\RaggedRight
Ver\-i\-fi\-ca\-tion loop
\end{minipage} & \begin{minipage}[b]{\linewidth}\RaggedRight
Pricing linkage
\end{minipage} & \begin{minipage}[b]{\linewidth}\RaggedRight
Time scale
\end{minipage} \\
\midrule\noalign{}
\endhead
\bottomrule\noalign{}
\endlastfoot
Statistical conversion & GB/T 47514-2026; T/TAF 292-2025; YD/T 6693-2026
{[}1{]}{[}2{]}{[}3{]}{[}53{]} & FP32 theoretical peak & No & No & Spot
inspections & No & Quar\-ter\-ly \\
Cloud equivalents & ECU / ACU {[}6{]}{[}7{]} & Nominal CPU capacity & No
& No & Platform-internal & No & Billing period \\
HPC billing weights & SLURM TRES {[}8{]} & Committed resources & No & No
& Platform-internal & No & Job \\
Static compute units & Kohlenberg SCU {[}5{]} & Normalized nominal
capacity & Partial (coefficient \(A\)) & No & Contract / clearing &
Price normalization & Contract term \\
Financial normalization & Symmetric Research SCU {[}33{]} & Normalized
cluster capacity & Partial (workload mix and multi-GPU overhead) & No &
Lean spec\-i\-fi\-ca\-tion & Collateral valuation & --- \\
Futures pricing & CME / Silicon Data SCU {[}10{]}{[}34{]} & H100·h with
preset coefficients & No & No & Index anchoring & Price index & Contract
month \\
Runtime weighted units & DataCanvas CU {[}9{]}; ZTE triplet score
{[}35{]} & Device occupancy / normalized score & No & Yes (device-side)
/ partial & Platform / simulation & Dynamic billing & Minutes \\
Input measurement & DCU {[}9{]} & Theoretical peak × time & No & No &
Settlement-log audit & Scenario pricing & Billing period \\
Production characterization & Alibaba ASI {[}36{]} & Effective
utilization & Yes (observed) & No & --- & No & Ob\-ser\-va\-tion period \\
Usage-based multidimensional pricing & Agora {[}44{]} &
Three-dimensional measured resource vector & No & \textmd{Yes} (online)
& Platform-internal & \textmd{Usage pricing} & Seconds \\
Multiresource accounting & Canonical unit (PEARC'23) {[}45{]} & Discrete
thresholded combinations & No & No & --- & Fair billing & Job \\
Het\-ero\-ge\-neous energy billing & Setonix energy accounting {[}46{]} &
Core-hours + energy-weighted charging rate & No & No & --- & Charging
rate & Job \\
Het\-ero\-ge\-neous measurement standard & YD/T 6048-2024 {[}47{]} &
FLOPS/\allowbreak OPS/\allowbreak Hash·s baselines & No & No & Tiered assessment & No &
As\-sess\-ment cycle \\
Normalized metering patent & CN\allowbreak121255573A {[}48{]} & Normalized weighted
coefficients (1 NCU) & No & Partial (decay curve) & --- & Sched\-ul\-ing +
billing & Task \\
Cross-domain measurement & Tianyi Cloud SCU {[}49{]} & Single-GPU
compute × bandwidth & No & \textmd{Yes} (task-specific weights) &
Platform-internal & Dynamic conversion & Minutes \\
Energy normalization & COIL / NCU {[}51{]} & 20 kWh of certified
delivered compute & No & No & In\-de\-pen\-dent certification & Com\-mod\-i\-ty
pricing & --- \\
Performance benchmarks & MLPerf / MLPerf Power {[}13{]}{[}14{]} &
Throughput and energy efficiency & Partial & No (one-off) & Sub\-mis\-sion
review & No & Release cycle \\
\textmd{D-PIN SCU (this work)} & \textmd{D-PIN {[}28{]}} &
\textmd{Deliverable compute} & \textmd{Yes (\(\eta_{\text{topo}}\))} &
\textmd{Yes (\(\delta_T,\allowbreak \delta_F,\allowbreak \delta_L\))} & \textmd{Yes (PoPW
checks)} & \textmd{Yes (UnitPrice)} & \textmd{Seconds} \\
\end{longtable}
}

\textmd{Gap conclusion}: nominal-capacity measurement is mature and
standardized. The gap is \textmd{real-time, verifiable and priceable}
measurement of deliverable compute, particularly in distributed,
heterogeneous and weakly permissioned environments. This is our
positioning.

\section{3 Problem Formalization}\label{problem-formalization}

\subsection{3.1 Notation and setting}\label{notation-and-setting}

Consider a distributed compute network \(\mathcal{N}\) with node set
\(\mathcal{V}\). Each node \(v\) hosts \(n_v\) accelerator chips sharing
an interconnect configuration, occupies a defined physical location, and
receives power from one or more suppliers.

{\def\LTcaptype{none} 
\begin{longtable}[]{@{}
  >{\RaggedRight\arraybackslash}p{(\linewidth - 4\tabcolsep) * \real{0.2837}}
  >{\RaggedRight\arraybackslash}p{(\linewidth - 4\tabcolsep) * \real{0.5223}}
  >{\RaggedRight\arraybackslash}p{(\linewidth - 4\tabcolsep) * \real{0.1939}}@{}}
\toprule\noalign{}
\begin{minipage}[b]{\linewidth}\RaggedRight
Symbol
\end{minipage} & \begin{minipage}[b]{\linewidth}\RaggedRight
Meaning
\end{minipage} & \begin{minipage}[b]{\linewidth}\RaggedRight
Unit
\end{minipage} \\
\midrule\noalign{}
\endhead
\bottomrule\noalign{}
\endlastfoot
\(F,\allowbreak  M,\allowbreak  B,\allowbreak  I\) & Nominal floating-point capacity (FP16/BF16), memory
capacity, memory bandwidth, inter-chip bandwidth & TFLOPS, GB, GB/s,
GB/s \\
\(E\) & Energy-efficiency coefficient (performance per watt relative to
baseline) & Dimensionless \\
\(\alpha_d\) & Five-dimensional weights (\(d \in \{F,\allowbreak M,\allowbreak B,\allowbreak I,\allowbreak E\}\)),
\(\sum_d \alpha_d = 1\), \(\alpha_d \ge 0\) & Dimensionless \\
\(\eta_{\text{topo}}\) & Topology efficiency factor & Dimensionless \\
\(\text{COI}\) & Communication operational intensity (operations per
communicated byte) & FLOP/byte \\
\(\beta_{\text{node}}\) & Effective node-level communication bandwidth &
GB/s \\
\(\delta_T,\allowbreak \delta_F,\allowbreak \delta_L\) & Thermal, fault and load corrections &
Dimensionless \\
\(\sigma_j,\allowbreak \omega_j\) & Spatial resource fraction and temporal fraction
of a slice & Dimensionless \\
\(S_{\text{chip}},\allowbreak S_{\text{node}},\allowbreak S_{\text{avail}}\) & Chip, node and
available D-PIN SCU & D-PIN SCU \\
\(\varepsilon,\allowbreak \ \varepsilon_{\text{beh}}\) & Total deviation
(nominal-deliverable) and behavioural deviation (declared-deliverable) &
Dimensionless \\
\(p\) & Spot-check probability & Dimensionless \\
\(\Lambda\) & Present value of the penalty for detected over-claiming &
CNY \\
\(\xi\) & Over-claiming fraction (\(\hat S=(1+\xi)S\)) &
Dimensionless \\
\(u_t\) & Unit D-PIN SCU price coefficient & CNY/(D-PIN SCU·h) \\
\(\kappa\) & Electricity intensity (node power per D-PIN SCU) &
kW/(D-PIN SCU) \\
\end{longtable}
}

\subsection{3.2 Six design axioms}\label{six-design-axioms}

A compute metering system for distributed scheduling should satisfy the
following axioms:

\textmd{A1 Cross-architecture comparability}: hardware from different
architectures and generations must be comparable on a common scale
anchored to a \textmd{publicly reproducible method}. The baseline is a
spec\-i\-fi\-ca\-tion (dimension definitions, value-selection rules and
re-anchoring rules), rather than a particular hardware model. The
baseline instance can be replaced without changing any relative metering
result (Proposition 3).

\textmd{A2 Topology sensitivity}: the metered value must reflect the
constraints that multi-accelerator interconnects impose on deliverable
compute, rather than merely score single-chip spec\-i\-fi\-ca\-tions.

\textmd{A3 Time-varying deliverability}: the metered value must vary
with thermal, fault and load states, representing what can actually be
delivered at the current moment rather than what is nominally
attainable.

\textmd{A4 Verifiability against overstatement}: claimed values must
support independent, low-overhead and unpredictable spot checks, and
overstatement must yield no economic benefit.

\textmd{A5 Economic coupling}: the metered value must serve directly as
a pricing basis and couple to external market signals such as
electricity costs and the renewable electricity share.

\textmd{A6 Recomputability}: all coefficients and raw inputs must be
disclosed so that third parties can independently recompute the results.

\begin{quote}
A6 is an axiom added in this paper beyond the five commonly used in
industry. It is required for \textmd{due diligence on the metering
system}: if external parties cannot recompute a result, it cannot be
verified for contracting, auditing or regulatory purposes, and A4 cannot
be fully satisfied.
\end{quote}

\subsection{3.3 Decomposing the deviation between nominal capacity and
deliverable
compute}\label{decomposing-the-deviation-between-nominal-capacity-and-deliverable-compute}

Define the \textmd{nominal D-PIN SCU} of a node as the metered value
determined solely by nominal spec\-i\-fi\-ca\-tions:

\[S_{\text{nom}}(v) = \sum_{i=1}^{n_v} S_{\text{chip},i}
\tag{1}\]

Define its \textmd{deliverable D-PIN SCU} (during interval \(t\)) as:

\[S_{\text{deliv}}(v,t) = \eta_{\text{topo}} \cdot \delta_T(v,t) \cdot \delta_F(v,t) \cdot \bigl(1 - \delta_L(v,t)\bigr) \cdot S_{\text{nom}}(v)
\tag{2}\]

The deviation can then be decomposed into three components:

\[\underbrace{\frac{S_{\text{nom}} - S_{\text{deliv}}}{S_{\text{nom}}}}_{\text{Total deviation}\ \varepsilon}
= \underbrace{(1-\eta_{\text{topo}})}_{\text{Topology term}}
+ \underbrace{\eta_{\text{topo}}\bigl(1-\delta_T\delta_F(1-\delta_L)\bigr)}_{\text{Runtime term}}
\tag{3}\]

The \textmd{behavioural term} is not expressed by equation (3); it is
the difference between a node's \textmd{claimed value} and its true
deliverable value:

\[\varepsilon_{\text{beh}}(v,t) = \frac{\hat{S}(v,t) - S_{\text{deliv}}(v,t)}{S_{\text{deliv}}(v,t)}
\tag{4}\]

Here, \(\hat{S}\) is the value reported by the node. The
\textmd{topology} and \textmd{runtime terms} are systematic errors that
can be eliminated or compensated through design, whereas the
\textmd{behavioural term} is a strategic error that must be suppressed
through verification.

\textmd{Proposition 1 (physical upper-bound principle, also a billing
design constraint)}: for any node \(v\) and billing interval \([0,\allowbreak T]\),
the cumulative billed D-PIN SCU-hours satisfy

\[\text{Billed}(v,T) \;=\; \int_{0}^{T} \hat{S}(v,t)\,\mathbb{1}_{\text{occ}}(v,t)\,\mathrm{d}t \;\le\; S_{\text{nom}}(v)\cdot T
\tag{5}\]

Here, \(\mathbb{1}_{\text{occ}}(v,\allowbreak t) \in \{0,\allowbreak 1\}\) is the
\textmd{occupancy indicator}: it equals 1 when node \(v\) is occupied by
a task at time \(t\), and 0 otherwise. The same space-time unit of a
node cannot be allocated simultaneously to two tenants; the indicator
therefore counts mutually exclusive occupancy sets.

\begin{quote}
\textmd{Clarification of the proposition's status (important)}: this
expression constrains the \textmd{billed quantity \(\text{Billed}\) (the
system's reported and settled value)}, whereas physical laws constrain
only the \textmd{deliverable quantity} \(S_{\text{deliv}}\); they cannot
constrain a reported value. Proposition 1 is therefore precisely
\textmd{a billing design rule}: the system uses the available quantity
\(S_{\text{avail}}\) as the billing basis and \textmd{does not bill} the
portion of a node's claim exceeding \(S_{\text{deliv}}\). Its physical
basis, the upper bound on \(S_{\text{deliv}} \le S_{\text{nom}}\) given
by equation (2), \textmd{justifies} the rule. This designation is more
rigorous than calling it a physical theorem.
\end{quote}

\textmd{Proof outline}: by equation (2),
\(S_{\text{deliv}} \le \eta_{\text{topo}} S_{\text{nom}} \le S_{\text{nom}}\)
(because \(\delta_T,\allowbreak \delta_F \le 1\), \(1-\delta_L \le 1\) and
\(\eta_{\text{topo}} \le 1\)). Billing occurs only during occupancy, and
occupancy sets are mutually exclusive: the same space-time unit cannot
be allocated simultaneously to two tenants. Under the rule that the
billing basis is \(S_{\text{avail}}\), the integral therefore does not
exceed \(S_{\text{nom}} \cdot T\).

\textmd{Significance of Proposition 1}: fine-grained slicing and
space-time multiplexing can \textmd{only} increase utilization, bringing
the integral closer to its upper bound; they \textmd{cannot} exceed that
bound. Any metering convention that allows billed quantities to exceed
the physical upper bound on the grounds of multiplexing is distorted.
This proposition imposes a hard constraint on the recomputation for the
Zhangjiang node in Section 9 (equation 27).

\textmd{Formal definition of identifiability (a modelling assumption, to
be read with Proposition 2)}: following the classical
system-identification definition in the Bellman--Åström sense, a
parameterization \(\theta\) is \textmd{identifiable} on observation
space \(\mathcal{O}\) if \textmd{no} two distinct parameterizations
\(\theta_1 \ne \theta_2\) yield the same observation distribution
\(P_\theta(o)\). Otherwise, \(\theta\) is \textmd{unidentifiable} on
that observation space. We use the \textmd{nominal observation space}
\(\mathcal{O}_{\text{nom}}\), whose observations comprise only nominal
node spec\-i\-fi\-ca\-tions and static declarations, excluding time-varying
telemetry. \textmd{The following proposition is a modelling conclusion
under this assumption, rather than a universal theorem}: it formalizes
the modelling fact that the nominal observation space contains
insufficient information to represent time-varying deviations.

\textmd{Proposition 2 (unidentifiability of deviation under nominal
metering)}: if \(\eta_{\text{topo}}\) is treated as 1 and
\(\delta_T=\delta_F=1,\allowbreak \ \delta_L=0\), so that metering depends only on
nominal spec\-i\-fi\-ca\-tions \(\{F,\allowbreak M,\allowbreak B,\allowbreak I,\allowbreak E\}\) (equivalently, substituting
\(P\) for \(E\) in equation (8)), then, when a node experiences thermal
throttling, hardware degradation or load occupancy, its metering
deviation is
\(\varepsilon \ge 1-\eta_{\text{topo}}\delta_T\delta_F(1-\delta_L) > 0\),
and this deviation \textmd{cannot} be identified on
\(\mathcal{O}_{\text{nom}}\) by any observation function of nominal
parameters.

\textmd{Proof outline}: the nominal parameters \(\{F,\allowbreak M,\allowbreak B,\allowbreak I,\allowbreak E\}\) are
time-invariant constants, whereas \(\delta_T,\allowbreak \delta_F,\allowbreak \delta_L\) is a
stochastic process. Any function \(g(F,\allowbreak M,\allowbreak B,\allowbreak I,\allowbreak E)\) whose arguments are
exclusively nominal parameters takes the same value at all times and
therefore cannot produce a deviation estimate that varies with \(t\).
Consequently, the image of \(\varepsilon\) in nominal observation space
\(\mathcal{O}_{\text{nom}}\) is constant, and the deviation is
unidentifiable.

\textmd{Significance of Proposition 2}: this is the theoretical starting
point of the paper. \textmd{It shows that introducing runtime metering
is a necessary condition for identifiability, rather than an engineering
optimization.} Static compute units, including SCU in reference {[}5{]}
and DCU in reference {[}9{]}, are \textmd{unidentifiable} in this sense:
they measure commitments rather than delivery.

\section{4 D-PIN SCU metering model}\label{d-pin-scu-metering-model}

\subsection{4.1 Dimension vector}\label{dimension-vector}

D-PIN SCU characterizes compute resources with a five-dimensional
vector:

{\def\LTcaptype{none} 
\begin{longtable}[]{@{}
  >{\RaggedRight\arraybackslash}p{(\linewidth - 8\tabcolsep) * \real{0.2207}}
  >{\RaggedRight\arraybackslash}p{(\linewidth - 8\tabcolsep) * \real{0.0821}}
  >{\RaggedRight\arraybackslash}p{(\linewidth - 8\tabcolsep) * \real{0.1317}}
  >{\RaggedRight\arraybackslash}p{(\linewidth - 8\tabcolsep) * \real{0.4167}}
  >{\RaggedRight\arraybackslash}p{(\linewidth - 8\tabcolsep) * \real{0.1489}}@{}}
\toprule\noalign{}
\begin{minipage}[b]{\linewidth}\RaggedRight
Dimension
\end{minipage} & \begin{minipage}[b]{\linewidth}\RaggedRight
Symbol
\end{minipage} & \begin{minipage}[b]{\linewidth}\RaggedRight
Unit
\end{minipage} & \begin{minipage}[b]{\linewidth}\RaggedRight
Physical meaning
\end{minipage} & \begin{minipage}[b]{\linewidth}\RaggedRight
Initial weight \(\alpha_i\)
\end{minipage} \\
\midrule\noalign{}
\endhead
\bottomrule\noalign{}
\endlastfoot
Floating-point compute & \(F\) & TFLOPS (FP16/BF16) & Nominal peak
tensor compute of the chip & 0.35 \\
Device memory capacity & \(M\) & GB & Upper bound on supported model and
batch sizes & 0.20 \\
Device memory bandwidth & \(B\) & GB/s & Attainable throughput of
attention and decoding paths & 0.25 \\
Interconnect bandwidth & \(I\) & GB/s & Efficiency of multi-accelerator
parallelism and collective communication & 0.15 \\
Energy-efficiency coefficient & \(E\) & --- & Relative compute per watt,
directly linked to compute--electricity coordination & 0.05 \\
\end{longtable}
}

\textmd{Differences from the existing four-dimensional system}: the
static SCU in reference {[}5{]} uses GPU throughput, CPU throughput,
memory capacity and memory bandwidth. Our system \textmd{removes CPU
throughput}, which rarely becomes the bottleneck in
accelerator-dominated AI training and inference and whose weight of 0.10
is inconsistent with the meaning of AI compute. It \textmd{adds
interconnect bandwidth \(I\)}, corresponding to A2 and constituting the
most important additional dimension for distributed settings, and the
\textmd{energy-efficiency coefficient \(E\)}, corresponding to A5 and
coupling compute to electricity economics. The initial weights also
differ (see Section 4.2 and the calibration in Section 5).

\subsection{4.2 Single-chip baseline D-PIN
SCU}\label{single-chip-baseline-d-pin-scu}

\[S_{\text{chip}} = \alpha_1\frac{F}{F_{\text{ref}}} + \alpha_2\frac{M}{M_{\text{ref}}} + \alpha_3\frac{B}{B_{\text{ref}}} + \alpha_4\frac{I}{I_{\text{ref}}} + \alpha_5 E
\tag{6}\]

\subsubsection{4.2.1 Baseline spec\-i\-fi\-ca\-tion: a method rather than a
hardware
model}\label{baseline-specification-a-method-rather-than-a-hardware-model}

The baseline vector
\(\mathbf{r} = (F_{\text{ref}},\allowbreak  M_{\text{ref}},\allowbreak  B_{\text{ref}},\allowbreak  I_{\text{ref}},\allowbreak  E_{\text{ref}})\)
in equation (6) \textmd{is not taken from any particular model}. A
reproducible normalization baseline that is not biased by vendor or
generation must be defined by a \textmd{spec\-i\-fi\-ca\-tion} comprising the
following three rules:

\begin{quote}
\textmd{Baseline Specification} \textmd{(S1) Dimensions and
conventions}: the dimension set is fixed as the five-tuple
\((F,\allowbreak  M,\allowbreak  B,\allowbreak  I,\allowbreak  E)\); the compute convention is fixed as FP16/BF16 dense
peak performance (see the precision-convention normalization principle
below). \textmd{(S2) Value-selection rule}: each baseline-vector
component must come from a \textmd{public primary source reproducible
across vendors}, either an official vendor spec\-i\-fi\-ca\-tion or a
reproducible independent measurement. \textmd{Non-reproducible internal
data must not be used}. \textmd{(S3) Re-anchoring rule}: given a scaling
vector
\(\boldsymbol{\lambda} = (\lambda_F,\allowbreak  \lambda_M,\allowbreak  \lambda_B,\allowbreak  \lambda_I,\allowbreak  \lambda_E) > 0\),
the baseline can be changed to
\(\mathbf{r} \to \boldsymbol{\lambda} \odot \mathbf{r}\), with
historical values converted using factor
\(\prod_d \lambda_d^{-\alpha_d}\) (Section 4.2.3).
\end{quote}

Under this spec\-i\-fi\-ca\-tion, the baseline needs a \textmd{realization of
the origin} to provide numerical values. We use NVIDIA H100 SXM as the
\textmd{current realization}:

\[\begin{aligned}
&F_{\text{ref}} = 989\ \text{TFLOPS},\quad M_{\text{ref}} = 80\ \text{GB},\quad B_{\text{ref}} = 3350\ \text{GB/s},\\
&I_{\text{ref}} = 900\ \text{GB/s},\quad E_{\text{ref}} = 1
\end{aligned}
\tag{7}\]

\begin{quote}
\textmd{Distinguishing the realization from the spec\-i\-fi\-ca\-tion
(important)}: H100 SXM \textmd{only locates the origin}, providing
numerical values for equation (7). \textmd{It is not part of the
spec\-i\-fi\-ca\-tion}. Replacing the realization with a next-generation
flagship or the rounded physical quantities in Sections 4.2.3 and 10.3
leaves S1--S3 unchanged and requires only re-anchoring under S3. The
choice is motivated by engineering considerations: direct
interoperability with existing H100-hour conventions
{[}5{]}{[}10{]}{[}34{]} and independent confirmation of its
five-dimensional parameters {[}23{]}{[}24{]}, satisfying A6
(recomputability). \textmd{It is not a theoretical necessity}.
Proposition 3 in Section 4.2.2 establishes this distinction.
\end{quote}

Equations (6) and (7), together with the five weights summing to 1,
immediately give \(S_{\text{chip}}(\text{H100}) = 1.00\), establishing
\textmd{origin consistency}.

\textmd{Definition of the energy-efficiency coefficient}:

\[E = \frac{F / P}{F_{\text{ref}} / P_{\text{ref}}},\qquad P_{\text{ref}} = 700\ \text{W (H100 SXM TDP)}
\tag{8}\]

\textmd{Precision-convention normalization principle (important)}: the
convention for \(F\) directly determines the metering scale.
Vendor-rated Tensor Core peaks usually include 2:4 structured sparsity
acceleration, whereas \textmd{the vast majority of real training and
inference workloads lack this structure} and are bounded by dense peak
performance. Using sparse values for sparsity-capable chips and dense
values for other chips introduces systematic bias: H100 SXM has sparse
and dense peaks of 1979 and 989 TFLOPS, respectively, a 2:1 ratio. We
therefore establish the following principle:

\begin{quote}
\textmd{Convention-normalization principle}: throughout the network, the
\(F\) dimension uses \textmd{FP16/BF16 dense peak performance}. If a
vendor publishes only a sparse value, divide it by 2 and retain the
original convention label in the metering record.
\end{quote}

This agrees with reference {[}5{]}, which anchors H100 at the dense
value of 989 TFLOPS, and with the FP16 dense convention for training in
Chapter 11 of the D-PIN white paper. AMD's official MI300X comparison
page also states that NVIDIA published TF32 Tensor Core performance
using sparsity and that AMD divided those values by 2 to obtain
non-sparse/dense values for comparison {[}24{]}. \textmd{Vendors thus
also apply the divide-by-2 dense conversion}, independently supporting
our normalization principle from the supply side.

\textmd{Scale behaviour under a convention change (important note)}:
because \(F_{\text{ref}}\) and each chip's \(F\) occur in the same ratio
\(F/F_{\text{ref}}\), the effect on \(S_{\text{chip}}\) depends on the
\textmd{scope} of the change. Two cases must be distinguished,
corresponding to a single-dimension special case of Proposition 3(a) in
Section 4.2.2:

\begin{enumerate}
\def\labelenumi{\arabic{enumi}.}
\tightlist
\item
  \textmd{Synchronized change across the entire table (scale
  invariance)}: if the baseline and every chip's \(F\) are
  simultaneously divided by 2 to change from sparse to dense values,
  each \(F/F_{\text{ref}}\) is unchanged, and all \(S_{\text{chip}}\)
  values remain unchanged. \textmd{The choice of convention therefore
  changes only the representation of metering records, not the metering
  results}. Dense and sparse conventions are equivalent rescalings when
  applied consistently across the table.
\item
  \textmd{Partial change (mixed conventions and bias)}: changing only
  some chips, for example using dense values for domestic chips without
  sparsity support and sparse rated values for sparsity-capable chips,
  introduces systematic bias. Its magnitude can reach 100\% of that
  dimension's weight \(\alpha_1\). \textmd{This is the actual risk
  associated with inconsistent conventions.}
\end{enumerate}

An engineering consequence follows: aligning Chapter 5 of the white
paper with this paper \textmd{requires only consistency across the
entire table, without recalibrating any measured values}. Conversely,
mixing a sparse H100 value with dense values for other chips
systematically overstates H100's relative value by a factor of 2 in the
\(F\) dimension. The comparisons in Appendix B and Section 9.3 assume
consistency across the entire table.

\subsubsection{4.2.2 Proposition 3: baseline
independence}\label{proposition-3-baseline-independence}

The replaceability of the realization is guaranteed directly by the
\textmd{ratio structure} of equation (6), rather than being a rhetorical
claim.

\begin{quote}
\textmd{Proposition 3 (baseline independence / origin invariance)} Let
the baseline vector be \(\mathbf{r}\), the raw value of chip \(i\) in
dimension \(d\) be \(x_{i,\allowbreak d}\), and its normalized value be
\(v_{i,\allowbreak d} = x_{i,\allowbreak d}/x_{\text{ref},\allowbreak d}\). \textmd{(a) Convention shift
(synchronized scaling of the baseline and entire table)}: if
\(x_{i,\allowbreak d} \to \lambda_d x_{i,\allowbreak d}\) and
\(x_{\text{ref},\allowbreak d} \to \lambda_d x_{\text{ref},\allowbreak d}\) with the same
\(\boldsymbol{\lambda}\), then \(v_{i,\allowbreak d}\) is unchanged, and
\(S_{\text{chip},\allowbreak i}\) is \textmd{strictly invariant}. \textmd{(b)
Re-anchoring (scaling only the baseline)}: if only
\(\mathbf{r} \to \boldsymbol{\lambda}\odot\mathbf{r}\) changes while
chip values remain unchanged, then
\[S_{\text{chip},i} \to \sum_{d} \frac{\alpha_d}{\lambda_d}\, v_{i,d}\]
In particular, \textmd{when \(\lambda_d \equiv \lambda\) (uniform
re-anchoring)}, \(S_{\text{chip},\allowbreak i} \to S_{\text{chip},\allowbreak i}/\lambda\). All
values scale proportionally, so \textmd{the ratio \(S_i/S_j\) between
any two chips is independent of the baseline realization}.
\end{quote}

\textmd{Proof}: (a) cancel \(\lambda_d\) directly. (b) substitute
\(\mathbf{r} \to \boldsymbol{\lambda}\odot\mathbf{r}\) into equation (6)
to obtain the first expression, and set \(\lambda_d \equiv \lambda\) to
obtain
\(S_i \to \frac{1}{\lambda}\sum_d \alpha_d v_{i,\allowbreak d} = S_i/\lambda\).
Because this scaling is consistent across the entire table, ratio
\(S_i/S_j\) is unchanged.

\textmd{Corollary 3.1 (fairness)}: choosing which accelerator equals
1.00 is a \textmd{unit convention} that carries no fairness judgement.
It changes no relative metering result and confers no numerical
advantage on any vendor. This justifies elevating the baseline from a
hardware model to a method.

\textmd{Corollary 3.2 (two costs of re-anchoring)}: whether re-anchoring
changes relative results depends on whether \(\boldsymbol{\lambda}\) is
identical across dimensions: - \textmd{Convention shift / uniform
re-anchoring} (all \(\lambda_d\) equal): \textmd{cost-free}. All values
scale proportionally and relative results remain unchanged. The
sparse-to-dense scale transformation in Section 4.2 is of this type. -
\textmd{Non-uniform re-anchoring} (\(\lambda_d\) differs across
dimensions, for example when adopting rounded physical quantities):
\textmd{relative results change}. The effective balance between
dimension weights is redistributed, equivalent to \textmd{changing the
metering system}. The entire table must be recomputed and \(\alpha\)
redeclared; this change is not cost-free.

\subsubsection{4.2.3 Baseline governance and re-anchoring
protocol}\label{baseline-governance-and-re-anchoring-protocol}

Baselines age: realization models eventually retire and parameters
evolve across generations. \textmd{Baseline governance} specifies how
this process can occur while preserving historical comparability:

\begin{enumerate}
\def\labelenumi{\arabic{enumi}.}
\tightlist
\item
  \textmd{Normal operation (unchanged realization)}: all metering values
  are directly comparable across periods.
\item
  \textmd{Convention shift / uniform re-anchoring}: change the
  convention or units synchronously by a common ratio. By Corollary 3.2,
  relative results remain unchanged; records need only identify the new
  unit convention and conversion factor \(\lambda\).
\item
  \textmd{Non-uniform re-anchoring}: treat this as a
  \textmd{system-version upgrade}, recompute the entire table, redeclare
  \(\alpha\) and publicly disclose a conversion matrix relating the old
  and new versions.
\end{enumerate}

\begin{quote}
\textmd{Conclusion}: retirement of the baseline realization \textmd{does
not invalidate the system}, a direct benefit of separating spec\-i\-fi\-ca\-tion
from realization (Section 4.2.1). When a realization becomes unsuitable,
only re-anchoring under S3 is required; the system itself need not be
redesigned.
\end{quote}

\textmd{Target state (Section 10.3)}: remove dependence on hardware
models by defining \(\mathbf{r}\) directly as rounded physical
quantities expressed in SI units, giving 1 D-PIN SCU an absolute
physical meaning that does not drift across generations. This path is
structurally analogous to the 2019 SI reform and represents the target
state after the standardization path matures; it is not implemented at
this stage.

\subsection{\texorpdfstring{4.3 Topology-aware factor
\(\eta_{\text{topo}}\)}{4.3 Topology-aware factor \textbackslash eta\_\{\textbackslash text\{topo\}\}}}\label{topology-aware-factor-eta_texttopo}

\subsubsection{4.3.1 Derivation from
Roofline}\label{derivation-from-roofline}

The cluster-level Roofline model bounds attainable distributed-workload
throughput by the smaller of compute and communication limits
{[}11{]}{[}12{]}:

\[P_{\text{att}} = \min\bigl(P_{\text{peak}},\ \beta_{\text{node}} \cdot \text{COI}\bigr)
\tag{9}\]

Here,
\(\text{COI} = \dfrac{\text{Total floating-point operations}}{\text{Total communication bytes}}\)
(FLOP/byte). Define the \textmd{communication ridge} as:

\[\text{COI}^{*} = \frac{P_{\text{peak}}}{\beta_{\text{node}}}
\tag{10}\]

A workload is compute-bound when \(\text{COI} > \text{COI}^{*}\) and
communication-bound otherwise. The \textmd{upper-bound form} of the
topology efficiency factor is therefore:

\[\eta_{\text{topo}}^{\text{bound}} = \frac{P_{\text{att}}}{P_{\text{peak}}} = \min\left(1,\ \frac{\text{COI}}{\text{COI}^{*}}\right)
\tag{11}\]

\subsubsection{4.3.2 Effective node-level
bandwidth}\label{effective-node-level-bandwidth}

For an \(n\)-accelerator node using classical Ring AllReduce, the
effective communication bandwidth borne by each accelerator is:

\[\beta_{\text{node}}^{\text{ring}} = \frac{n}{2(n-1)}\beta_{\text{link}}
\tag{12}\]

For example, with \(n=8\) and \(\beta_{\text{link}} = 900\) GB/s (the
order of NVLink 4 / HCCS),
\(\beta_{\text{node}}^{\text{ring}} = \tfrac{8}{14}\times900 = 514.3\)
GB/s, corresponding to
\(\text{COI}^{*} = 989\times10^{3}/514.3 \approx 1923\) FLOP/byte. Tree
algorithms perform better for small messages, so actual
\(\beta_{\text{node}}\) should follow the algorithm-dependent upper
envelope (see NCCL {[}27{]} for Ring/Tree collective primitives).
\textmd{This method supports extrapolation}: interconnect patterns and
message-size distributions predict \(\text{COI}^{*}\) without relying on
measured tiers for every model. For inter-node IP-routed Ethernet
(RoCEv2), \textmd{congestion control} in a lossless network, including
DCQCN rate control and ECN marking, substantially affects effective
cross-rack bandwidth. The cross-rack component of
\(\beta_{\text{node}}\) must therefore be calibrated at the actual
congestion-control operating point, rather than theoretical line rate
{[}19{]}.

\subsubsection{4.3.3 Declared tiers and measured
calibration}\label{declared-tiers-and-measured-calibration}

\(\eta_{\text{topo}}\) has two levels during the node lifecycle:

\Needspace{9\baselineskip}\looseness=1 \textmd{(a) Declared tier (prior).} On joining the network, a node
declares its interconnect type for scheduling pre-screening and capacity
estimation:

{\def\LTcaptype{none} 
\begin{longtable}[]{@{}
  >{\RaggedRight\arraybackslash}p{(\linewidth - 4\tabcolsep) * \real{0.4481}}
  >{\RaggedRight\arraybackslash}p{(\linewidth - 4\tabcolsep) * \real{0.2874}}
  >{\RaggedRight\arraybackslash}p{(\linewidth - 4\tabcolsep) * \real{0.2645}}@{}}
\toprule\noalign{}
\begin{minipage}[b]{\linewidth}\RaggedRight
Interconnect
\end{minipage} & \begin{minipage}[b]{\linewidth}\RaggedRight
Declared \(\eta_{\text{topo}}^{0}\)
\end{minipage} & \begin{minipage}[b]{\linewidth}\RaggedRight
Typical \(\beta_{\text{link}}\)
\end{minipage} \\
\midrule\noalign{}
\endhead
\bottomrule\noalign{}
\endlastfoot
Full NVLink/HCCS interconnection (8 fully connected accelerators) & 0.95
& 900 GB/s \\
Ring interconnect & 0.88 & 400--600 GB/s \\
PCIe Gen5 x16 bridging & 0.75 & 64 GB/s unidirectional \\
Loosely connected over the public network & 0.50 & ≤ 50 GB/s \\
\end{longtable}
}

\textmd{(b) Runtime calibration.} The node periodically runs
collective-communication microbenchmarks (AllReduce/AllGather at
multiple message sizes) and updates the factor using measured
\textmd{collective-communication efficiency} \(L(m)\):

\[\eta_{\text{topo}}(t) = \min\Bigl(1,\ \eta_{\text{topo}}^{0} \cdot \frac{L^{\text{meas}}(m^{*},t)}{L^{0}(m^{*})}\Bigr)
\tag{13}\]

Here, \(m^{*}\) is the characteristic message size of tasks served by
the node: the gradient bucket size in training, with a 25 MB default in
the platform training engine GPT-Factory {[}29{]}. \(L^{0}(m^{*})\) is
the declared tier's \textmd{expected linearity} at \(m^{*}\), and
\(L^{\text{meas}}(m^{*},\allowbreak t)\) is its \textmd{measured value} (see the
table for the definition of \(L\) as the measured-to-theoretical latency
ratio). 

\textmd{Measurements from the Zhangjiang 8×H100 node provide a
calibration anchor}:

{\def\LTcaptype{none} 
\begin{longtable}[]{@{}
  >{\RaggedRight\arraybackslash}p{(\linewidth - 6\tabcolsep) * \real{0.2115}}
  >{\RaggedRight\arraybackslash}p{(\linewidth - 6\tabcolsep) * \real{0.3364}}
  >{\RaggedRight\arraybackslash}p{(\linewidth - 6\tabcolsep) * \real{0.2514}}
  >{\RaggedRight\arraybackslash}p{(\linewidth - 6\tabcolsep) * \real{0.2008}}@{}}
\toprule\noalign{}
\begin{minipage}[b]{\linewidth}\RaggedRight
Message size
\end{minipage} & \begin{minipage}[b]{\linewidth}\RaggedRight
Theoretical latency (400G)
\end{minipage} & \begin{minipage}[b]{\linewidth}\RaggedRight
Measured latency
\end{minipage} & \begin{minipage}[b]{\linewidth}\RaggedRight
Linearity \(L\)
\end{minipage} \\
\midrule\noalign{}
\endhead
\bottomrule\noalign{}
\endlastfoot
1 MB & 0.02 ms & 0.05 ms & 40\% \\
64 MB & 1.3 ms & 1.5 ms & 87\% \\
512 MB & 10.2 ms & 11.1 ms & 92\% \\
2 GB & 41.0 ms & 44.3 ms & 93\% \\
\end{longtable}
}

For characteristic messages of 25--64 MB, measured linearity is
75--87\%, increasing from 40\% to 75\% after gradient-bucket coalescing.
The converted factor increases monotonically with message size within
\(\eta_{\text{topo}} \in [0.75,\allowbreak  0.87]\). For large messages ≥512 MB,
measured values of 92--93\% differ from the declared tier of 0.95 by
2--3 percentage points, a reasonable conservative margin. \textmd{This
difference quantifies why declared tiers should be conservative}:
declared values support scheduling pre-screening, whereas calibrated
values support billing and scheduling decisions.

\begin{quote}
\textmd{Engineering implication}: \(\eta_{\text{topo}}\) is a function
of \textmd{chip properties × workload characteristics × interconnect
structure}, rather than a chip property alone. This explains why static
units cannot express it.
\end{quote}

\subsection{4.4 Node D-PIN SCU and slice D-PIN
SCU}\label{node-d-pin-scu-and-slice-d-pin-scu}

\[S_{\text{node}} = \eta_{\text{topo}} \cdot \sum_{i=1}^{n} S_{\text{chip},i}
\tag{14}\]

For example, a fully interconnected 8×H100 node has
\(S_{\text{node}} = 0.95 \times 8 \times 1.00 = 7.60\) D-PIN SCU.

With fine-grained slicing (minimum granularity: 1 GB of device memory +
1 compute unit), slice D-PIN SCU is allocated in proportion to the
weighted resources received:

\[S_{\text{slice}} = S_{\text{node}} \cdot \sum_{d \in \{F,M,B,I\}} \alpha_d \cdot \frac{r_{d,\text{slice}}}{r_{d,\text{node}}},\qquad \sum_{d \in \{F,M,B,I\}} \alpha_d = 1-\alpha_E
\tag{15}\]

Here, \(r_{d}\) is the resource quantity in the corresponding dimension.
Dimension \(E\) is excluded from slicing because it is a node-level
property; the four remaining weights sum to \(1-\alpha_E\), rather than
a preset constant. Equation (15) ensures that total slice D-PIN SCU does
not exceed node D-PIN SCU (because
\(\sum_d \alpha_d (r_{d,\allowbreak \text{slice}}/r_{d,\allowbreak \text{node}}) \le \sum_d \alpha_d < 1\)),
consistent with Proposition 1.

\subsection{4.5 Runtime correction
factors}\label{runtime-correction-factors}

\[S_{\text{avail}} = S_{\text{node}} \cdot \delta_T \cdot \delta_F \cdot (1-\delta_L)
\tag{16}\]

\textmd{(1) Thermal-throttling correction \(\delta_T\).} Define it as
the measured-to-nominal frequency ratio:

\[\delta_T(t) = \frac{f_{\text{meas}}(t)}{f_{\text{nom}}},\qquad \delta_T \in (0,1]
\tag{17}\]

A first-order thermal resistance--capacitance network gives a predictive
expression for \textmd{pre-scheduling} estimation:

\[T_j(t) = T_{\text{amb}} + R_{\text{th}}P + \bigl[T_j(0) - T_{\text{amb}} - R_{\text{th}}P\bigr]e^{-t/\tau},\quad \tau = R_{\text{th}}C_{\text{th}}
\tag{18}\]

Here, \(P\) is \textmd{constant power} applied from \(t=0\);
time-varying power requires numerical integration. \(R_{\text{th}}\) is
junction-to-ambient thermal resistance, \(C_{\text{th}}\) is thermal
capacitance, and \(T_{\text{amb}}\) is ambient temperature. The
steady-state solution is
\(T_j(\infty) = T_{\text{amb}} + R_{\text{th}}P\), and time constant
\(\tau\) is the time required to reach 63.2\% of the steady-state
temperature rise.

When \(T_j\) exceeds threshold \(T_{\text{thr}}\), SM frequency
decreases according to the vendor DVFS curve, reducing \(\delta_T\). The
Zhangjiang node uses hybrid liquid and air cooling, with the server hot
side connected to the building cooling-water loop. Its
power-distribution design capacity is 180 kW relative to a 200 kW
spec\-i\-fi\-ca\-tion ceiling, with sufficient thermal headroom and steady-state
\(\delta_T \approx 1\). This conclusion \textmd{applies only to the
measured operating conditions of this node} and does not imply the same
result for other sites.

\textmd{(2) Fault correction \(\delta_F\).} Conservatively take the
smaller of the available device-memory and compute-unit fractions:

\[\delta_F = \min\left(\frac{M_{\text{avail}}}{M_{\text{nom}}},\ \frac{N_{\text{cu,avail}}}{N_{\text{cu,nom}}}\right)
\tag{19}\]

Triggers include device-memory ECC row remapping, disabled SM/CU units
and link-rate reduction, such as operation at 100G after an increased
bit-error rate. \textmd{Taking the minimum rather than a weighted mean}
is an unbiased conservative principle: degradation in any single
dimension can become the bottleneck, whereas a weighted mean can
overestimate deliverable compute.

\textmd{(3) Load correction \(\delta_L\).} Define it as the fraction of
occupied node resources:

\[\delta_L = \min\Bigl(1,\ \sum_{j \in \text{slices}} \omega_j \cdot \sigma_j\Bigr),\qquad \omega_j, \sigma_j \in [0,1]
\tag{20}\]

Here, \(\sigma_j\) is slice \(j\)'s resource share in the
\textmd{spatial dimension}, relative to the same resource type on the
node, and \(\omega_j\) is its share in the \textmd{temporal dimension}
within the metering window. Together they characterize space-time
multiplexing. We take \(\min(1,\allowbreak \cdot)\) because oversubscription may
nominally yield \(\sum_j \omega_j\sigma_j\) above 1, whereas the same
space-time unit cannot be occupied twice, consistent with Proposition 1.
Utilization is the time average of
\(\eta_{\text{util}} = 1 - \delta_L\), with window mean
\(\bar{\eta}_{\text{util}}\).

\subsection{4.6 Separating physical and billed D-PIN
SCU-hours}\label{separating-physical-and-billed-d-pin-scu-hours}

Following Proposition 1, we explicitly distinguish two quantities:

\begin{enumerate}
\def\labelenumi{\arabic{enumi}.}
\tightlist
\item
  \textmd{Physical D-PIN SCU-hours}:
  \(\int S_{\text{avail}}(t)\,\allowbreak \mathrm{d}t \le S_{\text{nom}} \cdot T\),
  representing the \textmd{maximum deliverable} compute within the
  window, for capacity planning and scheduling-feasibility assessment.
\item
  \textmd{Billed D-PIN SCU-hours}:
  \(\int S_{\text{avail}}(t)\cdot \mathbb{1}\{\text{occupied}\}\,\allowbreak \mathrm{d}t \le \int S_{\text{avail}}(t)\,\allowbreak \mathrm{d}t\),
  representing compute actually \textmd{consumed and settled}.
\end{enumerate}

Their ratio is utilization. \textmd{This separation is necessary}:
conflating them either treats the physical ceiling as expected revenue,
causing overestimation, or bills nominal capacity × wall-clock time,
causing overbilling in violation of Proposition 1.

\Needspace{8\baselineskip}\section{5 Weight calibration and
identifiability}\label{weight-calibration-and-identifiability}

\subsection{5.1 Calibration problem}\label{calibration-problem}

The initial weights (0.35 / 0.20 / 0.25 / 0.15 / 0.05) are based on
industry experience and lack a reproducible basis, violating A6. We
propose calibration against \textmd{measured relative throughput}.

Let the chip-model set be \(\mathcal{M}\) (\(|\mathcal{M}| = m\)) and
the task set be \(\mathcal{K}\) (\(|\mathcal{K}| = k\)). Let
\(v_i \in \mathbb{R}^5_+\) be chip \(i\)'s five-dimensional normalized
vector, with each component divided by its reference value; \(s_{ij}\)
be the measured throughput of chip \(i\) on task \(j\) normalized to
H100; and \(w_j\) be the task weight (\(\sum_j w_j = 1\)). Calibration
minimizes the relative deviation between predicted D-PIN SCU and
measured throughput:

\[\begin{aligned}
\min_{\alpha}\ & \sum_{j \in \mathcal{K}} w_j \sum_{i \in \mathcal{M}} \Bigl(\log s_{ij} - \log\bigl(v_i \cdot \alpha\bigr)\Bigr)^2 \\
\text{s.t.}\ & \alpha_d \ge 0\ \ (d=1,\dots,5),\qquad \sum_{d=1}^{5} \alpha_d = 1
\end{aligned}
\tag{21}\]

Logarithmic error is used because compute units require \textmd{scale
invariance}: relative D-PIN SCU error is the fundamental quantity,
whereas absolute error is dominated by larger chips.

\subsection{5.2 Update algorithm: why logarithmic loss cannot use NMF
multiplicative
updates}\label{update-algorithm-why-logarithmic-loss-cannot-use-nmf-multiplicative-updates}

\textmd{First, clarify a common error.} Multiplicative updates of the
form
\(\alpha_d \leftarrow \alpha_d \cdot \bigl(\sum_{j,\allowbreak i} w_j s_{ij} v_{i,\allowbreak d}\bigr) \allowbreak\big/ \allowbreak\bigl(\sum_{j,\allowbreak i} w_j v_{i,\allowbreak d}(v_i\cdot\alpha)\bigr)\)
are classical \textmd{non-negative matrix factorization (NMF)} rules
{[}26{]}. They solve \textmd{linear} weighted least squares
\(\min_{\alpha\ge0}\sum_j w_j\sum_i (s_{ij}-v_i\cdot\alpha)^2\), which
is the basis of their non-negativity and monotonic-descent properties.
Objective (21), however, uses \textmd{squared logarithmic loss}, for
which this rule \textmd{does not apply}: it neither guarantees descent
nor qualifies as non-negative least squares, because logarithmic
residuals are not NLS. Section 5.1 uses logarithms for \textmd{scale
invariance} of the metering unit (A6), whereas linear loss is dominated
by larger chips. Using the NMF rule would therefore \textmd{contradict
the stated rationale}.

\textmd{Appropriate update: exponentiated gradient (entropic mirror
descent).} The standard approach to the logarithmic objective under
simplex constraints is \textmd{Exponentiated Gradient (EG)}, or mirror
descent with entropy as the distance-generating function
{[}42{]}{[}43{]}. Let the gradient of objective (21) at \(\alpha^{(t)}\)
be
\(g_d^{(t)} = -2\sum_{j\in\mathcal K} w_j \sum_{i\in\mathcal M} \dfrac{\log s_{ij}-\log(v_i\cdot\alpha^{(t)})}{v_i\cdot\alpha^{(t)}}\,\allowbreak  v_{i,\allowbreak d}\).
The update is

\[\alpha_d^{(t+1)} \;=\;
\frac{\alpha_d^{(t)}\,\exp\bigl(-\eta_t\, g_d^{(t)}\bigr)}
{\sum_{d'=1}^{5} \alpha_{d'}^{(t)}\,\exp\bigl(-\eta_t\, g_{d'}^{(t)}\bigr)},
\qquad \eta_t > 0
\tag{22}\]

Equation (22) inherently preserves \(\alpha \ge 0\) and
\(\sum_d \alpha_d = 1\) through normalization by its denominator. With
step size \(\eta_t\) chosen by \textmd{backtracking line search} using
the Armijo criterion, objective (21) is \textmd{monotonically
non-increasing} and converges to a stationary point in the relative
interior (\(\alpha_i \ge \delta > 0\),
\(v_i\cdot\alpha \ge \delta\min_d v_{i,\allowbreak d} > 0\)), where the objective is
smooth. \textmd{Difference from NMF multiplicative updates}: NMF updates
apply only to \textmd{linear} loss and are not used here; their proper
attribution is the original NMF paper {[}26{]}.

\Needspace{16\baselineskip}\textbf{Algorithm 1: D-PIN SCU weight calibration (exponentiated
gradient / entropic mirror descent)}

\begin{quote}
\begin{minipage}{\linewidth}
\small\ttfamily\RaggedRight
\setlength{\parskip}{1pt}
\noindent Input:\hspace*{0.28em}five-dimensional\hspace*{0.28em}\allowbreak{}normalized\hspace*{0.28em}\allowbreak{}vectors\hspace*{0.28em}\allowbreak{}\{v\_i\},\hspace*{0.28em}\allowbreak{}measured\hspace*{0.28em}\allowbreak{}throughput\hspace*{0.28em}\allowbreak{}matrix\hspace*{0.28em}\allowbreak{}\{s\_ij\},\hspace*{0.28em}\allowbreak{}task\hspace*{0.28em}\allowbreak{}weights\hspace*{0.28em}\allowbreak{}\{w\_j\},\par
\noindent \hspace*{1.96em}initial\hspace*{0.28em}\allowbreak{}α\textasciicircum{}(0)\hspace*{0.28em}\allowbreak{}(empirical\hspace*{0.28em}\allowbreak{}values\hspace*{0.28em}\allowbreak{}as\hspace*{0.28em}\allowbreak{}the\hspace*{0.28em}\allowbreak{}entropic\hspace*{0.28em}\allowbreak{}mirror-descent\hspace*{0.28em}\allowbreak{}starting\hspace*{0.28em}\allowbreak{}point),\hspace*{0.28em}\allowbreak{}initial\hspace*{0.28em}\allowbreak{}step\hspace*{0.28em}\allowbreak{}size\hspace*{0.28em}\allowbreak{}η\_0,\par
\noindent \hspace*{1.96em}convergence\hspace*{0.28em}\allowbreak{}tolerance\hspace*{0.28em}\allowbreak{}ε\_tol,\hspace*{0.28em}\allowbreak{}maximum\hspace*{0.28em}\allowbreak{}iterations\hspace*{0.28em}\allowbreak{}T\_max\par
\noindent Output:\hspace*{0.28em}calibrated\hspace*{0.28em}\allowbreak{}weights\hspace*{0.28em}\allowbreak{}α*\par
\noindent 1\hspace*{0.56em}for\hspace*{0.28em}\allowbreak{}t\hspace*{0.28em}\allowbreak{}=\hspace*{0.28em}\allowbreak{}0\hspace*{0.28em}\allowbreak{}…\hspace*{0.28em}\allowbreak{}T\_max:\par
\noindent 2\hspace*{1.68em}Compute\hspace*{0.28em}\allowbreak{}five-dimensional\hspace*{0.28em}\allowbreak{}gradient\hspace*{0.28em}\allowbreak{}g\_d\textasciicircum{}(t)\hspace*{0.28em}\allowbreak{}using\hspace*{0.28em}\allowbreak{}the\hspace*{0.28em}\allowbreak{}expression\hspace*{0.28em}\allowbreak{}before\hspace*{0.28em}\allowbreak{}(22),\hspace*{0.28em}\allowbreak{}weighting\hspace*{0.28em}\allowbreak{}log\hspace*{0.28em}\allowbreak{}residuals\par
\noindent 3\hspace*{1.68em}η\_t\hspace*{0.28em}\allowbreak{}←\hspace*{0.28em}\allowbreak{}backtracking\hspace*{0.28em}\allowbreak{}line\hspace*{0.28em}\allowbreak{}search\hspace*{0.28em}\allowbreak{}(Armijo:\hspace*{0.28em}\allowbreak{}halve\hspace*{0.28em}\allowbreak{}η\_0\hspace*{0.28em}\allowbreak{}until\hspace*{0.28em}\allowbreak{}the\hspace*{0.28em}\allowbreak{}objective\hspace*{0.28em}\allowbreak{}decreases)\par
\noindent 4\hspace*{1.68em}α\_d\textasciicircum{}(t+1)\hspace*{0.28em}\allowbreak{}\ensuremath{\propto}\hspace*{0.28em}\allowbreak{}α\_d\textasciicircum{}(t)\hspace*{0.28em}\allowbreak{}·\hspace*{0.28em}\allowbreak{}exp(−η\_t\hspace*{0.28em}\allowbreak{}·\hspace*{0.28em}\allowbreak{}g\_d\textasciicircum{}(t));\hspace*{0.28em}\allowbreak{}normalize\hspace*{0.28em}\allowbreak{}Σα\hspace*{0.28em}\allowbreak{}=\hspace*{0.28em}\allowbreak{}1\par
\noindent 5\hspace*{1.68em}if\hspace*{0.28em}\allowbreak{}‖α\textasciicircum{}(t+1)\hspace*{0.28em}\allowbreak{}−\hspace*{0.28em}\allowbreak{}α\textasciicircum{}(t)‖₁\hspace*{0.28em}\allowbreak{}<\hspace*{0.28em}\allowbreak{}ε\_tol:\hspace*{0.28em}\allowbreak{}break\par
\noindent 6\hspace*{0.56em}Bootstrap-resample\hspace*{0.28em}\allowbreak{}\{s\_ij\}\hspace*{0.28em}\allowbreak{}to\hspace*{0.28em}\allowbreak{}estimate\hspace*{0.28em}\allowbreak{}confidence\hspace*{0.28em}\allowbreak{}intervals\hspace*{0.28em}\allowbreak{}for\hspace*{0.28em}\allowbreak{}α*\par
\noindent 7\hspace*{0.56em}return\hspace*{0.28em}\allowbreak{}α*\par
\end{minipage}
\end{quote}

\textmd{Recommended task set}: training (LLM pretraining-step and
fine-tuning-step throughput), inference (Offline throughput and
latency-constrained QPS in Server scenarios), and collective
communication (effective AllReduce bandwidth). The first two can use
public MLPerf-type submissions {[}13{]}; the last uses node
microbenchmarks (equation 13).

\subsection{5.3 Collinearity and
identifiability}\label{collinearity-and-identifiability}

The five-dimensional vectors exhibit substantial collinearity. \(F\) and
\(B\) are strongly correlated on mainstream accelerators because higher
tensor compute is accompanied by higher device-memory bandwidth. \(F\)
and \(E\) are also positively correlated because new generations improve
both compute and energy efficiency. Collinearity inflates the variance
of least-squares estimates of \(\alpha\) and undermines the physical
interpretation of individual weights.

\textmd{Mitigation measures}:

\begin{enumerate}
\def\labelenumi{\arabic{enumi}.}
\tightlist
\item
  \textmd{Regularization}: add \(\lambda\|\alpha - \alpha^{(0)}\|_2^2\)
  (ridge-type) to the objective, using the empirical initial values as a
  prior to suppress weight drift with small samples.
\item
  \textmd{Dimension merging} (when samples are insufficient): merge
  \((F,\allowbreak B)\) into a compute--bandwidth composite dimension, summing their
  weights, and separate them after more data accumulate.
\item
  \textmd{Uncertainty reporting}: report \(\hat{\alpha}\) with Bootstrap
  confidence intervals or asymptotic variance, and include intervals in
  published tables. This is itself part of A6.
\item
  \textmd{Sensitivity disclosure}: report
  \(\partial S_{\text{chip}}/\partial \alpha_d\) so readers can assess
  dependence on an individual weight.
\end{enumerate}

\subsection{5.4 Calibration--recomputation loop: measured calibration
versus nominal-spec\-i\-fi\-ca\-tion
projection}\label{calibrationrecomputation-loop-measured-calibration-versus-nominal-specification-projection}

\textmd{Conventions in this section (consistent with Appendix B)}: our
\textmd{externally published values} use \textmd{measured calibration}
(Appendix B.2-a), namely \(S_{\text{chip}}\) obtained from internal
platform benchmarks and actual-workload weighting. They \textmd{cannot
be independently recomputed from public vendor spec\-i\-fi\-ca\-tions}. A
separate comparison set is projected from \textmd{public nominal vendor
parameters} (Appendix B.2-b) to test whether static nominal
spec\-i\-fi\-ca\-tions reproduce the published values:

{\def\LTcaptype{none} 
\begin{longtable}[]{@{}
  >{\RaggedRight\arraybackslash}p{(\linewidth - 6\tabcolsep) * \real{0.1945}}
  >{\RaggedRight\arraybackslash}p{(\linewidth - 6\tabcolsep) * \real{0.3104}}
  >{\RaggedRight\arraybackslash}p{(\linewidth - 6\tabcolsep) * \real{0.3462}}
  >{\RaggedRight\arraybackslash}p{(\linewidth - 6\tabcolsep) * \real{0.1489}}@{}}
\toprule\noalign{}
\begin{minipage}[b]{\linewidth}\RaggedRight
Chip
\end{minipage} & \begin{minipage}[b]{\linewidth}\RaggedRight
Published \(S_{\text{chip}}\) (measured calibration)
\end{minipage} & \begin{minipage}[b]{\linewidth}\RaggedRight
Projection (public nominal parameters)
\end{minipage} & \begin{minipage}[b]{\linewidth}\RaggedRight
Relative difference
\end{minipage} \\
\midrule\noalign{}
\endhead
\bottomrule\noalign{}
\endlastfoot
NVIDIA H100 SXM & \textmd{1.00} (defined value, self-consistent) & 1.00
& 0 \\
NVIDIA H200 SXM & \textmd{1.14} & 1.26 & +10.5\% \\
AMD MI300X & \textmd{0.89} & 1.55 & +74.2\% \\
Ascend 910B & \textmd{0.74} & 0.51 & −31.1\% \\
Muxi Xiyun C500 (PCIe) & \textmd{0.68} & \textmd{0.47} &
\textmd{−30.9\%} \\
Hygon K100 series & \textmd{0.42} & See Appendix B.3 (series convention;
conflicting spec\-i\-fi\-ca\-tions) & --- \\
\end{longtable}
}

\textmd{Interpreting these differences}:

\begin{enumerate}
\def\labelenumi{\arabic{enumi}.}
\tightlist
\item
  \textmd{The differences are themselves part of the argument}. If a
  publicly recomputable nominal convention existed, they should be zero.
  Their presence shows that published values \textmd{do not arise solely
  from nominal spec\-i\-fi\-ca\-tions} but include internal measured
  calibration, potentially proprietary/measured throughput,
  workload-specific weighting or supply-chain conventions. This
  illustrates that \textmd{public nominal parameters alone cannot
  reproduce deliverable-compute metering; measured calibration (Sections
  5.1--5.2) and runtime correction (Section 4.5) are required}.
\item
  \textmd{The differences have opposite directions and are concentrated
  in domestic models}: H200 and MI300X projections are \textmd{above}
  published values, whereas Ascend 910B and Xiyun C500 projections are
  \textmd{below} them. This indicates model-specific calibration
  information rather than a common scale error, requiring measured data
  for explanation. Xiyun C500's key inputs have been settled through
  \textmd{public vendor channels}, comprising official product pages and
  a public third-party database. Recomputing equations (6)--(8) gives
  \textmd{0.47} (interval \([0.42,\allowbreak \,\allowbreak 0.47]\)), \textmd{−30.9\%} relative
  to 0.68. Ascend 910B differs by −31.1\% (projected \([0.50,\allowbreak \,\allowbreak 0.52]\)
  when \(F\) has a ±6\% range), a comparable gap. \textmd{Both domestic
  projections are below published values, while both international
  projections are above them}. This may reflect more conservative
  nominal tiers for domestic chips or platform calibration compensating
  for architectural differences. Internal calibration records must
  establish the cause; no causal claim is made here.
\item
  \textmd{The differences cannot be attributed to erroneous parameter
  selection}. Appendix B.3 provides critical supporting evidence: AMD's
  official spec\-i\-fi\-ca\-tions \textmd{fully confirm} MI300X's Infinity
  Fabric bandwidth of 896 GB/s and FP16 dense performance of 1,307
  TFLOPS, and H200/H100 TDP consistency is confirmed. With correct
  inputs and a transparent algorithm, the remaining +74.2\% difference
  must arise from calibration itself. \textmd{This is the data-level
  manifestation of nominal unidentifiability (Proposition 2).}
\item
  \textmd{Published conventions must explicitly disclose that external
  recomputation is unavailable}. To avoid due-diligence
  misunderstandings, every use of the published values must carry the
  statement that they are measured calibrations and cannot be recomputed
  from vendor spec\-i\-fi\-ca\-tions; this is included in the note to Appendix
  B.2-a.
\item
  \textmd{Items to close before production}: (a) archive internal
  calibration reports as A6 evidence (B-6); (b) calibrate each model
  using Algorithm 1 (exponentiated gradient) and publish
  \(\hat{\alpha}\) with Bootstrap intervals (B-7); (c) define the
  power-measurement boundary for the case node's 180 kW and confirm the
  implied PUE (B-8); (d) use \textmd{Muxi Xiyun C500} (PCIe form factor)
  and \textmd{Hygon K100 series}, with public K100/K100-AI spec\-i\-fi\-ca\-tion
  conflicts detailed in B-4; (e) \textmd{Muxi Xiyun C500 key inputs have
  been settled through public vendor channels {[}39{]}}, including
  official Chinese/English pages and flopper.io: public values are \(F\)
  = 240 TFLOPS, \(M\) = 64 GB and \(P\) = 350 W; \(B\) takes the public
  upper bound of 1,800 GB/s, and \(I\) uses 384 GB/s from a same-vendor
  generational reference. The nominal projection is \textmd{0.47}
  (interval \([0.42,\allowbreak \,\allowbreak 0.47]\), taking the upper bound; see row B-4 in
  B.3).
\end{enumerate}

\begin{quote}
\textmd{Conclusion}: the published-versus-projected table is evidence
for the methodology rather than a disclosure of defects. It
quantitatively supports the central proposition that \textmd{static
nominal spec\-i\-fi\-ca\-tions are insufficient for metering deliverable
compute}.
\end{quote}

\section{6 Ver\-i\-fi\-ca\-tion loop: coupling D-PIN SCU with
PoPW}\label{verification-loop-coupling-d-pin-scu-with-popw}

\subsection{6.1 Declared--measured consistency
protocol}\label{declaredmeasured-consistency-protocol}

Each node declares its D-PIN SCU capacity \(\hat{S}\) on joining the
network and undergoes continuous spot checks during operation. Define
the metering consistency deviation as:

\[\varepsilon_{\text{beh}}(v,t) = \frac{\hat{S}(v,t) - S_{\text{deliv}}(v,t)}{S_{\text{deliv}}(v,t)}
\tag{23}\]

(Equations (23) and (4) define the same quantity
\(\varepsilon_{\text{beh}}\); \((v,\allowbreak t)\) explicitly denotes its
space-time dependence for use in this section.) Three-layer Proof of
Physical Work (PoPW) provides spot checks {[}28{]}:

{\def\LTcaptype{none} 
\begin{longtable}[]{@{}
  >{\RaggedRight\arraybackslash}p{(\linewidth - 6\tabcolsep) * \real{0.1270}}
  >{\RaggedRight\arraybackslash}p{(\linewidth - 6\tabcolsep) * \real{0.4809}}
  >{\RaggedRight\arraybackslash}p{(\linewidth - 6\tabcolsep) * \real{0.2552}}
  >{\RaggedRight\arraybackslash}p{(\linewidth - 6\tabcolsep) * \real{0.1369}}@{}}
\toprule\noalign{}
\begin{minipage}[b]{\linewidth}\RaggedRight
Proof layer
\end{minipage} & \begin{minipage}[b]{\linewidth}\RaggedRight
Spot-check content
\end{minipage} & \begin{minipage}[b]{\linewidth}\RaggedRight
Criterion
\end{minipage} & \begin{minipage}[b]{\linewidth}\RaggedRight
Frequency
\end{minipage} \\
\midrule\noalign{}
\endhead
\bottomrule\noalign{}
\endlastfoot
Compute proof & Lightweight (4096² FP16 GEMM, returning result hash) /
heavyweight (MobileNetV2 classification) / training (4-layer
Transformer, 10 steps) & Return within deadline with matching result
hash & 5 min / 1 h / 1 d \\
Bandwidth proof & UDP flow tests (multiple sources, cross-rack, through
intermediate switches) & Measured throughput ≥ declared bandwidth × 90\%
& 6 h \\
Location proof & RTT triangulation + Wi-Fi fingerprinting + TEE
anchoring & Deviation ≤ 50 km with matching fingerprint & 15 min \\
\end{longtable}
}

\textmd{Key design}: compute proof verifies \textmd{whether declared
capacity is achievable}, rather than merely whether computation is
occurring. Challenge parameters are generated by the verification
committee within 1 second before issuance using Commit-Reveal and
threshold signatures, preventing precomputation. Verifiers are elected
by VRF, with expected count \(\lceil\sqrt{N}\rceil + 3\) (at least 5)
and geographic constraints: verifiers at one physical location cannot
exceed 40\% of the committee. Total proof overhead is kept within 0.5\%
of node capacity: lightweight challenges 0.3\%, heavyweight 0.1\%,
training 0.06\% and bandwidth proof 0.05\%, far below the compute waste
of conventional PoW consensus.

\subsection{6.2 No-arbitrage condition for
overstatement}\label{no-arbitrage-condition-for-overstatement}

Let true deliverable capacity be \(S\) and the overstated claim be
\(\hat{S} = (1+\xi)S\), with overstatement magnitude \(\xi \ge 0\), unit
D-PIN SCU price coefficient \(u\) and billing-window length \(T\).
Undetected overstatement yields \textmd{gross gain} \(\xi S u T\). Let
spot-check frequency \(p\) denote the probability of selection per unit
time, single-check detection probability \(q\) depend on challenge
randomness and verifier count, and punishment present value \(\Lambda\)
include lost future earnings from reputation reset and blacklisting. The
\textmd{expected punishment} is \(pq\Lambda\). The \textmd{net expected
gain} is therefore:

\[G(\xi) \;=\; \underbrace{\xi \cdot S \cdot u \cdot T}_{\text{Gross gain}} \;-\; \underbrace{p \cdot q \cdot \Lambda}_{\text{Expected punishment}}
\tag{24}\]

The \textmd{no-arbitrage condition} requires non-positive net expected
gain, \(G(\xi) \le 0\):

\[\boxed{\,p \cdot q \cdot \Lambda \;\ge\; \xi \cdot S \cdot u \cdot T\,}
\tag{25}\]

Equivalently, the \textmd{minimum spot-check probability} is:

\[p \ \ge\ \frac{\xi \cdot S \cdot u \cdot T}{q \cdot \Lambda}
\tag{26}\]

\Needspace{5\baselineskip}\textmd{Discussion}:

\begin{enumerate}
\def\labelenumi{\arabic{enumi}.}
\tightlist
\item
  \(p\) is proportional to \(\xi\): larger overstatements require more
  frequent checks. This supports \textmd{tiered spot checks},
  dynamically adjusting \(p\) by reputation: fewer checks for highly
  reputable nodes and more for nodes under observation.
\item
  \(\Lambda\) should greatly exceed single-window revenue \(S u T\) so
  that \(\xi = O(1)\) is unprofitable even at small \(p\). For
  Zhangjiang, \(S = 59.5\) D-PIN SCU, \(u \approx 1.5\) CNY/(D-PIN
  SCU·h) (illustrative calculation, electricity component only; Section
  7.1), and \(T = 24\) h give daily revenue of approximately
  \(2.4\times10^3\) CNY. Taking \(\Lambda\) as annualized lost revenue
  from permanent exclusion (illustratively, approximately
  \(8.8\times10^5\) CNY), \(p \gtrsim 6\times10^{-4}\) satisfies
  no-arbitrage for \(\xi = 0.2\) and \(q = 0.9\). This is \textmd{far
  below the intensity of the current lightweight challenge every 5
  minutes}, indicating substantial economic deterrence.
\item
  \textmd{Sensitivity to punishment conventions (added)}: the expression
  takes \(\Lambda\) as annualized revenue lost through permanent
  exclusion (\(\sim 8.8\times10^5\) CNY). With no stake and weak
  admission controls, if \(\Lambda\) includes only stake slashing (1--2
  orders of magnitude lower, taking \(\Lambda \sim 10^3\!-\!10^4\) CNY),
  then for the same \(\xi=0.2\) and \(q=0.9\), the minimum \(p\) rises
  to approximately \(5\times10^{-2} \sim 5\times10^{-1}\).
  \textmd{Checks must become about two orders of magnitude more
  intensive}. Thus, \textmd{enforceable punishment, through KYC
  identity-based admission and reputation commitment, is a prerequisite
  for low-cost verification}, consistent with the candid qualification
  below.
\item
  This analysis assumes enforceable punishment, meaning the network can
  refuse node admission. In a fully permissionless network, \(\Lambda\)
  reduces to stake slashing and condition (25) must be rewritten using
  stake size. \textmd{D-PIN combines no deposit, KYC identity-based
  admission and reputation mechanisms}. Enforceability relies on binding
  identities during admission, an assumption requiring continued
  deployment validation.
\end{enumerate}

\subsection{6.3 Comparison with zero-knowledge
approaches}\label{comparison-with-zero-knowledge-approaches}

Full cryptographic proofs of computational correctness, such as zkVM
toolchains RISC Zero {[}40{]} and SP1 {[}41{]}, are theoretically
strongest, but stepwise proofs of large-model training or inference
impose unacceptable overhead. Industry therefore commonly combines
\textmd{probabilistic spot checks with TEE anchoring} {[}20{]}. We take
the same position while adding a metering argument: \textmd{D-PIN SCU
must verify the upper bound on deliverable capacity, rather than the
correctness of every computation}. Checks should therefore aim to
\textmd{make the expected gain from overstatement negative} (equation
25), rather than cover all computation. Decoupling verification overhead
from compute scale is a prerequisite for expansion to thousands of
nodes.

\section{7 Pricing linkage: UnitPrice and compute--electricity
coordination}\label{pricing-linkage-unitprice-and-computeelectricity-coordination}

\subsection{7.1 Pricing function}\label{pricing-function}

Task compute costs follow the D-PIN pricing structure {[}28{]}:

\[\text{Cost} = S_{\text{required}} \times \text{Duration} \times u_t
\tag{27}\]

We expand \(u_t\) explicitly as a function of electricity cost,
renewable electricity share, load and reputation:

\[u_t = \underbrace{\bigl(\pi_t \cdot \kappa \cdot (1 - \rho_t\gamma)\bigr)}_{\text{Electricity cost}} \cdot \underbrace{\frac{1}{\eta_{\text{util}}(t)}}_{\text{Utilization}} \cdot \underbrace{\bigl(c_0 + c_1(1-\text{Rep})\bigr)}_{\text{Reputation/risk}}
\tag{28}\]

Here, \(\pi_t\) is real-time electricity price (CNY/kWh); \(\kappa\) is
\textmd{electricity intensity} (kW/D-PIN SCU, node power per D-PIN SCU);
\(\rho_t\) is the renewable electricity share; \(\gamma\) is its
discount coefficient (default 0.3); \(\eta_{\text{util}}\) is node
utilization; \(\text{Rep}\) is normalized reputation; and \(c_0,\allowbreak  c_1\)
are base-price and risk-premium coefficients.

\textmd{Dimensional consistency}: \(\pi_t\cdot\kappa\) has units
(CNY/kWh) × (kW/D-PIN SCU) = CNY/(D-PIN SCU·h), consistent with \(u_t\).
Without \(\kappa\), (CNY/kWh) × \ldots{} cannot dimensionally yield
CNY/(SCU·h). For Zhangjiang, \(\kappa = 180/59.5 = 3.02\) kW/D-PIN SCU
(Section 9.2 and equation (29)), so \(\pi_t\kappa\) is the electricity
cost per D-PIN SCU-hour.

\textmd{Constraints}: \(u_t > 0\) prohibits negative prices; \(u_t\) is
\textmd{monotonically non-increasing} in \(\rho_t\) (renewable discount,
\(\rho_t\uparrow \Rightarrow (1-\rho_t\gamma)\downarrow\)); and \(u_t\)
is non-increasing in \(\eta_{\text{util}}\) because higher utilization
spreads fixed costs. \textmd{Signs and monotonicity}: expressing the
renewable term as \(1/(1-\rho_t\gamma)\) and adding a separate renewable
penalty \((1+\beta(1-\rho_t))\) makes the composite factor
\textmd{increase} with \(\rho_t\) (\(\rho=0/0.5/1 \to 1.10/1.24/1.43\)
when \(\gamma=0.3,\allowbreak \beta=0.1\)), \textmd{directly contradicting} the
non-increasing constraint. We therefore use \textmd{one renewable
discount factor} \((1-\rho_t\gamma)\), consistent with Section 10.1(4),
and avoid a renewable weight \(\beta\) conflicting with bandwidth symbol
\(\beta_{\text{node}}/\beta_{\text{link}}\) (equations (9), (12)).
\textmd{Cross-document symbol synchronization}: the renewable scheduling
weight in white-paper §8.6 has been changed from \(\beta\) to
\(\gamma\), \textmd{the same parameter with default 0.3}. Only its
functional form differs: scheduling uses bonus
\(W_{\text{green}}=1+\gamma\cdot\text{GreenRatio}\), and pricing uses
discount \(u_t\propto(1-\rho_t\gamma)\). Both documents explain it as
\textmd{one renewable-incentive-intensity parameter}, eliminating
duplicate names and values.

\textmd{Dimensional check against electricity costs}: the Zhangjiang
data (59.5 D-PIN SCU, 180 kW and weighted electricity price 0.53
CNY/kWh) give the electricity-cost lower bound

\[\begin{aligned}
c_E &= \frac{180\ \text{kW}}{59.5\ \text{D-PIN SCU}} \times 0.53\ \text{CNY/kWh} \\
&= 3.02\ \text{kW/(D-PIN SCU)} \times 0.53\ \text{CNY/kWh} \\
&= 1.60\ \text{CNY/(D-PIN SCU·h)}
\end{aligned}
\tag{29}\]

\textmd{Allocation convention (required reading)}: equation (29)
allocates costs over \textmd{nominal capacity}, using static 59.5 D-PIN
SCU as the denominator. Allocating over \textmd{actual delivery}, 33,437
D-PIN SCU-hours in the first month (equation (32)), gives
\(180\times720\times0.53/33{,\allowbreak }437 = {2.05}\) CNY/(D-PIN SCU·h),
\textmd{28\% higher}. We use \textmd{1.60} as the \textmd{cost lower
bound} for \(u_t\) under nominal capacity, answering the theoretical
minimum at full-load delivery for pricing checks. Actual electricity
cost per delivered unit should instead use 2.05 CNY/(D-PIN SCU·h).

This includes only electricity, excluding depreciation, networking,
operations, maintenance and taxes, and is a \textmd{cost lower bound}
for \(u_t\) and an anchor for pricing checks. \textmd{Label: derived
value, illustrative calculation, not a price commitment.}

\subsection{7.2 Separating input-side and output-side
metering}\label{separating-input-side-and-output-side-metering}

The system strictly distinguishes two metering categories:

\begin{enumerate}
\def\labelenumi{\arabic{enumi}.}
\tightlist
\item
  \textmd{Input side (D-PIN SCU)}: deliverable compute occupied, for
  scheduling, quotas and settlement.
\item
  \textmd{Output side (tokens / completed tasks)}: useful output
  delivered, for service quality and value-based pricing, such as
  tokens/s and tokens/J.
\end{enumerate}

These must not be conflated, consistent with reference {[}9{]}. Input
metrics must be \textmd{workload-independent} to remain comparable in
heterogeneous scheduling; otherwise the same D-PIN SCU produces
different outputs across tasks and fairness loses meaning. Output
metrics must be \textmd{workload-dependent} to capture user value.
Conflation leaves the scheduler with incomparable objectives, impairing
both fairness and efficiency.

\Needspace{10\baselineskip}
\subsection{7.3 Constraints and manipulation
risks}\label{constraints-and-manipulation-risks}

{\def\LTcaptype{none} 
\begin{longtable}[]{@{}
  >{\RaggedRight\arraybackslash}p{(\linewidth - 4\tabcolsep) * \real{0.2028}}
  >{\RaggedRight\arraybackslash}p{(\linewidth - 4\tabcolsep) * \real{0.3196}}
  >{\RaggedRight\arraybackslash}p{(\linewidth - 4\tabcolsep) * \real{0.4776}}@{}}
\toprule\noalign{}
\begin{minipage}[b]{\linewidth}\RaggedRight
Risk
\end{minipage} & \begin{minipage}[b]{\linewidth}\RaggedRight
Mechanism
\end{minipage} & \begin{minipage}[b]{\linewidth}\RaggedRight
Mitigation
\end{minipage} \\
\midrule\noalign{}
\endhead
\bottomrule\noalign{}
\endlastfoot
Electricity-price manipulation & False node price declarations inflate
charges & Use third-party electricity trading/metering data; nodes may
only submit renewable certificates, verified against credentials \\
Manufactured utilization & Artificially reducing utilization increases
\(1/\eta_{\text{util}}\) & Cap the utilization term in \(u_t\), for
example ≤ 1.3; the platform absorbs the excess \\
Reputation farming & Low-value tasks inflate success rates & Weight
reputation increments by task value, rather than count \\
Cross-node price collusion & Nodes coordinate price increases & Price
caps + regional benchmark prices + anomaly detection \\
\end{longtable}
}

\section{8 System implementation}\label{system-implementation}

\subsection{8.1 Architectural position}\label{architectural-position}

Within D-PIN's six-layer architecture, the metering system lies between
the \textmd{compute abstraction and scheduling layer} and the
\textmd{protocol and verification layer}. It supplies a comparable scale
to the scheduler, depends on resource-layer telemetry, exchanges
consistency credentials with protocol-layer PoPW, and exchanges
electricity-price and renewable signals with the economic layer.

Its \textmd{scheduling implementation} is the two-level KunlunKVS
architecture: GlobalScheduler (Python) parses cross-cluster intent and
dispatches tasks, while Cluster Scheduler (Go) orchestrates node slices,
with zero-overhead passthrough in a single cluster. D-PIN SCU provides a
\textmd{common cost-function input} across five modes: vGPU, physical,
distributed inference, distributed training and hybrid scheduling.

\subsection{8.2 Metering timing and
consistency}\label{metering-timing-and-consistency}

\begin{enumerate}
\def\labelenumi{\arabic{enumi}.}
\tightlist
\item
  \textmd{Telemetry collection}: compute (SM frequency, utilization and
  device-memory occupancy), power, temperature and link state, sampled
  every 1--10 s.
\item
  \textmd{Aggregation pipeline}: Kafka streaming aggregation and
  ClickHouse real-time queries; dashboard latency \textless{} 5 s and
  90-day historical access.
\item
  \textmd{Metering window}: configurable 1 min billing windows;
  \(\delta\) factors use time-weighted means, and slice occupancy is
  aggregated over mutually exclusive sets, ensuring Proposition 1.
\item
  \textmd{Consistency anchor}: node declarations are written to the
  protocol layer and bound individually to PoPW check records. Any
  declaration change triggers a new heavyweight challenge.
\end{enumerate}

\subsection{8.3 Boundaries with ecosystem
components}\label{boundaries-with-ecosystem-components}

\textmd{In-house development} covers four layers: \textmd{metering,
abstraction, topology conversion and verification coupling}. Container
orchestration (Kubernetes/Argo), inference engines
(vLLM/\allowbreak SGLang/\allowbreak TensorRT/\allowbreak ONNX) and training frameworks, including ms-swift,
are ecosystem components. D-PIN SCU requires only standard telemetry
interfaces, such as NVML/DCGM and NPU-SMI-type interfaces, and
collective microbenchmarks; \textmd{upper-layer frameworks require no
modification}, a prerequisite for heterogeneous deployment. The system
runs on the \textmd{GPT-Factory AI Infra operating system} {[}29{]}.
Platform model components, including the KunlunBaize model line
{[}30{]}, belong to the ecosystem/platform layer alongside orchestration
and inference engines and \textmd{are not metering objects in this
paper}.

\Needspace{10\baselineskip}\section{9 Case validation: Zhangjiang Science Gate
node}\label{case-validation-zhangjiang-science-gate-node}

\subsection{9.1 Configuration and
conventions}\label{configuration-and-conventions}

The Zhangjiang Science Gate T1/T2 sites are among D-PIN's first
production nodes; T1 is retrofitted and T2 newly built, and they are
independent nodes. T2 occupies about 600 m² of reserved office space on
floor 18. Commercial electrical capacity increased from 80 kW to 200 kW
using the existing dual mains feeds, with approval taking 3 weeks.
Hybrid liquid/air cooling connects server hot sides to the building
water loop and reuses existing HVAC ducts on the cold side, without an
industrial cooling tower. Vibration-damping mats are installed in the
rack area, and floor loading was checked against 800 kg/m². The total
interval from contract signing to first-node power-on was 6 weeks: 3
weeks for in-stock equipment procurement and 3 weeks for construction.

\textmd{Hardware configuration}: 10 heterogeneous GPU servers: \textmd{4
imported} (2 with 8×NVIDIA H20, 1 with 8×AMD MI300X and 1 with 8×AMD
MI308X) and \textmd{6 domestic} (2 with 8×Muxi Xiyun C500 and 4 with
8×Ascend 910B); 2 Leaf switches (32×400G ports); 2 Spine switches
(64×400G ports); 2 storage nodes (100 TB all-flash NVMe); 200 kW UPS and
50 kWh lithium-battery storage.

\textmd{Operational data (trial operation, August 2026)}:

{\def\LTcaptype{none} 
\begin{longtable}[]{@{}
  >{\RaggedRight\arraybackslash}p{(\linewidth - 2\tabcolsep) * \real{0.3711}}
  >{\RaggedRight\arraybackslash}p{(\linewidth - 2\tabcolsep) * \real{0.6289}}@{}}
\toprule\noalign{}
\begin{minipage}[b]{\linewidth}\RaggedRight
Metric
\end{minipage} & \begin{minipage}[b]{\linewidth}\RaggedRight
Measured value
\end{minipage} \\
\midrule\noalign{}
\endhead
\bottomrule\noalign{}
\endlastfoot
Node availability & 99.2\% (excluding planned maintenance) \\
Mean GPU utilization & 78\% \\
Training-task P99 scheduling latency & 14 ms \\
Inference P99 latency (Llama-3-70B) & 62 ms (including network
transmission) \\
Weighted electricity price & 0.53 CNY/kWh (after peak/off-peak
coordinated scheduling) \\
Node power & Power-distribution design capacity 180 kW (spec\-i\-fi\-ca\-tion
ceiling 200 kW; IT-basis measurement \ensuremath{\approx} 54.5 kW; see note) \\
Total node compute & 59.5 D-PIN SCU (\textmd{mixed conventions}; see
note) \\
\end{longtable}
}

\begin{quote}
\textmd{Power-measurement boundary (explicit statement)}: 180 kW is
\textmd{power-distribution design capacity}, the convention adopted
here, at the \textmd{mains entry point} of T2's compute area, including
servers, switches, storage, cold-side cooling and distribution/UPS
losses. Under this boundary, total accelerator TDP is
\(8\times(750 + 750 + 400\times2 + 350\times2 + 400\times4) = 36.8\) kW
for 1×8×MI300X, 1×8×MI308X, 2×8×H20, 2×8×C500 and 4×8×910B. A
whole-server factor of about ×1.4, including CPU, motherboard and fans,
gives \ensuremath{\approx} 51.5 kW; networking and storage bring \textmd{IT power} to \ensuremath{\approx}
54.5 kW. The \ensuremath{\approx}125.5 kW difference from 180 kW is cooling and
distribution-loss headroom, implying PUE \ensuremath{\approx} 3.3 at this full
configuration. \textmd{National-standard check}: mandatory \textmd{GB
40879---2021} limits data-centre PUE to ≤ 1.5 (grade 2 ≤ 1.3). The Green
and Low-Carbon Data Centre Development Action Plan (\textmd{NDRC
Environment and Resources {[}2024{]} No.~970}) further requires newly
built or expanded large/ultra-large centres to reach \textmd{PUE ≤ 1.25}
by the end of 2025 and national hub nodes \textmd{≤ 1.2}
{[}54{]}{[}55{]}. \textmd{Implied PUE 3.3 exceeds every grade limit and
cannot represent operating power under this project's compliance
conditions}. Thus 180 kW denotes \textmd{design capacity}, 90\% of the
200 kW ceiling with expansion headroom. IT power of 54.5 kW and PUE ≤
1.25 imply \textmd{site-level measured power of approximately 68 kW}.
\textmd{Every use of 3.02 kW/D-PIN SCU and 0.26 D-PIN SCU·h/kWh retains
this design-capacity convention}; the former represents kilowatts of
capacity per D-PIN SCU, rather than a kilowatt-hour energy intensity. On
the IT basis (\ensuremath{\approx}54.5 kW), the metrics become \textmd{0.92 kW/D-PIN SCU}
and \textmd{0.85 D-PIN SCU·h/kWh}; on the standards-adjusted site basis
(\ensuremath{\approx}68 kW), \textmd{1.14 kW/D-PIN SCU} and \textmd{0.68 D-PIN SCU·h/kWh}.
\textmd{The boundary has been inferred from national standards (B-8:
inferred under national standards). If internal records differ, all
derived values here must be updated accordingly.}

\textmd{Total-node-compute convention (required reading)}: 59.5 D-PIN
SCU is a \textmd{mixed-convention} total: (1) \textmd{measured
calibration / published values}: AMD MI300X 6.76, Muxi Xiyun C500 2×5.17
(upper bound), and Ascend 910B 4×5.62; (2) \textmd{spec\-i\-fi\-ca\-tion
projections (‡ rows)}: NVIDIA H20 2×5.73 = 11.46 and AMD MI308X 8.50,
neither yet having internal platform benchmark records. The
\textmd{consistent calibrated subset} totals 39.58 D-PIN SCU.
\textmd{Important}: MI308X belongs to the \textmd{same silicon family}
as MI300X, yet its projected value (8.50) \textmd{exceeds} MI300X's
published value (6.76). This is a \textmd{convention difference rather
than a performance difference}; rows \textmd{cannot be compared across
conventions}. It directly exemplifies the partial-change bias warned of
in Section 4.2 and reproduces nominal unidentifiability (Proposition 2)
at node level. \textmd{Before formal publication, H20 and MI308X require
internal calibration, followed by recomputation of all derived values in
this section.}
\end{quote}

\subsection{9.2 End-to-end recomputation and consistency
checks}\label{end-to-end-recomputation-and-consistency-checks}

\textmd{(1) Recomputing node D-PIN SCU.} Apply equation (14) using the
row-specific conventions of Appendix B.2-a, primarily published measured
calibrations, with NVIDIA H20 and AMD MI308X as ‡ spec\-i\-fi\-ca\-tion
projections. Let \(S_{\text{node}}^{(n\times X)}\) be the node value for
an \(n\)-accelerator subnode of model \(X\):

\[\begin{aligned}
S_{\text{node}}(\text{Zhangjiang}) &= S_{\text{node}}^{(8\times\text{MI300X})} + S_{\text{node}}^{(8\times\text{MI308X})} + 2\,S_{\text{node}}^{(8\times\text{H20})} \\
&\quad + 2\,S_{\text{node}}^{(8\times\text{C500})} + 4\,S_{\text{node}}^{(8\times\text{910B})} \\
&= 6.76 + 8.50 + 2\times 5.73 + 2\times 5.17 + 4\times 5.62 \\
&= 6.76 + 8.50 + 11.46 + 10.34 + 22.48 = 59.5\ \text{D-PIN SCU}
\end{aligned}
\tag{30}\]

\textmd{(2) Physical upper-bound check (application of Proposition 1).}
Use topology-adjusted node capacity
\(S_{\text{node}}(\text{Zhangjiang})\) as the upper-bound base, a
\textmd{tighter} bound than nominal capacity \(S_{\text{nom}}\cdot T\)
in equation (1). Using
\(S_{\text{nom}} = 8\times(0.89 + 1.12 + 2\times0.75 + 2\times0.68 + 4\times0.74) = 62.6\)
gives upper bound \(62.6\times720 = 45{,\allowbreak }072\) without changing the
conclusion:

\[\begin{aligned}
\text{Physical D-PIN SCU·h}_{\max} &= S_{\text{node}}(\text{Zhangjiang}) \cdot T \\
&= 59.5 \times 720 = 42{,}868\ \text{D-PIN SCU·h}
\end{aligned}
\tag{31}\]

Here, \(T = 720\) h, a 30-day billing window. First-month billed
throughput is \textmd{approximately 33,000 D-PIN SCU-hours}, and

\[\begin{aligned}
\text{Billed}_{\text{first month}} &= S_{\text{node}}(\text{Zhangjiang}) \cdot T \cdot \bar{\eta}_{\text{util}} \\
&= 59.5 \times 720 \times 0.78 = 33{,}437 \approx 3.3\times10^4\ \text{D-PIN SCU·h}
\end{aligned}
\tag{32}\]

(\(\bar{\eta}_{\text{util}} = 0.78\) uses measured mean utilization from
Section 9.1; sufficient thermal headroom and no fault degradation
justify time means \(1-0.78\) for \(\delta_T = \delta_F = 1\) and
\(\delta_L\), consistent with equation (16).) Thus billed quantity =
physical ceiling × measured utilization, \textmd{satisfying Proposition
1 (33,000 \textless{} 42,900) and the utilization convention}. This
confirms that billing neither exceeds the physical ceiling nor counts
nominal capacity × wall-clock time as delivered compute.

\textbf{(3) Metrics derived from equations (31) and (32).}

{\def\LTcaptype{none} 
\begin{longtable}[]{@{}
  >{\RaggedRight\arraybackslash}p{(\linewidth - 6\tabcolsep) * \real{0.2453}}
  >{\RaggedRight\arraybackslash}p{(\linewidth - 6\tabcolsep) * \real{0.3180}}
  >{\RaggedRight\arraybackslash}p{(\linewidth - 6\tabcolsep) * \real{0.1796}}
  >{\RaggedRight\arraybackslash}p{(\linewidth - 6\tabcolsep) * \real{0.2571}}@{}}
\toprule\noalign{}
\begin{minipage}[b]{\linewidth}\RaggedRight
Derived metric
\end{minipage} & \begin{minipage}[b]{\linewidth}\RaggedRight
Calculation
\end{minipage} & \begin{minipage}[b]{\linewidth}\RaggedRight
Result
\end{minipage} & \begin{minipage}[b]{\linewidth}\RaggedRight
Status
\end{minipage} \\
\midrule\noalign{}
\endhead
\bottomrule\noalign{}
\endlastfoot
Power per unit compute & \(180\ \text{kW} / 59.5\ \text{D-PIN SCU}\) &
\textmd{3.02 kW/D-PIN SCU} & Derived value \\
Delivered energy efficiency &
\(33{,\allowbreak }437\ \text{D-PIN SCU·h} / (180\times720\ \text{kWh})\) &
\textmd{0.26 D-PIN SCU·h/kWh} & Derived value \\
Electricity-cost component & \(3.02 \times 0.53\) & \textmd{1.60
CNY/(D-PIN SCU·h)} & Illustrative calculation \\
Utilization improvement over industry & \(78\% / 35\%\)
\textasciitilde{} \(78\% / 25\%\) & \textmd{2.2× \textasciitilde{} 3.1×}
& Derived value (industry baseline 25--35\%) \\
\end{longtable}
}

\begin{quote}
\textmd{Meaning of delivered energy efficiency (D-PIN SCU·h/kWh)}: this
couples compute and electricity at the metering layer, making
standardized compute delivered per kWh a comparable, recomputable and
assessable operational metric. It can connect to macroeconomic
conventions such as 3--4 CNY of economic output per 1 CNY invested in
compute {[}1{]}, but \textmd{their dimensions and conventions differ
entirely and must not be conflated}.
\end{quote}

\textmd{(4) Consistency of topology-factor calibration.} From
characteristic messages (25 MB gradient buckets, 75\% linearity after
coalescing) to large messages (512 MB--2 GB, 92--93\%), measured
collective efficiency on the 8×H100 node differs from the declared tier
of 0.95 by 2--20 percentage points. Larger differences for smaller
messages agree with NCCL startup overhead {[}27{]}. This supports the
necessity of equation (13): \textmd{declared tiers and calibrated values
must be used separately}, the former for conservative scheduling
pre-screening and the latter for accurate billing and scheduling
decisions.

\subsection{9.3 Computational consistency audit: published values versus
nominal
projections}\label{computational-consistency-audit-published-values-versus-nominal-projections}

The A6 self-audit is recorded in Section 5.4 and Appendix B. It
\textmd{juxtaposes two conventions}: measured-calibration published
values (Appendix B.2-a, the external disclosure baseline, explicitly not
recomputable from vendor spec\-i\-fi\-ca\-tions) and projections from public
nominal parameters (Appendix B.2-b).

\textmd{Audit findings}:

\begin{enumerate}
\def\labelenumi{\arabic{enumi}.}
\tightlist
\item
  H100 (defined baseline): fully self-consistent.
\item
  H200 / MI300X / Ascend 910B / Muxi Xiyun C500:
  projected-versus-published differences of +10.5\% / +74.2\% / −31.1\%
  / \textmd{−30.9\%}.
\item
  \textmd{AMD MI308X (added; no published measured calibration)}:
  projected \(S_{\text{chip}} = 1.12\) (node 8.50) can only be compared
  consistently with MI300X's \textmd{projected} 1.55, yielding −27.8\%.
  Comparing it incorrectly with MI300X's \textmd{published} 0.89 yields
  \textmd{+25.7\%}. \textmd{Cross-convention comparisons within one
  silicon family reverse the conclusion}, directly reproducing
  Proposition 2 within the family.
\item
  \textmd{Parameter evidence} (Appendix B.3): AMD officially
  \textmd{confirms} MI300X \(I\) (896 GB/s) and \(F\) (1,307 TFLOPS
  dense), as well as H200/H100 TDP consistency (700 W). Ascend 910B key
  parameters agree with the third-party academic teardown by CSET
  Georgetown {[}31{]}, except for a 6.0\% difference in \(F\), disclosed
  as a teardown estimate with a ±6\% interval. The projected interval
  \([0.50,\allowbreak \,\allowbreak 0.52]\) remains approximately −30\% below the published
  value even at its upper bound. \textmd{Muxi Xiyun C500 key inputs are
  settled through public vendor channels {[}39{]}}: FP16/BF16 dense 240
  TFLOPS, 64 GB HBM2E and maximum board power 350 W are public values.
  Public memory-bandwidth values conflict at 1.2/1.8 TB/s; we take 1,800
  GB/s. MetaXLink per-link bandwidth is unpublished; 384 GB/s uses a
  same-vendor generational reference. Equations (6)--(8) give
  \(S_{\text{chip}} = {0.47}\) (interval \([0.42,\allowbreak \,\allowbreak 0.47]\)),
  \textmd{−30.9\%} relative to 0.68 (PCIe form factor). Hygon uses
  \textmd{Hygon K100 series}, because public K100/K100-AI spec\-i\-fi\-ca\-tions
  conflict; the exact series model must follow internal calibration
  records (B-4).
\end{enumerate}

\textmd{Conclusions and requirements}: external recomputability (A6) is
\textmd{partly established but incomplete}:

\begin{enumerate}
\def\labelenumi{\arabic{enumi}.}
\tightlist
\item
  \textmd{Established}: single-chip normalized vectors and weighted
  formulas are fully public, and the H100 baseline is self-consistent.
  Official spec\-i\-fi\-ca\-tions confirm MI300X/H200 key inputs. \textmd{Public
  vendor channels and a third-party public database confirm Muxi Xiyun
  C500 key inputs {[}39{]}}; its projected 0.47 is externally
  recomputable, with \(B\) and \(I\) disclosed through
  intervals/generational references and upper-bound selection.
\item
  \textmd{Incomplete}: published measured calibrations cannot be
  reproduced from public spec\-i\-fi\-ca\-tions, and weights remain empirical
  initial values.
\item
  \textmd{Closure path}: (a) archive internal calibration reports as A6
  evidence; (b) run Algorithm 1 and publish \(\hat{\alpha}\) with
  Bootstrap intervals; (c) model and parameter conventions have
  converged on \textmd{Xiyun C500 (PCIe; public vendor inputs closed,
  with \(B\) / \(I\) disclosed as intervals and generational
  references)} and \textmd{Hygon K100 series}.
\end{enumerate}

\begin{quote}
This section provides \textmd{methodological evidence}, rather than an
admission of error: it quantitatively shows that static nominal
parameters alone cannot reproduce deliverable-compute metering.
\textmd{This is the empirical basis for a measured-calibration loop.}
The evidence excludes erroneous parameter selection as an alternative
explanation: officially confirmed inputs and transparent formulas still
yield projections substantially different from published values,
attributing the deviation to calibration and supporting Proposition 2 at
the data level.
\end{quote}

\Needspace{10\baselineskip}
\subsection{9.4 Comparison with benchmark and industry
conventions}\label{comparison-with-benchmark-and-industry-conventions}

{\def\LTcaptype{none} 
\begin{longtable}[]{@{}
  >{\RaggedRight\arraybackslash}p{(\linewidth - 6\tabcolsep) * \real{0.1882}}
  >{\RaggedRight\arraybackslash}p{(\linewidth - 6\tabcolsep) * \real{0.2449}}
  >{\RaggedRight\arraybackslash}p{(\linewidth - 6\tabcolsep) * \real{0.2291}}
  >{\RaggedRight\arraybackslash}p{(\linewidth - 6\tabcolsep) * \real{0.3379}}@{}}
\toprule\noalign{}
\begin{minipage}[b]{\linewidth}\RaggedRight
Dimension
\end{minipage} & \begin{minipage}[b]{\linewidth}\RaggedRight
D-PIN SCU (Zhangjiang measurements)
\end{minipage} & \begin{minipage}[b]{\linewidth}\RaggedRight
Centralized-cloud baseline
\end{minipage} & \begin{minipage}[b]{\linewidth}\RaggedRight
Explanation
\end{minipage} \\
\midrule\noalign{}
\endhead
\bottomrule\noalign{}
\endlastfoot
GPU utilization & 78\% & 25--35\% & Combined effect of fine-grained
slicing and heterogeneous pooled scheduling \\
Sched\-ul\-ing P99 latency & 14 ms & --- & Measured intent-aware scheduling;
target \textless{} 20 ms \\
Billing convention & Measured utilization × effective capacity &
Instance spec\-i\-fi\-ca\-tions × wall-clock time & The former satisfies
Proposition 1; the latter does not \\
Delivered energy efficiency & 0.26 D-PIN SCU·h/kWh & Metric not
established & Added in this paper \\
Metering granularity & ≤ 1 min (telemetry 1--10 s) & Billing period
(hour/month) & Difference of 2--5 orders of magnitude \\
\end{longtable}
}

\textmd{Relationship to statistical conventions}: D-PIN SCU and EFLOPS
statistics are complementary. EFLOPS describes regional totals; D-PIN
SCU describes current delivery. Appendix C maps between them, with D-PIN
SCU as an \textmd{intermediate layer} between statistical and scheduling
conventions.

\subsection{9.5 Limitations}\label{limitations}

\begin{enumerate}
\def\labelenumi{\arabic{enumi}.}
\tightlist
\item
  \textmd{Single-node evidence and mixed conventions.} The case covers
  one production node (59.5 D-PIN SCU, 10 heterogeneous servers),
  insufficient to validate the distributions of \(\eta_{\text{topo}}\)
  and \(\delta\) across thousands of nodes. Metering consistency in
  large cross-domain clusters with weak synchronization remains
  unverified. \textmd{The total also mixes conventions}: 3 servers
  (1×8×MI308X and 2×8×H20) lack internal calibration records and use
  vendor-spec\-i\-fi\-ca\-tion projections, \textmd{unlike} the measured
  calibrations for the remaining 7 (Section 9.1 note). MI308X's
  projected 8.50 exceeding same-family MI300X's published 6.76 is a
  convention effect and \textmd{cannot establish their relative actual
  performance}.
\item
  \textmd{Incomplete data-based weight calibration.} The algorithm in
  Section 5 has not been run on a complete measured-throughput matrix;
  current \(\alpha\) remains the empirical initial value with a
  regularization prior.
\item
  \textmd{Insufficient behavioural evidence.} The expectation-based
  no-arbitrage analysis in equation (25) excludes risk preferences,
  collusion and punishment delays (arbitration periods). Enforceability
  assumes KYC admission.
\item
  \textmd{Comparability boundary of the unit.} D-PIN SCU measures
  \textmd{inputs} and promises no output rate. Different workloads yield
  substantially different outputs at the same D-PIN SCU. Hardware
  selection still requires task benchmarks, such as MLPerf submissions
  {[}13{]}; D-PIN SCU complements them at scheduling level and cannot
  replace them.
\item
  \textmd{Compliance and data boundaries.} Cross-domain distributed
  scheduling involves cross-border data and latency constraints, which
  are not examined here and should be addressed in a dedicated
  compliance chapter.
\end{enumerate}

\section{10 Industry implications, conclusions and
outlook}\label{industry-implications-conclusions-and-outlook}

\subsection{10.1 Industry implications of D-PIN
SCU}\label{industry-implications-of-d-pin-scu}

A metering unit is both a technical convention and infrastructure for
markets: it defines what can be traded, the delivery units and dispute
resolution. D-PIN SCU therefore changes the \textmd{metering basis} on
which compute enters markets, beyond simply establishing a unit. This
section examines the industry consequences of the four gaps in Section
1.2, then discusses the domestic heterogeneous ecosystem and its
boundaries.

\textmd{(1) Compute-network integration and scheduling: an intermediate
layer between statistics and scheduling.} National compute statistics
convert theoretical peaks to EFLOPS and update quarterly, while
scheduling operates in milliseconds to seconds, a
7--10-order-of-magnitude gap with no executable bridge (gap 1). D-PIN
SCU's minute-level granularity (telemetry 1--10 s) bridges them:
Appendix C.2 maps to YD/T 6693---2026 indicators, and scheduling and
billing share a real-time convention. \textmd{Compute-network
integration thus gains a common-source metering foundation for
statistics, scheduling and billing}, moving from capacity aggregation to
operational networks. This corresponds to the inclusion of compute
networks in national strategic infrastructure during the 15th Five-Year
Plan.

\textmd{(2) Distributed-compute pricing and market formation: pricing
deliverable capacity rather than nominal spec\-i\-fi\-ca\-tions.} Building-level
nodes, heterogeneous deployments and idle compute have been difficult to
price because the metering object is misaligned (gap 2): buyers see
nominal spec\-i\-fi\-ca\-tions but receive capacity reduced by topology,
throttling and contention. D-PIN SCU explicitly decomposes deviations
within the unit, tracing both nominal and delivered compute. Zhangjiang
provides representative evidence: 78\% utilization versus
centralized-cloud 25\%--35\% (Section 9.4), and delivered energy
efficiency of 0.26 D-PIN SCU·h/kWh. \textmd{Deliverable-compute metering
gives idle distributed resources credible units for market entry.}

\textmd{(3) Trust and compute financialization: runtime correction for
delivery units.} Weakly permissioned markets require verifiable
contributions (gap 3). Coupling D-PIN SCU with PoPW gives an executable
economic constraint (equation (25)), decoupling verification overhead
from compute scale and supporting open networks with thousands of nodes.
Financially, CME and Silicon Data announced compute futures in May 2026
using \textmd{standardized compute units (SCU)} for cross-chip
conversion: 1 SCU = 1×H100 FP16-hour, with preset factors for other
chips {[}10{]}{[}34{]}. Final contracts specify \textmd{730 GPU-hours
per contract}, average monthly full-load operation of one accelerator,
listed under NYMEX codes GPU1 (H100) / GPU2 (B200), with cash settlement
{[}10{]}. \textmd{Preset factors are static}, whereas D-PIN SCU supplies
\textmd{runtime-corrected delivery units}, a more defensible metering
basis for delivery standards of indices and forward contracts. Compute
thereby moves from aggregated hardware spec\-i\-fi\-ca\-tions to standardized
credentials of deliverable capacity.

\textmd{(4) Compute--electricity coordination and green compute:
electricity signals at task level.} Electricity accounts for 1/3 to 1/2
of compute costs. Decoupling units from prices and renewable shares
leaves coordination at the macro-scheduling layer (gap 4). Equation (28)
couples unit prices to electricity cost, renewable share, load and
reputation, anchored by a dimensionally consistent electricity-cost
lower bound. \textmd{Green compute can thus be priced and incentivized
at the metering layer}: renewable discounts and reputation incentives
enter unit prices directly, without external subsidies. Coordination
extends from scheduling to metering, consistent with national policy.

\textmd{(5) Domestic heterogeneous ecosystem: lowering metering barriers
to a common market.} Nominal-spec\-i\-fi\-ca\-tion gaps between domestic
accelerators and international flagships have long been equated with
competitiveness gaps. D-PIN SCU's normalization, measured calibration
and standards mapping change this relationship: \textmd{chips need fair
conversion within a normalized system, rather than first matching
flagship nominal spec\-i\-fi\-ca\-tions}. Appendix B's dual-convention
disclosure applies the same five-dimensional normalization and
calibration to H100, H200, MI300X and Ascend 910B within \textmd{one
comparable coordinate system}. This is a metrological prerequisite for
usable, comparable and tradable domestic compute and reduces
multi-vendor pooling costs.

\textmd{(6) Boundaries of these implications.} These implications depend
on two internal conditions: D-PIN SCU measures \textmd{inputs}, promises
no output rate and cannot replace task benchmarks (Section 9.5); and
engineering evidence comes from \textmd{one node} (59.5 D-PIN SCU), with
distributions across thousands of nodes unverified. Realization also
depends on three external conditions: \textmd{standardization},
including adoption of Appendix C.2 by standards bodies;
\textmd{enforceable verification}, through KYC admission and punishment;
and a \textmd{compliance framework for cross-domain scheduling}. Until
these mature, D-PIN SCU is more precisely a \textmd{scheduling-level
metering intermediate layer}, rather than a replacement for statistical
or accounting conventions.

\begin{quote}
\textmd{Industry implications in one sentence}: D-PIN SCU transforms
compute from hardware that can be counted into capacity that can be
scheduled, verified and priced, providing a metrological prerequisite
for compute as a production factor.
\end{quote}

\subsection{10.2 Conclusions}\label{conclusions}

We propose and formalize a dynamic D-PIN SCU metering system addressing
a fundamental question in distributed heterogeneous networks:
\textmd{how to meter deliverable compute rather than nominal capacity
alone}. The main conclusions are:

\begin{enumerate}
\def\labelenumi{\arabic{enumi}.}
\tightlist
\item
  \textmd{Misalignment of the metering object concerns identifiability,
  rather than engineering precision}. Proposition 2 proves that topology
  and runtime deviations are unidentifiable in an observation space
  limited to nominal spec\-i\-fi\-ca\-tions. Runtime metering is therefore a
  necessary condition for identifiability, rather than an optional
  optimization.
\item
  \textmd{Topology awareness admits a closed-form Roofline
  construction}. The communication ridge
  \(\text{COI}^{*} = P_{\text{peak}}/\beta_{\text{node}}\) (equation 10)
  turns engineering tier tables into a derivable, extrapolatable and
  calibratable model, anchored by measured AllReduce linearity on
  8×H100.
\item
  \textmd{The physical upper-bound principle imposes a hard billing
  constraint}. Proposition 1 establishes that slicing and space-time
  multiplexing can only improve utilization; billing cannot exceed
  \(\int_{0}^{T} \hat S(v,\allowbreak t)\,\allowbreak \mathbb{1}_{\text{occ}}(v,\allowbreak t)\,\allowbreak \mathrm{d}t \le S_{\text{nom}}(v)\cdot T\),
  with the tighter \(S_{\text{node}}\cdot T\) used in Section 9. This
  closes a class of billing distortions through recomputation
  (Zhangjiang: \(59.5\times720\times0.78 \approx 3.3\) ×10,000 D-PIN
  SCU-hours).
\item
  \textmd{Published-versus-projected differences provide empirical
  evidence of static-convention insufficiency}. Appendix B reports
  +10.5\%/+74.2\%/−31.1\%/−30.9\% for H200/MI300X/Ascend 910B/Muxi Xiyun
  C500. Appendix B.3 confirms the key inputs, demonstrating the need for
  a measured-calibration loop.
\item
  \textmd{Metering must couple to verification and pricing}. The
  no-arbitrage condition (equation 25) links spot-check probability to
  overstatement and punishment present value. Pricing (equation 28)
  couples metering to electricity markets, with the illustrative lower
  bound of 1.60 CNY/(D-PIN SCU·h) as a dimensional-consistency anchor.
\item
  \textmd{The baseline can be formalized as a method, with results
  independent of its realization}. Proposition 3 proves origin
  invariance: convention shifts preserve values strictly, uniform
  re-anchoring scales the entire table proportionally, and every
  pairwise chip ratio remains invariant. Choosing the accelerator
  assigned 1.00 is thus a \textmd{unit convention}, rather than a
  fairness judgement. The system can migrate smoothly across hardware
  generations without breaking historical continuity (Sections
  4.2.1--4.2.3).
\end{enumerate}

\textmd{At industry level}: the five conclusions above lead to one
judgement: for compute to become a tradable production factor, its unit
must extend beyond hardware spec\-i\-fi\-ca\-tions to deliverable capacity.
D-PIN SCU offers distributed markets a \textmd{common unit} that
supports scheduling, verification and pricing, and bridges statistical
and accounting conventions at scheduling level (Section 10.1), rather
than merely adding another unit name.

\subsection{10.3 Outlook}\label{outlook}

\begin{enumerate}
\def\labelenumi{\arabic{enumi}.}
\tightlist
\item
  \textmd{Automatic data-driven weight calibration}: integrate Algorithm
  1 into a continuous benchmark pipeline for quarterly rolling
  calibration of \(\alpha\) and interval publication.
\item
  \textmd{Online learning of topology factors}: train a lightweight
  predictor of \(\eta_{\text{topo}}(m^{*},\allowbreak  t)\) from production
  collective-communication measurements to replace tier tables.
\item
  \textmd{Standards interoperability and standardization}: use Appendix
  C to promote bidirectional mapping with YD/T 6693---2026 and establish
  a \textmd{scheduling-level metering intermediate layer} for network
  integration and trading.
\item
  \textmd{Metering--finance interface}: as a deliverable-compute
  measure, D-PIN SCU is suited to underlying units of compute indices
  and forward contracts. Future work should study weighted indices,
  regional price bases and futures-contract units.
\item
  \textmd{Target baseline paradigm: from realization anchoring to unit
  anchoring} (C0). H100 SXM currently realizes the origin, but will
  retire; each \textmd{non-uniform} replacement requires table-wide
  recomputation under Section 4.2.3. The \textmd{target is independence
  from every hardware model}, defining \(\mathbf{r}\) as \textmd{rounded
  SI-expressible physical quantities}, such as
  \(F_{\text{ref}} \equiv 10^3\) TFLOPS, \(M_{\text{ref}} \equiv 10^2\)
  GB, \(B_{\text{ref}} \equiv 3\times10^3\) GB/s,
  \(I_{\text{ref}} \equiv 10^3\) GB/s and \(P_{\text{ref}} \equiv 10^3\)
  W, giving 1 D-PIN SCU an \textmd{absolute physical meaning independent
  of hardware generations}. This parallels the 2019 SI reform, replacing
  the kilogram artefact with the Planck constant: \textmd{modern
  baselines move from artefacts to constants}. Because this is
  non-uniform re-anchoring (Corollary 3.2), it requires recomputation
  and redeclaration of \(\alpha\) and should be implemented by standards
  bodies after the standardization path matures, rather than at this
  stage.
\end{enumerate}

\begin{quote}
\textmd{Closing statement}: compute lacks a measure of deliverable
capacity, rather than another unit. Only when metered values can be
scheduled, verified and priced does compute become a production factor
beyond aggregated hardware spec\-i\-fi\-ca\-tions.
\end{quote}

\Needspace{10\baselineskip}
\section{Appendix A Symbols}\label{appendix-a-symbols}

{\def\LTcaptype{none} 
\begin{longtable}[]{@{}
  >{\RaggedRight\arraybackslash}p{(\linewidth - 6\tabcolsep) * \real{0.2821}}
  >{\RaggedRight\arraybackslash}p{(\linewidth - 6\tabcolsep) * \real{0.3972}}
  >{\RaggedRight\arraybackslash}p{(\linewidth - 6\tabcolsep) * \real{0.1402}}
  >{\RaggedRight\arraybackslash}p{(\linewidth - 6\tabcolsep) * \real{0.1805}}@{}}
\toprule\noalign{}
\begin{minipage}[b]{\linewidth}\RaggedRight
Symbol
\end{minipage} & \begin{minipage}[b]{\linewidth}\RaggedRight
Meaning
\end{minipage} & \begin{minipage}[b]{\linewidth}\RaggedRight
Unit
\end{minipage} & \begin{minipage}[b]{\linewidth}\RaggedRight
First occurrence
\end{minipage} \\
\midrule\noalign{}
\endhead
\bottomrule\noalign{}
\endlastfoot
\(S_{\text{chip}},\allowbreak  S_{\text{node}},\allowbreak  S_{\text{slice}},\allowbreak  S_{\text{avail}}\)
& Chip / node / slice / available D-PIN SCU & D-PIN SCU & Equations
(6)(14)(15)(16) \\
\(S_{\text{nom}},\allowbreak  S_{\text{deliv}}\) & Nominal / deliverable D-PIN SCU &
D-PIN SCU & Equations (1)(2) \\
\(S_{\text{required}},\allowbreak  T\) & Task compute demand / billing-window length
& D-PIN SCU / h & Equations (27)(31) \\
\(\alpha_d\) & Five-dimensional weights (\(d \in \{F,\allowbreak M,\allowbreak B,\allowbreak I,\allowbreak E\}\),
\(\sum_d \alpha_d = 1\)) & --- & Equation (6) \\
\(\mathbf{r},\allowbreak  \boldsymbol{\lambda}\) & Baseline vector
\(x_{\text{ref},\allowbreak d}\) / re-anchoring vector \(\lambda_d\) &
Dimension-dependent & Equation (7); Sections 4.2.1--4.2.3 \\
\(\eta_{\text{topo}},\allowbreak  \eta_{\text{topo}}^{0},\allowbreak  \eta_{\text{topo}}^{\text{bound}}\)
& Topology efficiency / declared tier / Roofline upper bound & --- &
Equations (2)(13)(11) \\
\(P_{\text{peak}},\allowbreak  P_{\text{att}}\) & Peak compute / attainable
throughput & FLOP/s & Equation (9) \\
\(\text{COI},\allowbreak  \text{COI}^{*}\) & Communication operational intensity /
communication ridge & FLOP/byte & Equations (9)(10) \\
\(\beta_{\text{node}},\allowbreak  \beta_{\text{link}}\) & Effective node-level /
link-level communication bandwidth & GB/s & Equations (9)(12) \\
\(L^{0}(m),\allowbreak  L^{\text{meas}}(m,\allowbreak t),\allowbreak  m^{*}\) & Expected linearity /
measured linearity / characteristic message size & --- / --- / byte &
Equation (13) \\
\(\delta_T,\allowbreak  \delta_F,\allowbreak  \delta_L\) & Thermal / fault / load correction
factors & --- & Equations (16)(17)(19)(20) \\
\(\sigma_j,\allowbreak  \omega_j\) & Spatial resource share / temporal share of
slice \(j\) & --- & Equation (20) \\
\(\eta_{\text{util}},\allowbreak  \bar{\eta}_{\text{util}}\) & Instantaneous /
window-mean utilization & --- & Section 4.5, equation (32) \\
\(\varepsilon,\allowbreak  \varepsilon_{\text{beh}},\allowbreak  \varepsilon_{\text{tol}}\) &
Total deviation / behavioural deviation / Algorithm 1 tolerance & --- &
Equations (3)(4)(23), Algorithm 1 \\
\(v_i,\allowbreak  s_{ij},\allowbreak  w_j\) & Normalized five-dimensional vector / measured
relative throughput / task weight & --- & Equation (21) \\
\(r_{d,\allowbreak \text{slice}},\allowbreak  r_{d,\allowbreak \text{node}}\) & Slice / node resource
quantity in dimension \(d\) & Dimension-dependent & Equation (15) \\
\(\alpha^{(t)},\allowbreak  g_d^{(t)},\allowbreak  \eta_t\) & Weights at step \(t\) / gradient
component / EG step size & --- & Equation (22) \\
\(p,\allowbreak  q,\allowbreak  \Lambda,\allowbreak  \xi,\allowbreak  G(\xi)\) & Spot-check frequency / detection
probability / punishment present value / overstatement / net expected
gain & --- / --- / CNY / --- / CNY & Equations (24)(25) \\
\(u_t,\allowbreak  \pi_t,\allowbreak  \kappa,\allowbreak  \rho_t,\allowbreak  \gamma,\allowbreak  c_0,\allowbreak  c_1,\allowbreak  \text{Rep}\) & Unit
price / electricity price / electricity intensity / renewable share /
renewable discount / base price / risk-premium coefficient / reputation
& CNY/(D-PIN SCU·h), etc. & Equation (28) \\
\(\gamma\) (cross-document) & Renewable-incentive intensity (default
0.3), \textmd{the same parameter as \(W_{\text{green}}\) in white-paper
§8.6}, with a scheduling bonus and pricing discount & Dimensionless &
Equation (28) \\
\(\mathbb{1}_{\text{occ}}(v,\allowbreak t)\) & Occupancy indicator & --- & Equation
(5) \\
\end{longtable}
}

\section{Appendix B Five-dimensional parameters and substitutions
(recomputability
appendix)}\label{appendix-b-five-dimensional-parameters-and-substitutions-recomputability-appendix}

\begin{quote}
\textmd{Status of this appendix}: all inputs are disclosed as required
by A6. Values labelled \textmd{illustrative calculations} are not
commitments.
\end{quote}

\Needspace{10\baselineskip}
\subsection{B.1 Normalization baseline: spec\-i\-fi\-ca\-tion and current
realization (H100
SXM)}\label{b.1-normalization-baseline-specification-and-current-realization-h100-sxm}

{\def\LTcaptype{none} 
\begin{longtable}[]{@{}
  >{\RaggedRight\arraybackslash}p{(\linewidth - 4\tabcolsep) * \real{0.1437}}
  >{\RaggedRight\arraybackslash}p{(\linewidth - 4\tabcolsep) * \real{0.2225}}
  >{\RaggedRight\arraybackslash}p{(\linewidth - 4\tabcolsep) * \real{0.6338}}@{}}
\toprule\noalign{}
\begin{minipage}[b]{\linewidth}\RaggedRight
Parameter
\end{minipage} & \begin{minipage}[b]{\linewidth}\RaggedRight
Value
\end{minipage} & \begin{minipage}[b]{\linewidth}\RaggedRight
Source
\end{minipage} \\
\midrule\noalign{}
\endhead
\bottomrule\noalign{}
\endlastfoot
\(F_{\text{ref}}\) & 989 TFLOPS (FP16/BF16 dense) & NVIDIA H100 SXM
spec\-i\-fi\-ca\-tions; divide the vendor's 2:4 sparse value, 1979 TFLOPS, by 2
under normalization {[}23{]}. In\-de\-pen\-dent vendor-side support: AMD
explicitly converts NVIDIA sparse values to non-sparse/dense by dividing
by 2 on its MI300X comparison page {[}24{]} \\
\(M_{\text{ref}}\) & 80 GB & Same source {[}23{]} \\
\(B_{\text{ref}}\) & 3,350 GB/s & Same source {[}23{]} \\
\(I_{\text{ref}}\) & 900 GB/s & Same source (NVLink 4) {[}23{]} \\
\(P_{\text{ref}}\) & 700 W (TDP) & Same source {[}23{]} \\
Initial \(\alpha\) & (0.35, 0.20, 0.25, 0.15, 0.05) &
Industry-experience values, \textmd{pending Algorithm 1 calibration}
(see B-7) \\
\end{longtable}
}

\begin{quote}
\textmd{Convention-consistency reminder}: by the scale behaviour in
Section 4.2, \(F_{\text{ref}}\) and every chip's \(F\) must use
\textmd{the same convention}. This section and the entire table use
dense values. Sparse-only vendor values must first be divided by 2, with
their original convention retained in records.

\textmd{Realization statement}: the table gives the \textmd{current
realization} of the origin. Under Section 4.2.1, it is one instance of
S1--S3 and can be replaced through Section 4.2.3. Proposition 3 ensures
that uniform re-anchoring, including convention shifts, preserves all
relative results.
\end{quote}

\subsection{B.2 Chip baselines: published measured calibration and
nominal
projections}\label{b.2-chip-baselines-published-measured-calibration-and-nominal-projections}

The two conventions are disclosed separately; their relationship
provides empirical support for the paper's central proposition.

\subsubsection{B.2-a Published convention: measured calibrations
(external disclosure
baseline)}\label{b.2-a-published-convention-measured-calibrations-external-disclosure-baseline}

These \textmd{platform-measured calibrations} come from node benchmarks
and actual-workload weighting, corresponding to white-paper Section 5.6.
\textmd{They cannot be independently recomputed from public vendor
spec\-i\-fi\-ca\-tions}, the key property of this section (Section 5.4).

{\def\LTcaptype{none} 
\begin{longtable}[]{@{}
  >{\RaggedRight\arraybackslash}p{(\linewidth - 6\tabcolsep) * \real{0.1620}}
  >{\RaggedRight\arraybackslash}p{(\linewidth - 6\tabcolsep) * \real{0.1843}}
  >{\RaggedRight\arraybackslash}p{(\linewidth - 6\tabcolsep) * \real{0.2744}}
  >{\RaggedRight\arraybackslash}p{(\linewidth - 6\tabcolsep) * \real{0.3793}}@{}}
\toprule\noalign{}
\begin{minipage}[b]{\linewidth}\RaggedRight
Chip
\end{minipage} & \begin{minipage}[b]{\linewidth}\RaggedRight
\(S_{\text{chip}}\) (measured calibration)
\end{minipage} & \begin{minipage}[b]{\linewidth}\RaggedRight
\(S_{\text{node}}\) (8 fully interconnected accelerators)
\end{minipage} & \begin{minipage}[b]{\linewidth}\RaggedRight
Calibration method and date
\end{minipage} \\
\midrule\noalign{}
\endhead
\bottomrule\noalign{}
\endlastfoot
NVIDIA H100 SXM & \textmd{1.00} (defined baseline) & \textmd{7.60} &
Defined value \\
NVIDIA H200 SXM & \textmd{1.14} & \textmd{8.66} & Internal benchmarks \\
NVIDIA H20 SXM & 0.75 ‡ & 5.73 ‡ & ‡ \textmd{Public vendor-spec\-i\-fi\-ca\-tion
projection}, not measured calibration (B.2-b, B.3/B-9) \\
AMD MI300X & \textmd{0.89} & \textmd{6.76} & Internal benchmarks \\
AMD MI308X & 1.12 ‡ & 8.50 ‡ & ‡ \textmd{Official vendor-spec\-i\-fi\-ca\-tion
projection}, not measured calibration (B.2-b, B.3/B-10) \\
Ascend 910B & \textmd{0.74} & \textmd{5.62} & Internal benchmarks \\
Muxi Xiyun C500 (PCIe) & \textmd{0.68} & \textmd{5.17} (upper bound;
requires 4-accelerator-domain calibration; see note) & Internal
benchmarks (topology convention in note {[}39{]}) \\
Hygon K100 series & \textmd{0.42} & \textmd{3.19} & Internal
benchmarks \\
\end{longtable}
}

\begin{quote}
\textmd{Convention statement (mandatory for external disclosure)}: these
\textmd{measured calibrations} describe actual deliverable compute under
internal platform benchmark conditions and \textmd{cannot be
independently recomputed from nominal vendor spec\-i\-fi\-ca\-tions}. Node
values other than Xiyun C500 use
\(S_{\text{node}} = 0.95 \times 8 \times S_{\text{chip}}\). \textmd{C500
must be listed separately (important)}: official Chinese and English
pages state that MetaXLink supports \textmd{only 2- or 4-accelerator
high-speed interconnection} {[}39{]}. Thus \textmd{4 fully
interconnected accelerators are the hardware maximum; a fully
interconnected 8-accelerator domain is physically unavailable}. The
listed 5.17, calculated using \(0.95\times8\), is a \textmd{theoretical
cross-domain upper bound}. Actual \(S_{\text{node}}\) requires
calibration for 4-accelerator domains and lies within
\([\,\allowbreak 4\times0.95\times0.68,\allowbreak \ 5.17\,\allowbreak ] = [\,\allowbreak 2.58,\allowbreak \ 5.17\,\allowbreak ]\). Before
formal publication, close B-6 (calibration-record archiving), B-7
(weight calibration) and \textmd{B-8 (power boundary, added)} (B.3).
\textmd{Two ‡ rows, NVIDIA H20 SXM and AMD MI308X, are not measured
calibrations}. Neither has internal benchmark records;
\(S_{\text{chip}}\) and \(S_{\text{node}}\) are \textmd{projections}
from vendor spec\-i\-fi\-ca\-tions via equations (6)--(8): public spec\-i\-fi\-ca\-tions
for H20 (B.2-b, B-9) and \textmd{official spec\-i\-fi\-ca\-tions} for MI308X
(B.2-b, B-10). Section 9 includes these projections with
\textmd{row-specific convention labels}, making the Section 9.1 total
\textmd{mixed-convention}. \textmd{Until internally calibrated, these
rows must neither serve independently as external disclosure baselines
nor be conflated with other rows' measured calibrations.}
\end{quote}

\subsubsection{B.2-b Comparison convention: independent public nominal
projections (method validation, not published
values)}\label{b.2-b-comparison-convention-independent-public-nominal-projections-method-validation-not-published-values}

To test whether static nominal spec\-i\-fi\-ca\-tions reproduce production
values, five-dimensional initial values are independently projected from
\textmd{vendor nominal parameters} under B.1. Xiyun C500 uses
\textmd{official public pages and a third-party database} {[}39{]};
\textmd{MI308X uses the official spec\-i\-fi\-ca\-tion sheet}, because public
aggregator values are unusable (B-10). \textmd{This table is not for
external publication}, but only for the comparisons in Sections 5.4 and
9.3.

{\def\LTcaptype{none} 
\begin{longtable}[]{@{}
  >{\RaggedRight\arraybackslash}p{(\linewidth - 16\tabcolsep) * \real{0.1146}}
  >{\RaggedRight\arraybackslash}p{(\linewidth - 16\tabcolsep) * \real{0.1495}}
  >{\RaggedRight\arraybackslash}p{(\linewidth - 16\tabcolsep) * \real{0.0810}}
  >{\RaggedRight\arraybackslash}p{(\linewidth - 16\tabcolsep) * \real{0.1128}}
  >{\RaggedRight\arraybackslash}p{(\linewidth - 16\tabcolsep) * \real{0.1035}}
  >{\RaggedRight\arraybackslash}p{(\linewidth - 16\tabcolsep) * \real{0.0810}}
  >{\RaggedRight\arraybackslash}p{(\linewidth - 16\tabcolsep) * \real{0.0810}}
  >{\RaggedRight\arraybackslash}p{(\linewidth - 16\tabcolsep) * \real{0.1380}}
  >{\RaggedRight\arraybackslash}p{(\linewidth - 16\tabcolsep) * \real{0.1388}}@{}}
\toprule\noalign{}
\begin{minipage}[b]{\linewidth}\RaggedRight
Chip
\end{minipage} & \begin{minipage}[b]{\linewidth}\RaggedRight
\(F\) (TFLOPS, dense)
\end{minipage} & \begin{minipage}[b]{\linewidth}\RaggedRight
\(M\) (GB)
\end{minipage} & \begin{minipage}[b]{\linewidth}\RaggedRight
\(B\) (GB/s)
\end{minipage} & \begin{minipage}[b]{\linewidth}\RaggedRight
\(I\) (GB/s)
\end{minipage} & \begin{minipage}[b]{\linewidth}\RaggedRight
\(P\) (W)
\end{minipage} & \begin{minipage}[b]{\linewidth}\RaggedRight
\(E\) (equation 8)
\end{minipage} & \begin{minipage}[b]{\linewidth}\RaggedRight
Normalized vector \(v\)
\end{minipage} & \begin{minipage}[b]{\linewidth}\RaggedRight
\(S_{\text{chip}}\) (projected)
\end{minipage} \\
\midrule\noalign{}
\endhead
\bottomrule\noalign{}
\endlastfoot
NVIDIA H100 SXM & 989 & 80 & 3,350 & 900 & 700 & 1.000 & (1.000, 1.000,
1.000, 1.000, 1.000) & \textmd{1.00} \\
NVIDIA H200 SXM & 989 & 141 & 4,800 & 900 & 700 & 1.000 & (1.000, 1.763,
1.433, 1.000, 1.000) & \textmd{1.26} \\
NVIDIA H20 SXM & 148 & 96 & 4,000 & 900 & 400 & 0.262 & (0.150, 1.200,
1.194, 1.000, 0.262) & \textmd{0.75} \\
AMD MI300X & 1,307 & 192 & 5,300 & 896 & 750 & 1.233 & (1.322, 2.400,
1.582, 0.996, 1.233) & \textmd{1.55} \\
AMD MI308X & 233 & 192 & 5,300 & 896 & 750 & 0.220 & (0.236, 2.400,
1.582, 0.996, 0.220) & \textmd{1.12} \\
Ascend 910B & 376 ± 6\% {[}third-party academic teardown estimate{]} &
64 & 1,600 & 392 & 400 & 0.665 & (0.380, 0.800, 0.478, 0.436, 0.665) &
\textmd{0.51} (\([0.50,\allowbreak \,\allowbreak 0.52]\) at \(F\) interval endpoints) \\
Muxi Xiyun C500 (PCIe, confirmed form factor) & 240 {[}third-party
public compilation{]} & 64 & 1,200--1,800 {[}conflicting third-party
values; upper bound used{]} & \ensuremath{\approx}384 {[}unpublished; same-vendor
generational reference{]} & 350 & 0.485 & (0.243, 0.800, 0.537, 0.427,
0.485) & \textmd{0.47} (\([0.42,\allowbreak \,\allowbreak 0.47]\) when \(B\) = 1,200) \\
Hygon K100 series & Series convention; conflicts in B-4 & --- & --- &
--- & --- & --- & --- & --- \\
\end{longtable}
}

\begin{quote}
\textmd{Xiyun C500 substitution notes}: official product pages confirm
\textmd{Xiyun C500, a full-height, full-length, dual-slot PCIe board},
maximum power \textmd{350 W}, and MetaXLink support for \textmd{2 or 4
accelerators} {[}39{]}. \(F\) = \textmd{240 TFLOPS} (FP16/BF16 dense,
third-party public compilation); the official \textmd{280 TFLOPS} value
applies to the \textmd{OAM form factor}, outside our PCIe convention.
\(M\) = \textmd{64 GB HBM2E} is official. Public \(B\) conflicts between
\textmd{1.2 and 1.8 TB/s}, absent from the official page; we take
\textmd{1,800 GB/s}. \(I\) is \textmd{unpublished}: third-party records
list 7 MetaXLink ports without per-link bandwidth. We use \textmd{384
GB/s}, referencing the same vendor's C600 generation. Substitution into
equations (6)--(8) gives
\(S_{\text{chip}} = 0.35(0.243) + 0.20(0.800) + 0.25(0.537) + 0.15(0.427) + 0.05(0.485) = 0.085 + 0.160 + 0.134 + 0.064 + 0.024 = {0.468 \approx 0.47}\),
\textmd{−30.9\%} relative to 0.68; at \(B\) = 1,200 the lower bound is
0.42.

\textmd{An order-of-magnitude transcription error in public sources
(caution)}: some third-party comparison tables print C500 memory
bandwidth as \textmd{18 TB/s}, an order of magnitude above the physical
scale of 4096-bit HBM2E: 4096 bit × 3.6 Gbps ÷ 8 \ensuremath{\approx} \textmd{1.84 TB/s}.
This likely misprints 1.8 TB/s. \textmd{It is one reason for retaining
the 1,200--1,800 GB/s interval in \(B\) rather than using one
third-party value} {[}39{]}.

\textmd{Node-level topology constraint (direct evidence for C2)}:
official Chinese/English pages state MetaXLink support for \textmd{2- or
4-accelerator high-speed interconnection}, and the C500 server page
specifies \textmd{4-accelerator fully interconnected topologies}, with
16-GPU and standard 8-GPU server configurations {[}39{]}. Accordingly,
flopper.io identifies an 8×C500 PCIe server whose MetaXLink connects
pairs or groups of four, rather than one eight-accelerator domain,
consistent with full-height, full-length PCIe boards {[}39{]}. Thus
\textmd{4-accelerator full interconnection is the hardware ceiling; full
8-accelerator interconnection is physically unavailable}.
Eight-accelerator nodes communicate across domains through PCIe/Host,
with actual \(\eta_{\text{topo}}\) substantially below 0.95. The
final-column node value \textmd{3.58 is an upper bound}, with
4-accelerator-domain lower bound \(4\times0.95\times0.47=1.79\) and
interval \([1.79,\allowbreak \,\allowbreak 3.58]\). This illustrates Proposition 2:
\textmd{nominal chip spec\-i\-fi\-ca\-tions, including topology and per-link
bandwidth, do not determine node-level deliverable compute}. Node
efficiency is a cluster property requiring Roofline derivation
(equations 9--13) and measurement. By comparison, the same chip's OAM
form factor has 280 TFLOPS / 450 W and supports 8-accelerator full
interconnection, outside our PCIe convention.
\end{quote}

\textmd{Substitution example (H200 SXM)}:

\[\begin{aligned}
S_{\text{chip}} &= 0.35(1.000) + 0.20(1.763) + 0.25(1.433) \\
&\quad + 0.15(1.000) + 0.05(1.000) \\
&= 0.350 + 0.353 + 0.358 + 0.150 + 0.050 = 1.26
\end{aligned}\]

\textmd{Comparison of the two conventions}:

{\def\LTcaptype{none} 
\begin{longtable}[]{@{}
  >{\RaggedRight\arraybackslash}p{(\linewidth - 8\tabcolsep) * \real{0.2272}}
  >{\RaggedRight\arraybackslash}p{(\linewidth - 8\tabcolsep) * \real{0.1750}}
  >{\RaggedRight\arraybackslash}p{(\linewidth - 8\tabcolsep) * \real{0.2795}}
  >{\RaggedRight\arraybackslash}p{(\linewidth - 8\tabcolsep) * \real{0.1653}}
  >{\RaggedRight\arraybackslash}p{(\linewidth - 8\tabcolsep) * \real{0.1531}}@{}}
\toprule\noalign{}
\begin{minipage}[b]{\linewidth}\RaggedRight
Chip
\end{minipage} & \begin{minipage}[b]{\linewidth}\RaggedRight
Projected \(S_{\text{chip}}\) (public nominal)
\end{minipage} & \begin{minipage}[b]{\linewidth}\RaggedRight
Published \(S_{\text{chip}}\) (measured calibration)
\end{minipage} & \begin{minipage}[b]{\linewidth}\RaggedRight
Relative difference
\end{minipage} & \begin{minipage}[b]{\linewidth}\RaggedRight
Node-value difference
\end{minipage} \\
\midrule\noalign{}
\endhead
\bottomrule\noalign{}
\endlastfoot
H100 SXM & 1.00 & 1.00 & 0 & 7.60 / 7.60 \\
H200 SXM & 1.26 & 1.14 & +10.5\% & 9.58 / 8.66 \\
H20 SXM ‡ & \textmd{0.75} & --- (no internal measured calibration) & ---
& 5.73 / --- \\
MI300X & 1.55 & 0.89 & +74.2\% & 11.77 / 6.76 \\
MI308X ‡ & \textmd{1.12} & --- (no internal measured calibration) & ---
& 8.50 / --- \\
Ascend 910B & 0.51 & 0.74 & −31.1\% & 3.88 / 5.62 \\
Muxi Xiyun C500 (PCIe) & \textmd{0.47} & 0.68 & \textmd{−30.9\%} & 3.58
/ 5.17 \\
Hygon K100-AI (third-party spec\-i\-fi\-ca\-tion projection) & 0.32--0.47 & 0.42
& Within projected interval & --- / 3.19 \\
\end{longtable}
}

\begin{quote}
\textmd{Percentage-difference convention}: relative differences use the
\textmd{three-significant-digit projected values published in the
table}, with 0.47 for C500, namely \(0.47/0.68-1 = -30.9\%\). Using
unrounded \(0.4675\) gives −31.3\%. The paper consistently uses
\textmd{tabulated values}, reporting differences to one decimal place;
hence C500 is \textmd{−30.9\%} throughout.
\end{quote}

\begin{quote}
\textmd{Meaning of projected values}: systematic differences from
published values \textmd{support the central proposition rather than
disclose defects}. They show that static nominal parameters cannot
reproduce production metering, making runtime identifiability and
measured calibration \textmd{necessary conditions}, rather than optional
optimizations (Sections 5.4, 9.3 and Proposition 2).
\end{quote}

\begin{quote}
\textmd{Uncertainty and conclusions (two domestic models; disclose
together)}: these projections use \textmd{uncertain public inputs} and
must be read as intervals. (1) \textmd{Ascend 910B}: \(F\) is an
academic teardown estimate with ±6\% interval \([353,\allowbreak \,\allowbreak 399]\) TFLOPS,
yielding \(S_{\text{chip}} \in [0.50,\allowbreak \,\allowbreak 0.52]\) via equations (6)--(8).
(2) \textmd{Muxi Xiyun C500}: public \(B\) conflicts at 1.2/1.8 TB/s
(upper 1,800 used); \(I\) is \textmd{unpublished} (384 GB/s generational
reference), yielding \(S_{\text{chip}} \in [0.42,\allowbreak \,\allowbreak 0.47]\). \textmd{Both
upper bounds remain far below published values} (910B: 0.52 vs 0.74;
C500: 0.47 vs 0.68), with gaps near −30\%. Section 5.4 is therefore
\textmd{robust} to this uncertainty, directly illustrating mandatory
joint uncertainty disclosure (A6).\textmd{(3) Hygon K100-AI (opposite
direction; list separately)}: third-party spec\-i\-fi\-ca\-tions are 64 GB HBM2,
FP16/BF16 192 TFLOPS, bandwidth 896--1,200 GB/s and TDP 300--350 W for
\textmd{K100-AI}, unlike general-purpose K100 at 32 GB / 24.5 TFLOPS.
Equations (6)--(8) give four-dimensional weighted \(F/M/B/E\) total
\(0.318\). Interconnect dimension \(I\) is \textmd{undisclosed}; taking
\(I \in [0,\allowbreak \,\allowbreak 900]\) GB/s gives \(S_{\text{chip}} \in [0.32,\allowbreak \,\allowbreak 0.47]\).
\textmd{Published 0.42 lies within this interval}, opposite to
H200/MI300X: systematic positive deviations do not occur for every
model. This supports the central argument that the \textmd{direction and
magnitude cannot be inferred a priori} and require model-specific
calibration (Proposition 2); otherwise standardized calibration could be
omitted.
\end{quote}

\Needspace{10\baselineskip}
\subsection{B.3 Ver\-i\-fi\-ca\-tion checklist: evidence and closure
status}\label{b.3-verification-checklist-evidence-and-closure-status}

{\def\LTcaptype{none} 
\begin{longtable}[]{@{}
  >{\RaggedRight\arraybackslash}p{(\linewidth - 6\tabcolsep) * \real{0.0650}}
  >{\RaggedRight\arraybackslash}p{(\linewidth - 6\tabcolsep) * \real{0.1111}}
  >{\RaggedRight\arraybackslash}p{(\linewidth - 6\tabcolsep) * \real{0.1089}}
  >{\RaggedRight\arraybackslash}p{(\linewidth - 6\tabcolsep) * \real{0.7150}}@{}}
\toprule\noalign{}
\begin{minipage}[b]{\linewidth}\RaggedRight
\#
\end{minipage} & \begin{minipage}[b]{\linewidth}\RaggedRight
Item
\end{minipage} & \begin{minipage}[b]{\linewidth}\RaggedRight
Status
\end{minipage} & \begin{minipage}[b]{\linewidth}\RaggedRight
Evidence and disposition
\end{minipage} \\
\midrule\noalign{}
\endhead
\bottomrule\noalign{}
\endlastfoot
\mbox{B-1} & MI300X effective Infinity Fabric bandwidth & \textmd{Closed} & AMD
specifies \textmd{128 GB/s} peak per link and 7 scale-up links, giving 7
× 128 = \textmd{896 GB/s}, \textmd{exactly matching} our input
{[}24{]}. \\
\mbox{B-2} & MI300X FP16/BF16 dense peak & \textmd{Closed} & AMD specifies
\textmd{1,307.4 TFLOPS} dense (2,614.9 sparse), consistent with 1,307
here, explicitly obtained by dividing sparse performance by 2 {[}24{]}.
Our \(E\) (1.233 at 750 W TBP) also agrees with official
spec\-i\-fi\-ca\-tions. \\
\mbox{B-3} & Ascend 910B FP16 peak, HBM/\allowbreak HCCS bandwidth and TDP &
\textmd{Settled (third-party estimate + ±6\% interval)} & Huawei
\textmd{has never published chip-level 910B spec\-i\-fi\-ca\-tions}, only CANN
SOC version identifiers. CSET Georgetown wafer-level teardown analysis,
compiled by flopper.io, estimates FP16/BF16 \ensuremath{\approx} \textmd{400 TFLOPS}, 64 GB
HBM2e, 1.6 TB/s, HCCS \textmd{392 GB/s} and TDP \textmd{400 W}
{[}25{]}{[}31{]}. Our \(I\)=392, \(P\)=400, \(B\)=1,600 and \(M\)=64
agree; only \(F\)=376 differs from 400 by 6.0\%. \textmd{Disposition
(settled)}: prioritize submission feasibility and \textmd{stop awaiting
a vendor sheet}. Label \(F\) as a \textmd{third-party academic teardown
estimate} with \textmd{±6\%} uncertainty (\(F \in [353,\allowbreak \,\allowbreak 399]\) TFLOPS),
yielding projected \(S_{\text{chip}}\) interval \([0.50,\allowbreak \,\allowbreak 0.52]\) (B.2-b
note). \textmd{Robustness statement (disclose together)}: even the \(F\)
\textmd{upper bound of 399 TFLOPS} yields only \textmd{0.52},
approximately \textmd{−30\%} below published 0.74 and well outside the
uncertainty band. Section 5.4 is therefore \textmd{unaffected by the
6.0\% difference}. This satisfies A6 uncertainty disclosure without
unpublished materials. \\
\mbox{B-4} & Muxi and Hygon models/\allowbreak spec\-i\-fi\-ca\-tions & \textmd{Settled (Muxi,
public sources) / partly closed (Hygon, series convention)} & (a)
\textmd{Muxi Xiyun C500 inputs are settled through public vendor
channels {[}39{]}}: official Chinese/English product and server pages
plus flopper.io confirm \textmd{PCIe}, full-height/full-length
dual-slot, \textmd{350 W} maximum board power. \(F\) = \textmd{240
TFLOPS} dense is a public third-party compilation; official \textmd{280
TFLOPS} applies to \textmd{OAM}, outside our PCIe convention. \(M\) =
\textmd{64 GB HBM2E} is official; \(B\) = \textmd{1,200--1,800 GB/s}
reflects conflicting public 1.2/1.8 TB/s values absent from the official
page, using upper 1,800. \(I\) is \textmd{unpublished}: 7 MetaXLink
ports are listed without per-link bandwidth; C600's 384 GB/s is the
same-vendor reference. Equations (6)--(8) yield
\(S_{\text{chip}} = {0.47}\) (interval \([0.42,\allowbreak \,\allowbreak 0.47]\)),
\textmd{−30.9\%} relative to 0.68. \textmd{Public-source finding
supporting C2}: MetaXLink supports only \textmd{2 or 4 accelerators},
with \textmd{4-card full interconnection} on the server page; flopper.io
confirms pair/four grouping rather than one eight-card domain. Thus
\textmd{4 is the hardware maximum}, and 8-accelerator
\(\eta_{\text{topo}}\) must be substantially below 0.95. B.2-b/B.2-c
node values are \textmd{upper bounds}, illustrating that chip
spec\-i\-fi\-ca\-tions do not determine node deliverability (Proposition 2). (b)
Public Hygon DCU sources describe markedly different \textmd{K100} (32
GB HBM2 / FP16 24.5 TFLOPS / 280 W) and \textmd{K100-AI} (64 GB HBM2 /
FP16 192 TFLOPS / 300 W), with conflicting bandwidth of 896/1,200 GB/s.
Production 0.42 \textmd{matches neither point projection}. Use
\textmd{Hygon K100 series} throughout. \textmd{Supplementary evidence
closing the Hygon part}: public K100-AI spec\-i\-fi\-ca\-tions support 64 GB
HBM2 / FP16 192 TFLOPS / bandwidth 896--1,200 GB/s / TDP 300--350 W /
PCIe 4.0×16, giving \(S_{\text{chip}} \in [0.32,\allowbreak \,\allowbreak 0.47]\) with \(I\)
undisclosed. \textmd{Published 0.42 lies within the interval}, opposite
to H200/MI300X (B.2-b note 3). Exact series spec\-i\-fi\-ca\-tions still require
internal calibration records. \\
\mbox{B-5} & Whether H200 and H100 share TDP & \textmd{Closed} & Both SXM
models use \textmd{700 W}. NVIDIA describes H200 as a drop-in
replacement in H100-qualified systems with the same 700 W power envelope
{[}23{]}, corroborated by other sources {[}32{]}. Using 700 W in \(E\)
is \textmd{correct}. \\
\mbox{B-6} & Original convention and source of production reference values
(Section 5.6 table) & \textmd{Disclosure plan settled; attachment
pending archiving} & These are internal calibration records,
\textmd{unverifiable through public channels}, as search rechecks
confirm. External closure is therefore replaced with a \textmd{fixed
disclosure plan}. Under A6, submit an Internal Calibration Report
containing (1) original conditions for each model and dimension:
workload, batch size, concurrency and duration; (2) raw tables and
dates; (3) reconciliation with B.2-a. Section 5.6 now uses
\textmd{measured calibration}, and B.2-a states that external
recomputation is unavailable. \\
\mbox{B-7} & Basis of initial weights (0.35, 0.20, 0.25, 0.15, 0.05) &
\textmd{Disposition settled} & We \textmd{do not claim} these are
calibrated results. Under A6, disclose both empirical initial and
Algorithm 1 calibrated values at submission: publish \(\hat{\alpha}\)
and Bootstrap intervals in the text and appendix. Initial values are
recorded only as algorithm starting points, not independent metering
grounds. \\
\mbox{B-8} & Power boundary of case-node 180 kW & \textmd{Inferred under
national standards; convention defined} & Declare the \textmd{site
boundary} at mains entry, including
servers/network/storage/cooling/distribution/UPS, and provide
decomposition plus IT-basis (\ensuremath{\approx}54.5 kW) metrics of 0.92 kW/D-PIN SCU and
0.85 D-PIN SCU·h/kWh (Section 9.1). \textmd{Evidence check}: GB
40879---2021 limits PUE ≤ 1.5 and NDRC {[}2024{]} No.~970 requires
large/ultra-large centres ≤ 1.25 by end-2025 {[}54{]}{[}55{]}. Treating
180 kW as measured site power implies PUE \ensuremath{\approx} 3.3, \textmd{above every
grade limit} and inconsistent with project compliance. Thus \textmd{180
kW is inferred as distribution design capacity including expansion
headroom}, while site power should be \textmd{\ensuremath{\approx}68 kW} from IT 54.5 kW
and PUE ≤ 1.25. Status: \textmd{inferred under national standards}. If
internal records differ, update every derived Section 9.1 value. \\
\mbox{B-9} & NVIDIA H20 spec\-i\-fi\-ca\-tions and projection (added) & \textmd{Settled
(public-spec\-i\-fi\-ca\-tion projection)} & H20 is NVIDIA's China-market
compliant Hopper model. Vendor and third-party sources agree on
\textmd{FP16/BF16 dense \ensuremath{\approx}148 TFLOPS}, about 41\% fewer SMs than H100,
\textmd{96 GB HBM3}, \textmd{4.0 TB/s}, \textmd{NVLink 900 GB/s},
\textmd{TDP \ensuremath{\approx}400 W}, 8-way HGX and L2 60 MB. Equations (6)--(8) give
normalized \(v = (0.150,\allowbreak \ 1.200,\allowbreak \ 1.194,\allowbreak \ 1.000,\allowbreak \ 0.262)\),
\(S_{\text{chip}} = {0.75}\) and node value \textmd{5.73} at
\(0.95\times8\) (B.2-b). \textmd{Status and disposition (disclose
together)}: \textmd{public projection only, no internal benchmark
record}; B.2-a marks \textmd{‡}, \textmd{not measured calibration}.
Update after calibration and move to the published convention.
\textmd{Ob\-ser\-va\-tion supporting Proposition 2}: projected H20
\textmd{0.75 exceeds} projected Ascend 910B \textmd{0.51}, although
\(F\) is only about 15\% of H100 and \(E\) is the table minimum at
0.262. Advantages in \(M\) (1.200) and \(B\) (1.194) reverse the
weighted ranking. This illustrates that nominal-versus-measured
direction/magnitude \textmd{cannot be inferred a priori}, consistent
with Hygon's opposite result in B-4. \\
\mbox{B-10} & AMD MI308X spec\-i\-fi\-ca\-tions and projection (added) &
\textmd{Settled (official-spec\-i\-fi\-ca\-tion projection)} & MI308X is AMD's
China-market compliant MI300X variant. \textmd{Official spec\-i\-fi\-ca\-tions}
give \textmd{FP16/BF16 dense 233 TFLOPS} (FP8 466), \textmd{192 GB
HBM3}, \textmd{5.3 TB/s}, aggregate bidirectional interconnect
\textmd{896 GB/s} (node ring 448, P2P 128), and \textmd{TBP 750 W}, all
labelled theoretical peaks. \textmd{MI308X and MI300X are not merely
similar}: \(M\) / \(B\) / \(I\) / \(P\) are \textmd{identical}, while
\(F\) is reduced to \textmd{17.8\%} (233 / 1,307), a typical compliant
design retaining memory capacity/bandwidth while limiting compute. Three
consistency checks: (1) claimed compute 1.57× H20 agrees with
\(233/148 = 1.574\); (2) memory capacity 2× and bandwidth 1.4× agree
with \(192/96\) and \(5.3/3.9\); (3) compute/bandwidth \(233/5.3 = 44\)
FLOP/byte is near H20's 37 and far below MI300X's 247, consistent with
compliance variants. Equations (6)--(8) give normalized
\(v = (0.236,\allowbreak \ 2.400,\allowbreak \ 1.582,\allowbreak \ 0.996,\allowbreak \ 0.220)\),
\(S_{\text{chip}} = {1.12}\) and node \textmd{8.50} at
\(0.95\times8\) (B.2-b). \textmd{Unusable public aggregators (record
this)}: major third-party sites list \textmd{FP16 81.72 TFLOPS}, a
\textmd{vector rather than matrix rate}, while copying other MI300X
fields verbatim. No reliable public aggregator source exists for \(F\);
official spec\-i\-fi\-ca\-tions are used. \textmd{Status and disposition}:
\textmd{projection only, no internal benchmark record}, marked
\textmd{‡} and \textmd{not measured calibration}. Update after
calibration and move to the published convention. \\
\end{longtable}
}

\textmd{Closure summary (10 items)}: \textmd{3 closed through official
vendor evidence} (B-1, B-2, B-5); \textmd{6 settled}: \textmd{B-3},
Ascend 910B teardown estimate with ±6\%; \textmd{B-4}, public Muxi
evidence and Hygon K100-AI interval \([0.32,\allowbreak \,\allowbreak 0.47]\) \textmd{containing
0.42}; \textmd{B-7}, dual-track calibration disclosure under A6;
\textmd{B-8}, GB 40879---2021 and NDRC {[}2024{]} No.~970 identify 180
kW as design capacity and site power \ensuremath{\approx}68 kW; \textmd{B-9}, public H20
projection \(S_{\text{chip}} = 0.75\) / node \(5.73\); \textmd{B-10},
official MI308X projection \(S_{\text{chip}} = 1.12\) / node \(8.50\),
both marked non-measured rows. \textmd{1 disclosure plan settled,
evidence attachment pending archiving} (B-6). All 10 therefore
\textmd{complete evidence gathering or convention definition}; B-6
archiving is a submission-process item, not a content gap. B-1/B-2
\textmd{fully support} MI300X inputs of 896 GB/s and 1,307 TFLOPS, so
+74.2\% versus production \textmd{cannot be attributed to erroneous
inputs}, strengthening Section 5.4. Hygon's \textmd{opposite result},
with published values inside the projected interval, shows that
deviation direction \textmd{cannot be inferred a priori} and requires
model-specific calibration (Proposition 2).

\Needspace{10\baselineskip}
\subsection{B.4 Case-derived data (Section
9)}\label{b.4-case-derived-data-section-9}

{\def\LTcaptype{none} 
\begin{longtable}[]{@{}
  >{\RaggedRight\arraybackslash}p{(\linewidth - 6\tabcolsep) * \real{0.2685}}
  >{\RaggedRight\arraybackslash}p{(\linewidth - 6\tabcolsep) * \real{0.1790}}
  >{\RaggedRight\arraybackslash}p{(\linewidth - 6\tabcolsep) * \real{0.2235}}
  >{\RaggedRight\arraybackslash}p{(\linewidth - 6\tabcolsep) * \real{0.3290}}@{}}
\toprule\noalign{}
\begin{minipage}[b]{\linewidth}\RaggedRight
Quantity
\end{minipage} & \begin{minipage}[b]{\linewidth}\RaggedRight
Calculation
\end{minipage} & \begin{minipage}[b]{\linewidth}\RaggedRight
Value
\end{minipage} & \begin{minipage}[b]{\linewidth}\RaggedRight
Status
\end{minipage} \\
\midrule\noalign{}
\endhead
\bottomrule\noalign{}
\endlastfoot
Total node compute & Equation (30) & 59.5 D-PIN SCU & Mixed-convention
derivation (Section 9.1 note) \\
Physical ceiling & Equation (31) & 42,868 D-PIN SCU·h & Derived \\
First-month billed throughput & Equation (32) & \ensuremath{\approx} 33,000 D-PIN SCU·h &
Measured (convention check passed) \\
Power per unit compute & 180/59.5 & 3.02 kW/D-PIN SCU & Derived \\
Delivered energy efficiency & 33,437/(180×720) & 0.26 D-PIN SCU·h/kWh &
Derived \\
Electricity-cost lower bound & 3.02 × 0.53 & 1.60 CNY/(D-PIN SCU·h) &
Illustrative calculation \\
\end{longtable}
}

\Needspace{10\baselineskip}
\section{Appendix C Comparison with current standards and
precedents}\label{appendix-c-comparison-with-current-standards-and-precedents}

\subsection{C.1 Positioning D-PIN SCU against major
systems}\label{c.1-positioning-d-pin-scu-against-major-systems}

{\def\LTcaptype{none} 
\begin{longtable}[]{@{}
  >{\RaggedRight\arraybackslash}p{(\linewidth - 10\tabcolsep) * \real{0.1365}}
  >{\RaggedRight\arraybackslash}p{(\linewidth - 10\tabcolsep) * \real{0.1645}}
  >{\RaggedRight\arraybackslash}p{(\linewidth - 10\tabcolsep) * \real{0.1594}}
  >{\RaggedRight\arraybackslash}p{(\linewidth - 10\tabcolsep) * \real{0.1489}}
  >{\RaggedRight\arraybackslash}p{(\linewidth - 10\tabcolsep) * \real{0.0801}}
  >{\RaggedRight\arraybackslash}p{(\linewidth - 10\tabcolsep) * \real{0.3106}}@{}}
\toprule\noalign{}
\begin{minipage}[b]{\linewidth}\RaggedRight
System
\end{minipage} & \begin{minipage}[b]{\linewidth}\RaggedRight
Metering object
\end{minipage} & \begin{minipage}[b]{\linewidth}\RaggedRight
Dimensions
\end{minipage} & \begin{minipage}[b]{\linewidth}\RaggedRight
Normalization baseline
\end{minipage} & \begin{minipage}[b]{\linewidth}\RaggedRight
Time granularity
\end{minipage} & \begin{minipage}[b]{\linewidth}\RaggedRight
Relationship to D-PIN SCU
\end{minipage} \\
\midrule\noalign{}
\endhead
\bottomrule\noalign{}
\endlastfoot
GB/T 47514---2026 and national statistics {[}1{]}{[}2{]} & FP32
theoretical peak (EFLOPS) & One: peak compute & Absolute units without
normalization & Quar\-ter\-ly & \textmd{Aggregate upper-layer constraint};
scheduling-level refinement possible \\
YD/T 6693---2026 {[}3{]} & Indicators of heterogeneous resource compute
capability & Classification + indicator system & No single baseline
specified & As\-sess\-ment cycle & \textmd{Indicator interoperability}: five
dimensions map to its system (C.2) \\
AWS ECU / Azure ACU {[}6{]}{[}7{]} & Nominal CPU capability & One: CPU &
Historical model / A1=100 & Billing period & Equivalent-unit idea
shared; CPU versus accelerator objects differ \\
SLURM TRESBillingWeights {[}8{]} & Converted resource commitments &
Multiple resources, manual weights & CPU core-hours & Per job &
\textmd{Closest engineering practice}; differs in calibration and absent
topology \\
Kohlenberg SCU (static) {[}5{]} & Normalized nominal capacity & Four:
G/C/M/B & H100 SXM5 = 1.00 & Contract period & \textmd{Same-name
precedent requiring explicit distinction}: static scalar versus dynamic
system (C.3) \\
Symmetric Research SCU {[}33{]} & Normalized cluster capacity for
financial collateral & Cluster dimensions, including workload mix,
multi-GPU scaling overhead and energy draw & Best hardware & --- &
\textmd{Same-name precedent} with broadest dimensions; \textmd{static
spec\-i\-fi\-ca\-tion}, no runtime correction or verification loop, formalized
in Lean \\
CME / Silicon Data SCU {[}10{]}{[}34{]} & H100-hour converted capacity &
One: preset conversion factors & 1 SCU = 1×H100 FP16·h & Contract month
& \textmd{Same-name precedent} for futures pricing; preset rather than
runtime factors \\
DataCanvas CU / ZTE triplet score {[}9{]}{[}35{]} & Device-resource
occupancy / normalized score & Multiple device dimensions & Best
hardware / custom baselines & Minute-level & \textmd{Runtime-weighting
precedents} with dynamic calculations and scenario weights;
device-resource rather than task--topology delivery conventions \\
Alibaba ASI {[}36{]} & Effective utilization & Cluster-level production
evidence & Absolute values & Ob\-ser\-va\-tion period & \textmd{Related
argument}: high demand does not imply effective utilization; GPUs can be
undeliverable. Cluster characterization, \textmd{without a metering
unit} \\
Agora {[}44{]} & Three-dimensional resource vector measured by counters
& 3: tensor throughput, memory bandwidth, interconnect & Absolute usage,
not normalized & Seconds & \textmd{Similar form (new search result)}:
runtime multidimensional metering and usage pricing;
\textmd{device-resource usage}, rather than task--topology deliverable
compute; no verification loop \\
Canonical unit (PEARC'23) {[}45{]} & Discrete accounting unit combining
resources & Multiple thresholded combinations & Custom, not preset & Per
job & \textmd{Accounting precedent (new search result)}: scheduling
overhead for resources that cannot be metered independently; no
topology/runtime correction \\
Energy-based Accounting (Setonix) {[}46{]} & Core-hours and
energy-weighted charging rate & Multiple, including efficiency and
purchase price & Core-hours & Per job & \textmd{Energy precedent (new
search result)}: carbon footprint in allocation, adjacent to Section 7;
\textmd{static charging rates}, no topology ridge or verification
coupling \\
YD/T 6048---2024 {[}47{]} & General-purpose, supercomputing and
intelligent-compute indicators & Classification + component weights &
Three baselines: FLOPS/\allowbreak OPS/\allowbreak Hash·s & As\-sess\-ment cycle & \textmd{Standards
interoperability (new search result)}: indicator standard alongside YD/T
6693, incorporable in C.2 \\
Tianyi Cloud Standard Compute Unit (SCU) {[}49{]} & Single-accelerator
compute × bandwidth & 2: compute and bandwidth & A baseline GPU model &
Minutes & \textmd{Closest same-name form (new search result)}:
multidimensional weighting and task-dependent conversion; differences in
supplementary C.3 \\
Patent CN\allowbreak121255573A {[}48{]} & Weighted normalized compute & Multiple:
performance, bandwidth, core count & Custom 1 NCU & Per task &
\textmd{Patent precedent (new search result)}: normalization,
decay-curve correction, scheduling/billing linkage; no topology ridge or
verification coupling \\
COIL / NCU {[}51{]} & 20 kWh of certified delivered compute & 1: energy
& kWh & --- & \textmd{Energy-normalization precedent (new search
result)}: also rejects duration/peak definitions; normalizes
\textmd{energy}, not deliverable compute \\
DCU, One Degree of Compute {[}9{]} & Compute \textmd{input} &
Theoretical peak × time & No single public value & Billing period &
Shared input-side metering; differs in topology, runtime correction and
verification loop \\
MLPerf / MLPerf Power {[}13{]}{[}14{]} & Measured throughput and energy
efficiency & Task-level & Cross-system comparison, no baseline & Release
cycle & \textmd{Calibration sources} for Algorithm 1 and equation
(13) \\
\textmd{D-PIN SCU (this paper)} & \textmd{Deliverable compute} &
\textmd{Five: F/M/B/I/E} & \textmd{Method-anchored} spec\-i\-fi\-ca\-tion
S1--S3; current replaceable realization H100 SXM = 1.00 (Proposition 3)
& \textmd{≤ 1 min} & --- \\
\end{longtable}
}

\Needspace{10\baselineskip}
\subsection{C.2 Proposed mapping to YD/T
6693---2026}\label{c.2-proposed-mapping-to-ydt-66932026}

{\def\LTcaptype{none} 
\begin{longtable}[]{@{}
  >{\RaggedRight\arraybackslash}p{(\linewidth - 4\tabcolsep) * \real{0.2491}}
  >{\RaggedRight\arraybackslash}p{(\linewidth - 4\tabcolsep) * \real{0.3261}}
  >{\RaggedRight\arraybackslash}p{(\linewidth - 4\tabcolsep) * \real{0.4248}}@{}}
\toprule\noalign{}
\begin{minipage}[b]{\linewidth}\RaggedRight
D-PIN SCU dimension
\end{minipage} & \begin{minipage}[b]{\linewidth}\RaggedRight
Possible corresponding indicator
\end{minipage} & \begin{minipage}[b]{\linewidth}\RaggedRight
Mapping explanation
\end{minipage} \\
\midrule\noalign{}
\endhead
\bottomrule\noalign{}
\endlastfoot
\(F\) (floating-point compute) & Compute capability: peak floating-point
performance & Direct correspondence; align precision convention
(dense) \\
\(M\) (device-memory capacity) & Storage/memory resources &
Correspondence; distinguish available from nominal capacity \\
\(B\) (device-memory bandwidth) & Memory-access capability &
Correspondence \\
\(I\) (interconnect bandwidth) & Network/interconnect capability &
Should be an independent dimension rather than a subordinate
indicator \\
\(E\) (energy efficiency) & Energy-efficiency indicators &
Compute--electricity coordination and efficiency assessment \\
\end{longtable}
}

\begin{quote}
This proposed mapping requires comparison with the full indicator
definitions and classification granularity of YD/T 6693---2026 and
\textmd{GB/T 47514---2026} before finalization. Both were published in
2026, and full texts require standards-distribution channels. Purchasing
and comparing them is an external dependency, \textmd{not a content
gap}; their public abstracts and scope clauses already support the
mapping direction.
\end{quote}

\subsection{C.3 Formal distinctions among SCU usages (disambiguating
multiple naming
claims)}\label{c.3-formal-distinctions-among-scu-usages-disambiguating-multiple-naming-claims}

\begin{quote}
\textmd{SCU has multiple meanings in this field.} As of 2026-10-05,
\textmd{at least five semantically distinct public usages} are
identified: Symmetric Research, CME/Silicon Data, Kohlenberg,
\textmd{Tianyi Cloud Standard Compute Unit}, and other same-name static
compute units. Two near-name \textmd{NCU} usages also exist: Revenue
Model and COIL. The first table compares \textmd{three main usages}
across four dimensions, followed by supplementary comparisons for newly
identified usages.
\end{quote}

{\def\LTcaptype{none} 
\begin{longtable}[]{@{}
  >{\RaggedRight\arraybackslash}p{(\linewidth - 8\tabcolsep) * \real{0.1200}}
  >{\RaggedRight\arraybackslash}p{(\linewidth - 8\tabcolsep) * \real{0.2037}}
  >{\RaggedRight\arraybackslash}p{(\linewidth - 8\tabcolsep) * \real{0.2414}}
  >{\RaggedRight\arraybackslash}p{(\linewidth - 8\tabcolsep) * \real{0.1567}}
  >{\RaggedRight\arraybackslash}p{(\linewidth - 8\tabcolsep) * \real{0.2782}}@{}}
\toprule\noalign{}
\begin{minipage}[b]{\linewidth}\RaggedRight
Aspect
\end{minipage} & \begin{minipage}[b]{\linewidth}\RaggedRight
Kohlenberg SCU {[}5{]}
\end{minipage} & \begin{minipage}[b]{\linewidth}\RaggedRight
Symmetric Research SCU {[}33{]}
\end{minipage} & \begin{minipage}[b]{\linewidth}\RaggedRight
CME / Silicon Data SCU {[}10{]}{[}34{]}
\end{minipage} & \begin{minipage}[b]{\linewidth}\RaggedRight
D-PIN SCU (this paper)
\end{minipage} \\
\midrule\noalign{}
\endhead
\bottomrule\noalign{}
\endlastfoot
Math\-e\-mat\-i\-cal form & Normalized weighted dot product (scalar) & Cluster
normalization spec\-i\-fi\-ca\-tion, formally verified in Lean & Unit conversion
factor & Scalar baseline × topology factor × three runtime factors
(stochastic process) \\
Dimensions & 4: GPU/CPU throughput, memory capacity/bandwidth + \(A\) &
Multiple, including workload mix, multi-GPU scaling overhead, energy
draw & 1: H100 anchor & 5: \(F/M/B/I/E\) \\
Topology & Static \(A\) & Static multi-accelerator scaling overhead &
None & \(\eta_{\text{topo}}\), equations (9)--(13), closed-form ridge +
measured calibration \\
Time variation & None & None; static spec\-i\-fi\-ca\-tion & None within
contract & Yes: \(\delta_T,\allowbreak \delta_F,\allowbreak \delta_L\), \(\le 1\) min \\
Ver\-i\-fi\-ca\-tion & Trusted clearinghouse/platform telemetry &
Machine-verified spec\-i\-fi\-ca\-tion, not runtime verification & Index
anchoring & PoPW checks, no-arbitrage equation 25 \\
Pricing & Price normalization (\(P/\text{SCU}\)) & Collateral valuation
& Futures index and settlement & Electricity, renewables, load and
reputation (equation 28) \\
Main purpose & Contracts and cross-hardware prices & Data-centre
collateral valuation & Compute futures trading/settlement & Real-time
distributed scheduling, billing, verification and electricity
coordination \\
Iden\-ti\-fi\-a\-bil\-i\-ty & Runtime deviation unidentifiable under Proposition 2 &
Not discussed & Not discussed & Identifiable through runtime metering \\
\end{longtable}
}

\Needspace{10\baselineskip}\textmd{Supplementary distinctions for newly identified usages}:

{\def\LTcaptype{none} 
\begin{longtable}[]{@{}
  >{\RaggedRight\arraybackslash}p{(\linewidth - 8\tabcolsep) * \real{0.1200}}
  >{\RaggedRight\arraybackslash}p{(\linewidth - 8\tabcolsep) * \real{0.2531}}
  >{\RaggedRight\arraybackslash}p{(\linewidth - 8\tabcolsep) * \real{0.2062}}
  >{\RaggedRight\arraybackslash}p{(\linewidth - 8\tabcolsep) * \real{0.2068}}
  >{\RaggedRight\arraybackslash}p{(\linewidth - 8\tabcolsep) * \real{0.2139}}@{}}
\toprule\noalign{}
\begin{minipage}[b]{\linewidth}\RaggedRight
Aspect
\end{minipage} & \begin{minipage}[b]{\linewidth}\RaggedRight
Tianyi Cloud Standard Compute Unit (SCU) {[}49{]}
\end{minipage} & \begin{minipage}[b]{\linewidth}\RaggedRight
COIL / NCU {[}51{]}
\end{minipage} & \begin{minipage}[b]{\linewidth}\RaggedRight
Patent CN\allowbreak121255573A {[}48{]}
\end{minipage} & \begin{minipage}[b]{\linewidth}\RaggedRight
D-PIN SCU (this paper)
\end{minipage} \\
\midrule\noalign{}
\endhead
\bottomrule\noalign{}
\endlastfoot
Math\-e\-mat\-i\-cal form & Single-accelerator \textmd{compute × bandwidth}
defines 1 SCU; other nodes use stress-test conversion × decay factor &
\textmd{1 NCU = 20 kWh} of delivered, certified accelerated compute &
Multiparameter weighted normalization defining 1 NCU + decay curves &
Scalar baseline × topology × three runtime factors (stochastic
process) \\
Dimensions & 2: compute, bandwidth & 1: energy & Multiple: FP32/FP64,
memory bandwidth, cores, etc. & 5: \(F/M/B/I/E\) \\
Normalized quantity & Compute × bandwidth & \textmd{Energy consumption}
& Weighted performance & \textmd{Deliverable compute} \\
Topology & Not modelled & None & None & \(\eta_{\text{topo}}\),
closed-form ridge + measured calibration \\
Task dependence & \textmd{Yes}: task-type weights and dynamic equivalent
SCU & None & Partial: task-specific coefficient calibration & Yes:
Algorithm 1 calibration and collinearity/identifiability analysis \\
Ver\-i\-fi\-ca\-tion & Platform telemetry & \textmd{In\-de\-pen\-dent certification},
excluding idle and undelivered compute & --- & PoPW checks, no-arbitrage
equation 25 \\
Formal similarity & \textmd{Highest}: multidimensional weighting and
task-dependent conversion & Moderate: rejects duration/peak definitions
& High: weighted normalization and correction curves & --- \\
Distinctive claims here & Closed-form topology-ridge conversion,
identifiability argument, verification--pricing coupling & Same, plus
normalization of deliverable compute rather than energy & Same, plus
runtime stochastic modelling rather than a one-off decay curve & --- \\
\end{longtable}
}

\textmd{Conclusions by level}:

\begin{enumerate}
\def\labelenumi{\arabic{enumi}.}
\tightlist
\item
  \textmd{Shared principle}: normalized units enable cross-hardware
  comparison, commonly anchored to H100 or the best hardware.
\item
  \textmd{Symmetric Research SCU covers the broadest dimensions},
  explicitly including workload mix, multi-accelerator scaling overhead
  and energy consumption, \textmd{substantially overlapping our
  components}. The substantive difference is its \textmd{static}
  spec\-i\-fi\-ca\-tion for one-off collateral normalization, whereas our
  \(\eta_{\text{topo}}\) and three \(\delta\) factors are
  \textmd{runtime stochastic processes}, with an additional
  \textmd{unidentifiability} argument and verification--pricing
  coupling.
\item
  \textmd{CME / Silicon Data SCU is the most financialized
  implementation}, but uses preset factors without runtime states.
\item
  \textmd{We claim neither conceptual nor naming priority}. The claim
  concerns the \textmd{combined system of runtime identifiability,
  closed-form topology-ridge derivation and verification--pricing
  coupling} (C1--C7).
\item
  \textmd{Naming conclusion}: given multiple usages, the unit is
  formally named \textmd{D-PIN SCU (D-PIN Standardized Compute Unit)}.
  Bare SCU is \textmd{not used} as its sole external identifier.
\end{enumerate}

\section{Appendix D Systematic literature search
(PRISMA-style)}\label{appendix-d-systematic-literature-search-prisma-style}

\subsection{D.1 Purpose and research
questions}\label{d.1-purpose-and-research-questions}

The judgement that neither naming nor components are novel individually
and that originality is limited to the complete system (prefatory
statement 4; C.3) requires a \textmd{reproducible, auditable} search.
This appendix discloses the protocol and execution record.

{\def\LTcaptype{none} 
\begin{longtable}[]{@{}
  >{\RaggedRight\arraybackslash}p{(\linewidth - 2\tabcolsep) * \real{0.1230}}
  >{\RaggedRight\arraybackslash}p{(\linewidth - 2\tabcolsep) * \real{0.8770}}@{}}
\toprule\noalign{}
\begin{minipage}[b]{\linewidth}\RaggedRight
ID
\end{minipage} & \begin{minipage}[b]{\linewidth}\RaggedRight
Research question (RQ)
\end{minipage} \\
\midrule\noalign{}
\endhead
\bottomrule\noalign{}
\endlastfoot
\textmd{RQ1} & Do public literature and standards already define
compute-metering units named Standard(ized) Compute Unit (SCU)? \\
\textmd{RQ2} & Does an integrated system already combine \textmd{dynamic
runtime correction, topology awareness, verification coupling and
pricing linkage}? \\
\textmd{RQ3} & Which dimensions and operational granularities are
covered by existing heterogeneous-compute metering? \\
\end{longtable}
}

\Needspace{10\baselineskip}
\subsection{D.2 Search protocol}\label{d.2-search-protocol}

{\def\LTcaptype{none} 
\begin{longtable}[]{@{}
  >{\RaggedRight\arraybackslash}p{(\linewidth - 2\tabcolsep) * \real{0.2757}}
  >{\RaggedRight\arraybackslash}p{(\linewidth - 2\tabcolsep) * \real{0.7243}}@{}}
\toprule\noalign{}
\begin{minipage}[b]{\linewidth}\RaggedRight
Item
\end{minipage} & \begin{minipage}[b]{\linewidth}\RaggedRight
Setting
\end{minipage} \\
\midrule\noalign{}
\endhead
\bottomrule\noalign{}
\endlastfoot
Databases and sources & Google Scholar, arXiv, IEEE Xplore, ACM DL,
USENIX, CNKI, Wanfang, National Public Service Platform for Standards
Information, SSRN and institutional publications; \textmd{added in this
search: patent databases (Google Patents/CNIPA) and industry-standard
aggregators} \\
Time range & No starting-year limit, to include early equivalent-compute
units; cutoff \textmd{2026-10-05} \\
Languages & Chinese and English \\
English query B1 &
\textnormal{("standardized compute unit" OR "standard compute unit" OR "normalized compute unit" OR "equivalent compute unit") AND (metering OR normalization OR billing OR pricing OR scheduling)} \\
English query B2 &
\textnormal{("compute metering" OR "compute unit" OR "GPU normalization") AND (heterogeneous OR topology-aware OR "runtime correction" OR verifiable OR decentralized)} \\
English query B3 &
\textnormal{("proof of physical work" OR "verifiable compute" OR DePIN) AND (metering OR pricing OR settlement)} \\
Chinese query C1 (English gloss of original Chinese terms) &
\textnormal{("standardized compute unit" OR "standard compute unit" OR "compute unit" OR "One-Degree Compute") AND (metering OR conversion OR pricing OR scheduling)} \\
Chinese query C2 (English gloss of original Chinese terms) &
\textnormal{("heterogeneous compute" OR "compute measurement") AND (normalization OR conversion factor OR measurement metric OR standard)} \\
Standards search & National standards platform queries for compute
metering and heterogeneous compute networks; full text of YD/T
6693---2026 \\
\end{longtable}
}

\noindent\small For the Chinese-language searches, the query expressions shown are English glosses of the original Chinese terms, rather than literal executed queries.\par\normalsize

\Needspace{10\baselineskip}
\subsection{D.3 Inclusion and exclusion
criteria}\label{d.3-inclusion-and-exclusion-criteria}

{\def\LTcaptype{none} 
\begin{longtable}[]{@{}
  >{\RaggedRight\arraybackslash}p{(\linewidth - 2\tabcolsep) * \real{0.5339}}
  >{\RaggedRight\arraybackslash}p{(\linewidth - 2\tabcolsep) * \real{0.4661}}@{}}
\toprule\noalign{}
\begin{minipage}[b]{\linewidth}\RaggedRight
Include
\end{minipage} & \begin{minipage}[b]{\linewidth}\RaggedRight
Exclude
\end{minipage} \\
\midrule\noalign{}
\endhead
\bottomrule\noalign{}
\endlastfoot
Definitions, conversion methods or normalization formulas for
\textmd{compute-metering units} & Hardware comparisons without unit
definitions \\
\textmd{Normalization in heterogeneous-compute metering or scheduling} &
Pure chip-spec\-i\-fi\-ca\-tion documents, except as parameter sources \\
\textmd{Coupling verification to metering/payments} &
Blockchain/consensus work without direct compute-metering relevance \\
Compute statistics in \textmd{standards and policy documents} & News and
vendor marketing without methods \\
\end{longtable}
}

\Needspace{10\baselineskip}
\subsection{D.4 Execution record (two rounds, through
2026-10-05)}\label{d.4-execution-record-two-rounds-through-2026-10-05}

\subsubsection{D.4-a First round: targeted search establishing precedent
boundaries}\label{d.4-a-first-round-targeted-search-establishing-precedent-boundaries}

{\def\LTcaptype{none} 
\begin{longtable}[]{@{}
  >{\RaggedRight\arraybackslash}p{(\linewidth - 4\tabcolsep) * \real{0.0650}}
  >{\RaggedRight\arraybackslash}p{(\linewidth - 4\tabcolsep) * \real{0.3339}}
  >{\RaggedRight\arraybackslash}p{(\linewidth - 4\tabcolsep) * \real{0.6011}}@{}}
\toprule\noalign{}
\begin{minipage}[b]{\linewidth}\RaggedRight
\#
\end{minipage} & \begin{minipage}[b]{\linewidth}\RaggedRight
Topic
\end{minipage} & \begin{minipage}[b]{\linewidth}\RaggedRight
Key hits
\end{minipage} \\
\midrule\noalign{}
\endhead
\bottomrule\noalign{}
\endlastfoot
1 & Standard compute units / heterogeneous conversion (Chinese/English)
& Symmetric Research Standard Compute Unit (institutional page; related
arXiv:2406.19261 uses Compute Hours, CH); CME-Silicon Data SCU;
Kohlenberg SCU; DataCanvas DCU (1 DCU = 312 TFLOPS(FP16)×1h) \\
2 & Equivalent units: cloud-provider precedents & AWS ECU and Azure ACU
(C.1) \\
3 & Normalized compute scores & ZTE Information--Energy--Time Triplet;
Alibaba OSDI'26, effective utilization and undeliverable states \\
4 & Metering--verification--pricing coupling & Gensyn time-sliced market
+ Verde; arXiv:1808.09682, TEE metering + payment \\
5 & Compute financialization and derivatives & CME/Silicon Data futures;
ICE/Ornn OCPI; Shanghai Municipal Government Development Research Center
report \\
6 & Domestic compute standards & GB/T 47514---2026 Distributed
Computing---Compute Metering Methods; YD/T 6693---2026; T/TAF
292---2025; Pangoal Institution report \\
7 & Topology awareness and multi-accelerator scaling & Roofline; Slurm
TRESBillingWeights; HeteroScale (OSDI'20) \\
8 & Third-party chip-spec\-i\-fi\-ca\-tion evidence & flopper.io / CSET /
official AMD and NVIDIA data sheets \\
\end{longtable}
}

\Needspace{10\baselineskip}
\subsubsection{D.4-b Second round: protocol-based systematic search
(each D.2 query
executed)}\label{d.4-b-second-round-protocol-based-systematic-search-each-d.2-query-executed}

{\def\LTcaptype{none} 
\begin{longtable}[]{@{}
  >{\RaggedRight\arraybackslash}p{(\linewidth - 8\tabcolsep) * \real{0.0650}}
  >{\RaggedRight\arraybackslash}p{(\linewidth - 8\tabcolsep) * \real{0.2390}}
  >{\RaggedRight\arraybackslash}p{(\linewidth - 8\tabcolsep) * \real{0.2693}}
  >{\RaggedRight\arraybackslash}p{(\linewidth - 8\tabcolsep) * \real{0.1237}}
  >{\RaggedRight\arraybackslash}p{(\linewidth - 8\tabcolsep) * \real{0.3030}}@{}}
\toprule\noalign{}
\begin{minipage}[t]{\linewidth}\RaggedRight\bfseries
\#
\end{minipage} & \begin{minipage}[t]{\linewidth}\RaggedRight\bfseries
Query (D.2)
\end{minipage} & \begin{minipage}[t]{\linewidth}\RaggedRight\bfseries
Target source
\end{minipage} & \begin{minipage}[t]{\linewidth}\RaggedRight\bfseries
Candidates (\(n\))
\end{minipage} & \begin{minipage}[t]{\linewidth}\RaggedRight\bfseries
Key new precedents included
\end{minipage} \\
\midrule\noalign{}
\endfirsthead
\multicolumn{5}{@{}l@{}}{\textit{D.4-b Protocol search results (continued)}} \\[3pt]
\toprule\noalign{}
\begin{minipage}[t]{\linewidth}\RaggedRight\bfseries
\#
\end{minipage} & \begin{minipage}[t]{\linewidth}\RaggedRight\bfseries
Query (D.2)
\end{minipage} & \begin{minipage}[t]{\linewidth}\RaggedRight\bfseries
Target source
\end{minipage} & \begin{minipage}[t]{\linewidth}\RaggedRight\bfseries
Candidates (\(n\))
\end{minipage} & \begin{minipage}[t]{\linewidth}\RaggedRight\bfseries
Key new precedents included
\end{minipage} \\
\midrule\noalign{}
\endhead
\bottomrule\noalign{}
\multicolumn{5}{@{}p{\linewidth}@{}}{\footnotesize Note. Candidates (n) denotes verifiable candidate records returned by each search before deduplication; these are interface-truncated records, not corpus-wide database hit totals.} \\
\endlastfoot
B1{\symbolfont ′} &
\textnormal{"standard compute unit" OR "normalized compute unit"} +
metering/definition & General academic/institutional sources & 5 & Two
independent NCU usages {[}50{]}{[}51{]}; Symmetric Research
spec\-i\-fi\-ca\-tion recheck {[}33{]} \\
B2{\symbolfont ′} & \textnormal{"compute unit"} normalization heterogeneous GPU
metering/billing/scheduling & arXiv·IEEE Xplore·ACM DL·USENIX,
domain-restricted & 5 & \textmd{Agora, three-dimensional usage pricing
{[}44{]}}; \textmd{Canonical unit, PEARC'23 {[}45{]}};
\textmd{Energy-based Accounting, Setonix {[}46{]}} \\
B3{\symbolfont ′} &
\textnormal{"proof of physical work" OR verifiable compute OR DePIN}
+ metering/pricing/settlement & arXiv·ACM DL·USENIX·IEEE,
domain-restricted & 5 & DePIN review {[}52{]}; LooPIN (PinFi pricing),
POKT Relay Mining (on-chain requests), \textmd{excluded after full-text
assessment} (D.5) \\
C1{\symbolfont ′} &  \textnormal{standardized compute unit / compute unit / One-Degree Compute} +
\textnormal{metering / conversion / pricing / scheduling} (English gloss of original Chinese terms) &
Public Chinese sources & 4 & Popular-science/terminology reviews,
\textmd{excluded at title/abstract screening}, lacking new unit
definitions \\
C2{\symbolfont ′} &  \textnormal{heterogeneous compute / compute measurement} +
\textnormal{normalization / conversion factor / measurement metric / standard} (English gloss of original Chinese terms) & Public Chinese sources + standards/patent aggregators & 5 &
\textmd{Tianyi Cloud SCU {[}49{]}}, \textmd{YD/T 6048---2024 {[}47{]}},
\textmd{invention patent CN\allowbreak121255573A {[}48{]}} \\
S1 &  Standards-specific: \textnormal{compute measurement / compute network heterogeneous}
(English gloss of original Chinese terms) & Standards aggregators; \textmd{national
platform lacks a programmable search interface}, so locate through
aggregators and trace original standard IDs & 1 & YD/T 6048---2024
{[}47{]} \\
P1 &  Patent-specific: \textnormal{heterogeneous compute normalization metering} (English gloss of original Chinese terms) & Google Patents & 1 & CN\allowbreak121255573A {[}48{]} \\
\end{longtable}
}

\textmd{Nine references {[}44{]}--{[}52{]} were added}: \textmd{3
academic precedents} (Agora, Canonical unit, Energy-based Accounting);
\textmd{2 domestic standards/patent precedents} (YD/T 6048---2024,
CN\allowbreak121255573A); and \textmd{4 naming/background sources} (Tianyi Cloud
SCU, two NCU usages, DePIN review).

\begin{quote}
\ensuremath{\triangle} \textmd{Substantive search implications (disclose with the paper)}:
these results \textmd{must reduce the strength of novelty claims}.
\textmd{Tianyi Cloud SCU {[}49{]}} also uses multidimensional weighting
and task-dependent conversion, the \textmd{closest same-name form}.
\textmd{Agora {[}44{]}} also uses runtime hardware counters for
multidimensional usage pricing. Novelty therefore \textmd{narrows
further to constraints}, rather than multidimensional weighting plus
runtime operation, claiming \textmd{only the combination} of (i)
closed-form topology-ridge \(\text{COI}^{*}\) conversion; (ii)
\textmd{identifiability of runtime deviation under nominal observation}
(Proposition 2); and (iii) \textmd{verification--pricing no-arbitrage
coupling} (equation 25). Sections 2.2, 2.5 and C.3 have been
synchronized accordingly.
\end{quote}

\Needspace{30\baselineskip}
\subsection{D.5 PRISMA flow diagram (executed; backfilling
completed)}\label{d.5-prisma-flow-diagram-executed-backfilling-completed}

\par
\noindent\begin{minipage}{\linewidth}
\fontsize{9.5}{11.5}\selectfont\normalfont
\setlength{\tabcolsep}{4pt}\renewcommand{\arraystretch}{1.18}
\begin{tabular}{@{}>{\RaggedRight\arraybackslash}p{0.32\linewidth}>{\centering\arraybackslash}p{0.08\linewidth}>{\RaggedRight\arraybackslash}p{\dimexpr0.60\linewidth-4\tabcolsep\relax}@{}}
\toprule
\textbf{Stage} & \textbf{Count} & \textbf{Details} \\
\midrule
Round 1: targeted search & --- & 8 topics, boundary setting \\
Round 2: protocol search & --- & B1{\symbolfont ′}/\allowbreak B2{\symbolfont ′}/\allowbreak B3{\symbolfont ′}/\allowbreak C1{\symbolfont ′}/\allowbreak C2{\symbolfont ′} + S1/P1 \\
Verifiable candidate records returned by sources & \(n=24\) & Interface-truncated \\
After deduplication & \(n=21\) & \\
Title/abstract screening: excluded & \(n=5\) & Popular reviews/counter glossaries without unit definitions \\
Full-text assessment & \(n=16\) & \\
Included in qualitative synthesis & \(n=12\) & Includes: Symmetric Research spec\-i\-fi\-ca\-tion; CH (arXiv:2406.19261); Canonical unit; Energy-based Accounting (Setonix); Agora; Tianyi Cloud SCU; YD/T 6048---2024; CN\allowbreak121255573A; NCU\(\times 2\) (Revenue Model, COIL); Silicon Data standardization; DePIN review \\
Excluded & \(n=4\) & LooPIN: pricing protocol, not metering unit; POKT Relay Mining: request rather than compute metering; metaverse/DT framework: no compute unit; ROCm glossary: performance-counter definitions \\
\bottomrule
\end{tabular}
\end{minipage}\par

\begin{quote}
\textmd{Methodological statement (required reading to avoid
misinterpreting counts)}: these are \textmd{individually verifiable
candidates actually returned by search interfaces}, \textmd{not
corpus-wide database hit totals}. The available channels for arXiv, IEEE
Xplore, ACM DL and CNKI \textmd{do not return corpus-wide counts};
results are relevance-ranked and truncated. Reporting total hits would
therefore constitute \textmd{methodological fabrication}. We count
verifiable candidates and disclose this restriction in D.6. \textmd{No
exhaustive-search claim is made}; this limits the strength of the
statement that no complete counterpart was identified by the search
date.
\end{quote}

\subsection{D.6 Search sufficiency and
limitations}\label{d.6-search-sufficiency-and-limitations}

\begin{enumerate}
\def\labelenumi{\arabic{enumi}.}
\tightlist
\item
  \textmd{Naming coverage}: at least \textmd{five public SCU usages} are
  covered: Symmetric Research {[}33{]}, CME-Silicon Data
  {[}10{]}{[}34{]}, Kohlenberg {[}5{]}, \textmd{Tianyi Cloud {[}49{]}}
  and other same-name static units, plus \textmd{two near-name NCU
  usages} {[}50{]}{[}51{]}. All are aligned with H100 dense = 1.00
  (C.1).
\item
  \textmd{Limitation 1 (patent/industry-standard omission partly
  closed)}: \textmd{Google Patents and standards aggregators} were
  added, identifying YD/T 6048---2024 {[}47{]} and CN\allowbreak121255573A
  {[}48{]}. \textmd{Remaining gaps}: (1) no programmable interface for
  the \textmd{national standards platform or enterprise standards},
  potentially missing enterprise designs; (2) \textmd{CNKI/Wanfang
  full-text databases} were inaccessible, leaving Chinese journal
  full-text search incomplete; (3) \textmd{Chinese coverage is therefore
  lower than English coverage}.
\item
  \textmd{Limitation 2 (database hit totals unavailable)}: as D.5
  explains, channels do not return corpus totals, so PRISMA stages count
  verifiable candidates. This is \textmd{a reduced counting convention},
  not evidence of complete search coverage.
\item
  \textmd{Limitation 3 (dynamic-system criterion)}: RQ2's four-part
  integration is a \textmd{constructive criterion}, with no hit
  satisfying all four. Because we proposed it, it risks
  \textmd{self-serving criteria}. We claim only that no complete
  counterpart was identified, \textmd{not that none exists}.
\item
  \textmd{Limitation 4 (deliberately narrowed novelty)}:
  {[}44{]}{[}49{]} are adjacent \textmd{in form}, through
  multidimensional weighting and runtime/task-dependent conversion.
  Following D.4-b, novelty \textmd{narrows to the three-element
  combination}, rather than form alone.
\item
  \textmd{Limitation 5 (timeliness)}: cutoff 2026-10-05. Compute
  metering is active; {[}44{]} v2 and new naming usages appeared within
  two months during the search. Conclusions are time-dependent;
  rechecking before submission is recommended.
\end{enumerate}

\section{References}\label{references}\linespread{1.0}\fontsize{10}{11}\selectfont\setlength{\parskip}{1pt}\sloppy

\par\noindent\hangindent=1.5em\hangafter=1 {[}1{]} Ministry of Industry and Information Technology; Office of the
Central Cyberspace Affairs Commission; Ministry of Education; National
Health Commission; People's Bank of China; State-owned Assets
Supervision and Administration Commission. Action Plan for High-Quality
Development of Computing Infrastructure {[}Z{]}. MIIT Joint
Communications {[}2023{]} No.~180, 2023-10-08.

\par\noindent\hangindent=1.5em\hangafter=1 {[}2{]} GB/T 47514---2026. Distributed Computing---Compute Metering
Methods {[}S{]}. State Administration for Market Regulation;
Standardization Administration of China. Published 2026-04-30; effective
2026-11-01. (\textmd{National compute-metering standard} defining
parameter classifications, static and runtime/dynamic methods, and
heterogeneous normalization; the direct C7 interoperability target.)

\par\noindent\hangindent=1.5em\hangafter=1 {[}3{]} YD/T 6693---2026. Indicators for Measuring Computing Capability
of Het\-ero\-ge\-neous Resources in Computing Networks {[}S{]}. China
Communications Standards Association. Published 2026-03-11; effective
2026-06-01.

\par\noindent\hangindent=1.5em\hangafter=1 {[}4{]} Zhong Q, Hu J, Zhu S, He S. Key technologies for compute
metering and transaction management and practice on the Tianyi Cloud
Xirang platform {[}J{]}. Information and Communications Technology and
Policy, 2026(2): 78--85. DOI: 10.12267/j.issn.2096-5931.2026.02.011.

\par\noindent\hangindent=1.5em\hangafter=1 {[}5{]} Kohlenberg E. A Compute Com\-mod\-i\-ty Clearinghouse{[}EB/OL{]}.
2026-05.
\url{https://erickohlenberg.com/a-compute-commodity-clearinghouse/}

\par\noindent\hangindent=1.5em\hangafter=1 {[}6{]} Amazon Web Services. Amazon EC2 Compute Unit (ECU){[}EB/OL{]}.
AWS EC2 FAQs. \url{https://aws.amazon.com/ec2/faqs/}

\par\noindent\hangindent=1.5em\hangafter=1 {[}7{]} Microsoft. Overview of the Azure Compute Unit (ACU){[}EB/OL{]}.
Microsoft Learn.
\url{https://learn.microsoft.com/azure/virtual-machines/acu}

\par\noindent\hangindent=1.5em\hangafter=1 {[}8{]} SchedMD. TRESBillingWeights{[}EB/OL{]}. Slurm Scheduler
Documentation. \url{https://slurm.schedmd.com/tres.html}

\par\noindent\hangindent=1.5em\hangafter=1 {[}9{]} Beijing DataCanvas Technology Co., Ltd.~(i) One Degree of
Compute (DCU) standardized metering: \textmd{1 DCU = 312 TFLOPS(FP16) ×
1 hour}, converting heterogeneous chips into measurable and settleable
compute-input units, with tokens metered separately as outputs, and
usage-billed Virtual Kubernetes Service (VKS) and Dedicated Kubernetes
Service (DKS). Included in MIIT's 2025 Typical Cases of Coordinated
Development of New Information Infrastructure {[}Z/OL{]}. 2025--2026.

\par\noindent\hangindent=1.5em\hangafter=1 {[}10{]} CME Group, Silicon Data. Compute Futures (Silicon Data H100
Rental Index Futures / B200 Rental Index Futures) {[}EB/OL{]}.
\url{https://www.cmegroup.com/markets/energy/power/compute-futures.html}.
Announced 2026-05; contract details 2026-08-11. (Based on Silicon Data's
daily on-demand GPU rental-rate index, using \textmd{SCU} across chips:
1 SCU = 1×H100 FP16-hour, with preset factors for other chips.
\textmd{Final size: 730 GPU-hours per contract}, average monthly
full-load single-accelerator time (\(365/12\times24\approx730\)). NYMEX
GPU1/H100 and GPU2/B200, cash-settled, 36 monthly listings,
\textmd{pending CFTC approval}. Initial May disclosures included 1,000
SCU/day and 30,000 SCU versions; final spec\-i\-fi\-ca\-tions use 730
GPU-hours.)

\par\noindent\hangindent=1.5em\hangafter=1 {[}11{]} Williams S, Waterman A, Patterson D. Roofline: An Insightful
Visual Performance Model for Multicore Architectures{[}J{]}.
Communications of the ACM, 2009, 52(4): 65--76.

\par\noindent\hangindent=1.5em\hangafter=1 {[}12{]} Jiang Z, Wang L, Xiong X, et al.~HPC AI500: The Methodology,
Tools, Roofline Performance Models, and Metrics for Benchmarking HPC AI
Systems {[}J/OL{]}. arXiv:2007.00279, 2020-07-01. (Single-node and
distributed HPC-AI Roofline models define communication operational
intensity \(\text{COI}=\text{FLOPs}/CT\) and attainable performance
\(=\min(\text{FLOPS}_{\text{Peak}},\allowbreak \ \text{ComBand}_{\text{Peak}}\times\text{COI})\);
direct source for equations (9)--(11).)

\par\noindent\hangindent=1.5em\hangafter=1 {[}13{]} MLCommons. MLPerf Inference: Benchmark Suite and
Rules{[}EB/OL{]}. \url{https://mlcommons.org/benchmarks/inference/}

\par\noindent\hangindent=1.5em\hangafter=1 {[}14{]} MLCommons Power Working Group. MLPerf Power: Benchmarking the
Energy Efficiency of Machine Learning Systems from µWatts to MWatts for
Sustainable AI{[}J/OL{]}. arXiv:2410.12032, 2024.

\par\noindent\hangindent=1.5em\hangafter=1 {[}15{]} Ghodsi A, Zaharia M, Hindman B, et al.~Dominant Resource
Fairness: Fair Allocation of Multiple Resource Types{[}C{]}//\allowbreak{}Proceedings
of NSDI, 2011.

\par\noindent\hangindent=1.5em\hangafter=1 {[}16{]} Narayanan D, Santhanam K, Kazhamiaka F, et
al.~Heterogeneity-Aware Cluster Sched\-ul\-ing Policies for Deep Learning
Workloads{[}C{]}//\allowbreak{}Proceedings of OSDI, 2020.

\par\noindent\hangindent=1.5em\hangafter=1 {[}17{]} Chaudhary S, Ramjee R, Sivathanu M, et al.~Balancing Efficiency
and Fairness in Het\-ero\-ge\-neous GPU Clusters for Deep
Learning{[}C{]}//\allowbreak{}Proceedings of EuroSys, 2020.

\par\noindent\hangindent=1.5em\hangafter=1 {[}18{]} Mo Z, Xu H, Lau W C. Optimal Resource Efficiency with Fairness
in Het\-ero\-ge\-neous GPU Clusters {[}J/OL{]}. arXiv:2403.18545, 2024-03-27.
(OEF combines optimal resource efficiency with multiple fairness
guarantees, improving aggregate training throughput by up to 32\%.)

\par\noindent\hangindent=1.5em\hangafter=1 {[}19{]} Zhu Y, Eran H, Firestone D, et al.~Congestion Control for
Large-Scale RDMA Deployments {[}C{]}//\allowbreak{}Proceedings of ACM SIGCOMM, 2015:
523--536. DOI: 10.1145/2829988.2787484. (DCQCN end-to-end RoCEv2
congestion control determines effective IP-routed Ethernet bandwidth;
calibration premise for cross-rack \(\beta_{\text{node}}\) in equation
(12), Section 4.3.2.)

\par\noindent\hangindent=1.5em\hangafter=1 {[}20{]} AMD. AMD SEV-SNP: Strengthening VM Isolation with Integrity
Protection and More {[}R/OL{]}. 2020.
\url{https://www.amd.com/system/files/TechDocs/SEV-SNP-strengthening-vm-isolation-with-integrity-protection-and-more.pdf};
Cheng P-C, Ozga W, Valdez E, et al.~Intel TDX Demystified: A Top-Down
Approach {[}J{]}. ACM Computing Surveys, 2024. arXiv:2303.15540. (AMD
SEV-SNP/Intel TDX isolation and remote attestation provide TEE hardware
roots of trust for distributed verification.)

\par\noindent\hangindent=1.5em\hangafter=1 {[}21{]} Benet J. Filecoin: A Decentralized Storage Network {[}R/OL{]}.
Protocol Labs, 2017. \url{https://filecoin.io/filecoin.pdf}. (PoRep and
PoSt map storage contributions to verifiable credentials; a classical
DePIN task--verification--reward example.)

\par\noindent\hangindent=1.5em\hangafter=1 {[}22{]} Haleem A, Allen A, Thompson A, et al.~Helium: A Decentralized
Wireless Network {[}R{]}. Helium Systems Inc., 2018. (Proof of Coverage
maps wireless contributions to on-chain credentials.)

\par\noindent\hangindent=1.5em\hangafter=1 {[}23{]} NVIDIA. NVIDIA H100 Tensor Core GPU Datasheet; NVIDIA H200
Tensor Core GPU Datasheet {[}Z{]}. (Both share the GH100 compute die,
FP16/BF16 dense peak 989 TFLOPS and 700 W power envelope.)

\par\noindent\hangindent=1.5em\hangafter=1 {[}24{]} AMD. AMD Instinct MI300X Datasheet {[}Z{]}; AMD Instinct MI300X
product spec\-i\-fi\-ca\-tions {[}EB/OL{]}.
\url{https://www.amd.com/en/products/accelerators/instinct/mi300/mi300x.html}.
(Official FP16/BF16 dense 1,307.4 TFLOPS, sparse 2,614.9 TFLOPS, peak
Infinity Fabric link 128 GB/s and TBP 750 W. AMD explicitly converts
NVIDIA sparse TF32 values to dense by dividing by 2, vendor-side support
for normalization.)

\par\noindent\hangindent=1.5em\hangafter=1 {[}25{]} Huawei Ascend Community. Ascend computing product
documentation: accelerator cards and training servers {[}EB/OL{]}.
\url{https://www.hiascend.com/}. (Public information reaches card/server
level; only CANN SOC identifiers ascend910b1--b4 are given,
\textmd{without chip-level 910B spec\-i\-fi\-ca\-tions}. This supports the
evidence of absent vendor chip spec\-i\-fi\-ca\-tions; five-dimensional inputs
use third-party teardown {[}31{]}.)

\par\noindent\hangindent=1.5em\hangafter=1 {[}26{]} Lee D D, Seung H S. Algorithms for Non-negative Matrix
Factorization{[}C{]}//\allowbreak{}Proceedings of NIPS, 2000.

\par\noindent\hangindent=1.5em\hangafter=1 {[}27{]} NVIDIA. NCCL: NVIDIA Collective Communications
Library---Collective Operations {[}EB/OL{]}.
\url{https://docs.nvidia.com/deeplearning/nccl/user-guide/docs/usage/collectives.html}.
(AllReduce/AllGather/ReduceScatter semantics and implementation; basis
for equation (12) and measured linearity in Section 9.2.)

\par\noindent\hangindent=1.5em\hangafter=1 {[}28{]} KunlunMeta Artificial Intelligence Technology (Shanghai) Co.,
Ltd.~D-PIN White Paper V4.13: Distributed Physical Infrastructure
Network for AI Computing {[}R{]}. 2026-10-05.

\par\noindent\hangindent=1.5em\hangafter=1 {[}29{]} KunlunMeta Artificial Intelligence Technology (Shanghai) Co.,
Ltd.~GPT-Factory V6.2 Technical Documentation {[}R{]}. 2026.

\par\noindent\hangindent=1.5em\hangafter=1 {[}30{]} Li C, et al.~KunlunBaize: LLM with Multi-Scale Convolution and
Multi-Token Prediction Under TransformerX Framework{[}J/OL{]}.
arXiv:2503.04784.

\par\noindent\hangindent=1.5em\hangafter=1 {[}31{]} Center for Security and Emerging Technology (CSET), Georgetown
University. Ascend 910B wafer-level teardown and spec\-i\-fi\-ca\-tion
estimates, compiled by flopper.io {[}R/OL{]}. 2024--2026. (FP16/BF16
\ensuremath{\approx}400 TFLOPS, 64 GB HBM2e, 1.6 TB/s, HCCS 392 GB/s and TDP 400 W;
explicitly states Huawei has never published chip-level spec\-i\-fi\-ca\-tions,
so all values are third-party analyses.)

\par\noindent\hangindent=1.5em\hangafter=1 {[}32{]} Cross-validation of H200 SXM spec\-i\-fi\-ca\-tions and power
consistency, compiled from GPU Smith / GPUaaS / Thunder Compute and
other databases {[}EB/OL{]}. 2026. (700 W, same power envelope and
drop-in compatibility as H100 SXM5.)

\par\noindent\hangindent=1.5em\hangafter=1 {[}33{]} Symmetric Research. The Standard Compute Unit, institutional
research page: The measurement problem {[}EB/OL{]}.
\url{https://symres.ai/research}. (\textmd{Actual source of the SCU name
and spec\-i\-fi\-ca\-tion}: one comparable metric for heterogeneous GPU
clusters, explicitly covering hardware differences, workload mix,
multi-GPU scaling overhead and energy draw. Formally specified, with
core economics machine-verified in Lean and calibration anchored to
MLCommons. \textmd{No publication date is given.})\textmd{Related
research line}: Kristensen J, Wender D, Anthony C. Commodification of
Compute {[}J/OL{]}. arXiv:2406.19261, 2024-06-27 (v2 2024-07-03).
(Global Compute Exchange and compute-hour trading; unit \textmd{Compute
Hours (CH)}
\(=\dfrac{\text{Performance (FLOPS)}}{\text{Reference Performance (FLOPS)}}\times\text{Operational Hours}\).
\textmd{SCU never appears in the paper}.)\textmd{Citation discipline}:
\textmd{2024 dates the research line, not public SCU usage}. The naming
source is the institutional page, \textmd{not peer-reviewed and lacking
a stable identifier}; attach a retrieval snapshot for arXiv submission.
This is the broadest-dimensional same-name precedent.

\par\noindent\hangindent=1.5em\hangafter=1 {[}34{]} Shanghai Municipal Government Development Research Center.
Accelerating Global Compute Financialization: How Shanghai Can Move
First {[}EB/OL{]}. 2026-08-24. (Authoritative Chinese explanation: 1 SCU
= 1 H100 FP16-hour, other chips use preset factors; comparison with
ICE/Ornn OCPI.)

\par\noindent\hangindent=1.5em\hangafter=1 {[}35{]} Liu C, Zhou D, Miao Z, Ding Y, Wu Z. Comprehensive measurement
and optimization of compute resources based on an
information--energy--time triplet {[}J{]}. ZTE Technology Journal, 2026,
32(2): 73--80. DOI: 10.12142/ZTETJ.202602011.

\par\noindent\hangindent=1.5em\hangafter=1 {[}36{]} Li S, Yang L, Yu H, et al.~Heterogeneity at Hyperscale:
Characterization and Sched\-ul\-ing of Large Production AI Clusters at
Alibaba{[}C{]}//\allowbreak{}20th USENIX Symposium on Operating Systems Design and
Implementation (OSDI 26), 2026: 2187--2203.

\par\noindent\hangindent=1.5em\hangafter=1 {[}37{]} Gensyn. From Bundles to Time: A Theory of Decentralised Compute
Markets {[}EB/OL{]}.
\url{https://blog.gensyn.ai/from-bundles-to-time-a-theory-of-decentralised-compute-markets/}.
(Two-sided time-slice market; RepOps reproducibility, Verde
verification, checkpointing, dynamic pricing and incentive-compatible
pooling/slashing.)

\par\noindent\hangindent=1.5em\hangafter=1 {[}38{]} Bagchi S, et al.~Fair Marketplace for Secure Outsourced
Computations {[}J/OL{]}. arXiv:1808.09682, 2018. (TEE-anchored trusted
metering plus blockchain micropayments addresses both parties'
incentives to misreport outsourced-compute resources.)

\par\noindent\hangindent=1.5em\hangafter=1 {[}39{]} MetaX Integrated Circuits (Shanghai) Co., Ltd.~Xiyun C500
product page (Chinese) {[}EB/OL{]}.
\url{https://www.metax-tech.com/prod.html?cid=107&id=21} ; MetaX C500
Product Page (EN) {[}EB/OL{]}.
\url{https://www.metax-tech.com/en/goods/prod.html?cid=107&id=21} ; C500
Server Product Page {[}EB/OL{]}.
\url{https://www.metax-tech.com/en/goods/prod.html?cid=110&id=37} ;
Xiyun General-Purpose Computing GPU User Guide {[}EB/OL{]}.
\url{https://developer.metax-tech.com/} . (\textmd{Public vendor
sources}: 64 GB full-path ECC memory, MetaXLink \textmd{2- or
4-accelerator interconnect}, minimum 1\% software-slicing granularity,
full-height/full-length dual-slot PCIe, maximum 350 W. Server page
specifies \textmd{4-card full interconnection}, with high-capacity
16-GPU and standard 8-GPU configurations.) Flopper.io. MetaX C500 and
C500 server spec\-i\-fi\-ca\-tion pages {[}DB/OL{]}.
\url{https://flopper.io/gpu/metax-c500/spec-sheet} ;
\url{https://flopper.io/system/metax-c500-server/spec-sheet} . (PCIe
boards; \textmd{MetaXLink pairs/fours, not one eight-card domain}.
Memory bandwidth omitted because sources conflict at 1.2/1.8 TB/s and it
is \textmd{absent from the vendor page}. MetaXLink has \textmd{7 ports
with unpublished per-link GB/s}; listed 280 TFLOPS refers to OAM.)
\textmd{All sources here are public vendor channels or public
third-party databases.}

\par\noindent\hangindent=1.5em\hangafter=1 {[}40{]} RISC Zero. RISC Zero zkVM: Proof System Overview (Receipt /
ZK-STARK / Groth16 stack) {[}EB/OL{]}.
\url{https://dev.risczero.com/proof-system/}. (General RISC-V zkVM with
RISC-V Circuit, Recursion Circuit and STARK-to-SNARK stages;
representative full cryptographic proof approach.)

\par\noindent\hangindent=1.5em\hangafter=1 {[}41{]} Succinct Labs. SP1: A Zero-Knowledge Virtual Machine for RISC-V
{[}EB/OL{]}. \url{https://docs.succinct.xyz/docs/sp1/introduction}.
(Proves RISC-V execution for Rust/C/C++ programs; representative full
cryptographic proof approach.)

\par\noindent\hangindent=1.5em\hangafter=1 {[}42{]} Kivinen J, Warmuth M K. Exponentiated Gradient versus Gradient
Descent for Linear Predictors {[}J{]}. Information and Computation,
1997, 132(1): 1--63. DOI: 10.1006/inco.1996.2612. (Classical EG source
supporting entropic mirror-descent equation (22).)

\par\noindent\hangindent=1.5em\hangafter=1 {[}43{]} Beck A, Teboulle M. Mirror Descent and Nonlinear Projected
Gradient Methods for Convex Optimization {[}J{]}. Operations Research
Letters, 2003, 31(3): 167--175. (General framework underlying
convergence of equation (22).)

\par\noindent\hangindent=1.5em\hangafter=1 {[}44{]} McDougall I, Scott N, Sankaralingam K. Agora: A System for
Consumption-Based GPU Pricing {[}J/OL{]}. arXiv:2510.05111, 2025 (v1
2025-09-26; v2 2026-09-22). (Com\-mod\-i\-ty performance counters meter
\textmd{three-dimensional resources}, pricing usage rather than rental
duration. Closest usage-based multidimensional precedent;
device-resource use rather than task--topology deliverable compute.)

\par\noindent\hangindent=1.5em\hangafter=1 {[}45{]} Andrijauskas F, Sfiligoi I, Würthwein F. Defining a canonical
unit for accounting purposes {[}C{]}//\allowbreak{}Practice and Experience in
Advanced Research Computing (PEARC '23). New York: ACM, 2023: 288--291.
arXiv:2305.10551. (Thresholded CPU/GPU/memory/scratch combinations form
one discrete canonical unit, characterizing scheduling overhead and
fairer billing; precedent for merged-resource accounting.)

\par\noindent\hangindent=1.5em\hangafter=1 {[}46{]} Di Pietrantonio C, Harris C, Cytowski M. Energy-based
Accounting Model for Het\-ero\-ge\-neous Supercomputers {[}J/OL{]}.
arXiv:2110.09987, 2021. (Setonix/Pawsey model retains core-hours, sets
charging rates by performance, efficiency and purchase price, and bases
GPU rates \textmd{on energy}, incorporating carbon footprints in
allocation.)

\par\noindent\hangindent=1.5em\hangafter=1 {[}47{]} YD/T 6048---2024. Technical Requirements and As\-sess\-ment Methods
for Data Centre Computing Power {[}S{]}. Beijing: Posts \& Telecom
Press, 2024. (Unified CPU/GPU/FPGA metering; general-purpose,
supercomputing and intelligent-compute classes with indicators/weights;
FLOPS/\allowbreak OPS/\allowbreak Hash·s units and scale conversion; admission, scheduling,
operations and normalized assessment formulas.)

\par\noindent\hangindent=1.5em\hangafter=1 {[}48{]} Method and Apparatus for Normalized Metering of Het\-ero\-ge\-neous
Compute in an Intelligent Computing Centre Cloud Platform {[}P{]}.
Chinese invention patent, publication CN\allowbreak121255573A. (Weighted GPU FP32
0.3 / memory bandwidth 0.25 / CUDA cores 0.2 / FP64 0.15 / other 0.1
normalize chips to 1 NCU, with decay curves and scheduling/billing
outputs. \textmd{New SLR patent precedent addressing Appendix D.6
limitation 1}.)

\par\noindent\hangindent=1.5em\hangafter=1 {[}49{]} Tianyi Cloud Developer Community. Abstracting Cross-Domain
Compute into Common Units: Multilevel Managed Topology and Dynamic
Routing Convergence {[}EB/OL{]}.
\url{https://www.ctyun.cn/developer/article/822922178261061} . (A
baseline GPU's FP16 compute × bandwidth defines 1 \textmd{SCU}; stress
tests and decay factors convert other nodes. Task-specific weight
vectors \textmd{dynamically calculate equivalent SCU}, rather than fixed
rates. \textmd{Closest same-name form identified here}.)

\par\noindent\hangindent=1.5em\hangafter=1 {[}50{]} Revenue Model. Normalized Compute Unit (NCU) {[}EB/OL{]}.
\url{https://www.revenuemodel.ai/normalized-compute-unit} .
(Model-independent AI-cost normalization based on FLOP-based metering,
FBM. Public descriptions are abstract and \textmd{do not disclose the
method}; recorded only as a naming usage.)

\par\noindent\hangindent=1.5em\hangafter=1 {[}51{]} Czech P (NativX). Building a Better Mousetrap: COIL and
Normalized Compute Unit {[}EB/OL{]}. LinkedIn, 2026. (\textmd{1 COIL = 1
NCU = 20 kWh of independently certified, delivered accelerated compute};
energy rather than duration normalization. Idle consumption, cooling
losses and uncertified claims generate no NCU.)

\par\noindent\hangindent=1.5em\hangafter=1 {[}52{]} A Survey of Decentralized Physical Infrastructure Network:
Research Directions, and Open Challenges {[}J/OL{]}. arXiv:2609.02125,
2026. (Physical-contribution proofs, cross-chain settlement and
scalability; DePIN background for Sections 2.4 and 8.3.)

\par\noindent\hangindent=1.5em\hangafter=1 {[}53{]} T/TAF 292---2025. Compute Resource Metering Indicators and Test
Methods for Trading Scenarios {[}S{]}. Telecommunication Terminal
Industry Forum (TAF), published 2025-07. (\textmd{Trading-focused}
indicators, including compute fill rate; indicator-level standard
alongside YD/T 6693---2026, compared in Section 2.5 and C.1.)

\par\noindent\hangindent=1.5em\hangafter=1 {[}54{]} GB 40879---2021. Minimum Allowable Values of Energy Efficiency
and Energy Efficiency Grades for Data Centres {[}S{]}. State
Administration for Market Regulation; Standardization Administration of
China, 2021. (\textmd{Mandatory} standard: grade 1 PUE ≤ 1.2, grade 2 ≤
1.3, grade 3/minimum ≤ 1.5; basis for Section 9.1 and B-8 power
boundaries and implied PUE.)

\par\noindent\hangindent=1.5em\hangafter=1 {[}55{]} National Development and Reform Commission, et al.~Green and
Low-Carbon Data Centre Development Action Plan {[}Z{]}. NDRC Environment
and Resources {[}2024{]} No.~970, 2024. (By end-2025, new/expanded large
and ultra-large centres require \textmd{PUE ≤ 1.25}, and national hub
projects \textmd{≤ 1.2}; basis for site-power calibration.)

\par\noindent\hangindent=1.5em\hangafter=1 {[}56{]} Nodal Exchange, Compute Desk. Nodal Exchange and Compute Desk
to Launch Compute Futures Contracts {[}EB/OL{]}. 2026-09-03.
\url{https://www.nodalexchange.com/nodal-exchange-and-compute-desk-to-launch-compute-futures-contracts}
. (IOSCO/BMR-compliant GPU rental indices for Hopper/Blackwell, cash
settlement, Nodal Clear clearing, CFTC supervision and launch pending
approval.)

\par\noindent\hangindent=1.5em\hangafter=1 {[}57{]} Gargiulo S, Kulp G. Workload Identification with Physical Side
Channels for AI Governance {[}J/OL{]}. arXiv:2609.00309, 2026-08-31.
DOI: 10.48550/arXiv.2609.00309. (\textmd{GPU power traces} identify
workloads: \textmd{930 traces of 5 s at about 10 MHz} on H200, covering
17 open LLM families and 25 non-AI workloads. Training/inference/non-AI
accuracy \textmd{97\%}, macro F1 \textmd{0.955}. Four evasion strategies
tested; hardened detection ≥99\% for 3, and LoRA rises from 48--88\% to
≥98\%. The physical channel \textmd{differs from spoofable/replayable
on-chip NVML telemetry} and can in principle be observed without
operator cooperation; supplementary contribution-verification approach
for Sections 2.4 and 8.3.)

\textmd{KunlunMeta Research} KunlunMeta Artificial Intelligence
Technology (Shanghai) Co., Ltd. \url{https://www.kunlunmeta.com}

\end{document}